\documentclass[a4paper,11pt]{article}
\usepackage{heppub}

\usepackage{xspace}
\usepackage{placeins}
\usepackage{wasysym}
\usepackage{slashed}
\usepackage{empheq}
\usepackage[usenames,dvipsnames]{xcolor}
\usepackage[normalem]{ulem}
\usepackage{graphicx} 
\usepackage{caption} 
\usepackage{subcaption} 
\usepackage{comment}
\usepackage{enumitem}

\newcommand{\nn}{\nonumber}

\newcommand{\as}{\alpha_s}
\newcommand{\nbar}{\bar n}
\newcommand{\Gcusp}{\Gamma{_\mathrm{cusp}}}
\newcommand{\eps}{\epsilon}

\newcommand{\cN}{\mathcal{N}}

\newcommand{\Tr}{\operatorname{Tr}}
\newcommand{\df}{\mathrm{d}}

\newcommand{\ord}[1]{\mathcal{O}(#1)}

\newcommand{\Mae}[3]{\big\langle#1\big\lvert#2\big\rvert#3\big\rangle}

\usepackage{marginnote}
\usepackage[normalem]{ulem}

\title{Strongly Coupled Soft Functions}
\author[a]{Bruno Scheihing-Hitschfeld}
\author[b,c]{and Zhiquan Sun}

\affiliation[a]{Kavli Institute for Theoretical Physics, University of California, Santa Barbara, California 93106, USA}
\affiliation[b]{Leinweber Institute for Theoretical Physics,\,University of California,\,Berkeley,\,CA\,94720,\,USA}
\affiliation[c]{Theory Group,\,Physics Division,\,Lawrence Berkeley National Laboratory,\,Berkeley,\,CA\,94720,\,USA}

\emailAdd{bscheihi@kitp.ucsb.edu}
\emailAdd{zqsun@berkeley.edu}

\date{August 10, 2026}

\abstract{%
The renormalization of operators built out of Wilson lines that meet at an angle (cusp) involve what is known as the cusp anomalous dimension, a universal object appearing in many processes in QCD due to the divergences from gluonic interactions. 
In this paper, we consider the vacuum expectation value of a pair of cusped Wilson lines separated in the transverse direction. 
This configuration can be related to a simple observable in heavy quark transverse momentum-dependent (TMD) fragmentation as well as the TMD soft function.
We compute the expectation values of Wilson loops in $\mathcal{N}=4$ super Yang-Mills theory at strong coupling via the AdS/CFT correspondence, an approach complementary to perturbative calculations of the cusp anomalous dimension. 
We first present a thorough analysis, directly in Minkowski signature, of the Nambu-Goto action and its saddle points for a Wilson line configuration with a single cusp. 
We find that there are two different classes of saddle points whose contributions dominate different regions of parameter space.
We calculate the cusp anomalous dimension $\Gamma_{\rm cusp}[\Delta\eta, \Delta\theta]$ for the whole range of $\Delta\eta$ from this setup, focusing on the cases $\Delta \theta = 0$ and $\Delta \theta = \pi$, and compare it with previous results in literature. 
We then use the techniques we developed to compute the expectation value of the two-cusp Wilson loop with transverse separation, 
and determine the profile of the transverse coordinate on the extremal surface in the large rapidity limit.
We discuss possible generalizations of our setup and results to soft functions in other gauge theories with a holographic dual.
}

\begin{document}

\maketitle

\section{Wilson loops: from heavy quark fragmentation to holography}

Recently, a simple observable for heavy quark transverse momentum-dependent (TMD) fragmentation process was proposed in terms of a vacuum matrix element of cusped Wilson lines~\cite{vonKuk:2023jfd,vonKuk:2024uxe},
\begin{align}
\label{eq:chi1_matrix_element}
    \chi_1(b_\perp)= \frac{1}{N_c} \Tr
\Mae{0}{W^\dagger(b_\perp) \,
   Y_v(b_\perp) \, Y_v^\dagger(0) \,
W(0)}{0} 
\,,\end{align}
where $W(x)$ ($Y_v (x)$) is the lightlike (timelike) Wilson line, defined as
\begin{align}
    W(x) &= \bar{\mathrm P} \Bigl[ \exp \Bigl(
   - i g \int_0^\infty \! \df s \, \nbar \cdot A(x + \nbar s)
\Bigr) \Bigr]
\,,
\\
Y_v(x) &= \bar{\mathrm P} \Bigl[ \exp \Bigl(
   -i g \int_0^\infty \! \df s \, v \cdot A(x + v s)
\Bigr) \Bigr]
\,,\end{align}
where $\nbar$ is the lightcone direction and $v$ is the direction of the heavy quark.
The anti-path ordering operator $\bar{\mathrm{P}}$ arranges the color matrices from the left to the right. Note that mathematically Hermitian conjugation changes anti-path ordering $\bar{\mathrm{P}}$ to path ordering $\mathrm{P}$ and $i$ to $-i$. 

\begin{figure}
    \centering
    \includegraphics[width=0.7\linewidth]{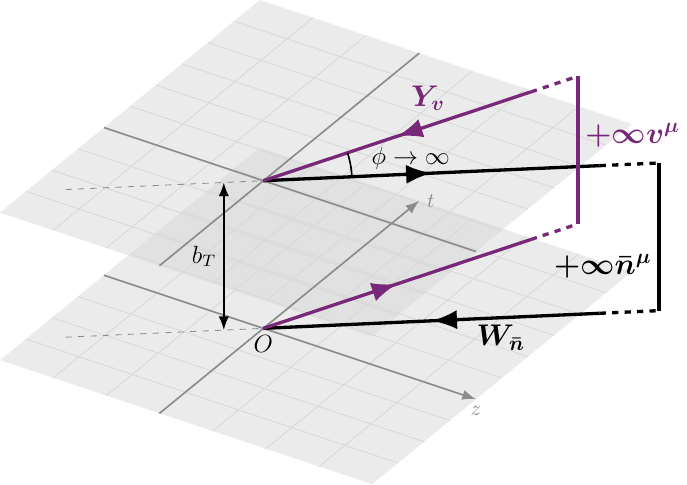}
    \caption{The Wilson line configuration that defines $\chi_1(b_T)$. We have two lightlike (timelike) Wilson lines separated by $b_T$ in the transverse direction, extending to $t=+\infty$ along $\nbar^\mu$ ($v^\mu$), and connected at $t=+\infty$ by a transverse Wilson line segment.
    Each $\nbar$-Wilson line forms a cusp with a $v$-Wilson line at $t=0$, with a cusp angle defined by $\cosh\phi = \frac{\nbar\cdot v}{\sqrt{\nbar^2 v^2}} = \infty$, 
    as lightlike $\nbar^2=0$.
    Therefore, we have two cusps transversely separated by $b_T$.
    The arrows on the Wilson lines indicate color flow, meaning the color matrices are arranged along the direction of the arrow.}
    \label{fig:Wilson_chi1}
\end{figure}

For gauge invariance, each Wilson lines and its conjugate must be connected at infinity by transverse gauge links~\cite{Boussarie:2023izj,Belitsky:2002sm}. 
We will discuss this subtlety in more detail in \sec{trans_gauge_link}.
Therefore the Wilson loop has the configuration of two infinitely long staples, each separated in the transverse direction by $b_\perp$ and connected at the origin to form a cusp angle defined by $\cosh \phi = \frac{\nbar \cdot v}{\sqrt{\nbar^2 v^2}}$, see~\fig{Wilson_chi1}.
Note that $\phi$ is equivalent to the change in rapidity $\Delta\eta$. 
When one of the Wilson lines is along the lightcone, we have a divergence in rapidity $\Delta\eta\to \infty$ as $\nbar^2 = 0$.
This rapidity divergence demands additional regularization in perturbation theory~\cite{Collins:2003fm,Collins:2008ht}, 
and induces what is known as the rapidity renormalization group evolution~\cite{Collins:2011ca,Chiu:2011qc},
which is ubiquitous in TMD physics (for a comprehensive review of TMD physics, see~\refcite{Boussarie:2023izj}), and generally Soft-Collinear Effective Theory (SCET)~\cite{Bauer:2000ew,Bauer:2000yr,Bauer:2001ct,Bauer:2001yt}. 
For a nice discussion of rapidity divergences, see e.g.~\refcite{Vladimirov:2017ksc}.
In what follows, we will effectively regulate this rapidity divergence using off-lightcone regularization in the timelike direction, i.e. we start with two timelike directions $v_1$ and $v_2$ and take the large rapidity limit $\Delta\eta \to \infty$ to recover the timelike-lightlike behavior. This is for example reminiscent of the heavy quark-antiquark form factor
proposed in \refcite{Ji:2019sxk} to extract the TMD soft function on the lattice.

Physically, $\chi_1(b_\perp)$ is the total fragmentation factor for a heavy quark $Q$ to hadronize into any heavy hadron $H_Q$ with transverse momentum $k_T \sim 1/b_T$ ($b_T$ is the magnitude of the Minkowskian vector $b_\perp$).
Our object directly relates to a physical observable,
i.e., the total TMD cross section for producing heavy hadrons in $e^+e^-$ collisions.
In particular, we may relate $\chi_1(b_\perp)$ to the energy-energy correlator (EEC) for heavy hadrons in the back-to-back limit, where the usual semi-inclusive sum over all final state hadrons is restricted to heavy hadrons $a, b \in \{H\}$ containing valence heavy quarks or antiquarks~\cite{vonKuk:2024uxe}:
\begin{align}
&\mathrm{EEC}_H(z_{\chi,H})
\equiv \frac{1}{\sigma_\mathrm{total}} \sum_{a,b\, \in \{H\}} \int \! \df \sigma_{ee \to abX}
\frac{E_a E_b}{Q^2} \delta \bigl( z_{\chi,H} - \frac{1-\cos\theta_{ab}}{2} \bigr)
\nn \\
&=\mathcal{H}_{ee\to Q\bar{Q}}(Q^2, \mu) \, C_m^2(m, \mu, Q^2)
Q^2 \!\! \int_0^\infty \! \df b_T \, b_T \, J_0\bigl( \sqrt{1-z_{\chi,H}} \, b_T Q \bigr)
\Bigl[ \chi_1\Bigl(b_T, \mu, \frac{Q}{m}\Bigr) \Bigr]^2
\,.\end{align}
Here, we have the UV- and rapidity-renormalized $\chi_1(b_\perp,\mu,\rho)$ with renormalization scale $\mu$ and the rapidity $\rho = \nbar \cdot v$, and $C_m(m, \mu, \sqrt{\zeta}=\rho m)$ is the perturbative coefficient generated by separately
matching collinear and soft modes at the scale $\mu \sim m$ onto HQET
and QCD with $n_\ell$ flavors, respectively~\cite{Hoang:2015vua}.
The renormalization properties of $\chi_1(b_\perp)$ are discussed in detail in~\refcite{vonKuk:2024uxe},
and we collect the evolution equations below:
\begin{align}
    \mu \frac{d}{d\mu} \ln \chi_1 &= - \frac{\as C_F}{2\pi} (4 \ln \rho -2 ) = \gamma_{\chi_1}(\as(\mu),\rho) \,,
\label{eq:chi1_RG1}
\\
\rho \frac{d}{d\rho}  \ln \chi_1  &= -\frac{\as C_F}{2\pi} L_b = \gamma_\zeta (b_T, \mu) \,,
\label{eq:chi1_RGrho}
\\
\mu \frac{d}{d\mu} \gamma_\zeta = \rho \frac{d}{d\rho} \gamma_{\chi_1} &= -2 \frac{\as C_F}{4\pi} \cdot 4 = -2 \bar\Gamma_\mathrm{cusp}[\as(\mu)]
\label{eq:chi1_RG2}
\,,\end{align}
where we have defined for convenience
\begin{align}
    \rho = \nbar  \cdot v, \quad 
    L_b = 2 \ln \frac{b_T \mu}{2e^{-\gamma_E}}
\,.\end{align}
The second evolution equation comes from the rapidity divergence and its regularization mentioned above,
and $\gamma_\zeta(b_T,\mu)$ is the Collins-Soper kernel which governs the rapidity dependence of $\chi_1$ on the boost parameter $\rho$.
$\bar\Gamma_\mathrm{cusp}[\as(\mu)] = \tfrac{\as}{4\pi} \, 4 C_F + \ord{\as^2}$ is the lightlike cusp anomalous dimension that is universal for many QCD distributions including TMD parton distribution functions and fragmentation functions. We note that the TMD soft function is a similar cusped Wilson loop to $\chi_1(b_\perp)$ with both staples along lightlike directions.

It is not surprising that the renormalization of $\chi_1(b_\perp)$ involves $\bar\Gamma_\mathrm{cusp}$, as it was realized long ago that the log-enhanced divergence generated by Wilson lines meeting at an angle (``cusp'') give rise to the universal cusp anomalous dimension which depends on the angle. 
Early illustrative examples include the calculation for the angle-dependent anomalous dimension for the heavy quark transition current~\cite{Falk:1990yz}.
We will review the literature on this topic in \sec{cusp_anom}.
The physics of cusped Wilson loops can also be related to a flux tube state, and the cusp anomalous dimension can be interpreted as the energy density of the flux tube~\cite{Alday:2007mf}.
The two Wilson lines in different directions can be thought of as energetic particles moving towards infinity along those directions, carrying gauge flux between them.
This interpretation nicely connects to the EEC in the back-to-back limit, and may provide insights into the breaking of QCD confining flux tubes~\cite{Jaarsma:2025tck,Alday:2007mf}.

The main objective of the current manuscript is to study $\chi_1(b_\perp)$ as a cusped Wilson loop in the strong coupling limit, 
with the above field theory motivations in mind.
In the rest of this section,
we review existing literature on the connection between cusped Wilson lines and the cusp anomalous dimension, and then briefly discuss the AdS/CFT formalism we will use to calculate the expectation value of Wilson loops.

\subsection{Cusp anomalous dimension}
\label{sec:cusp_anom}

It was first pointed out by Polyakov~\cite{Polyakov:1980ca}
that the vacuum expectation value of Wilson lines with a cusp exhibits extra ultraviolet logarithmic divergences,
\begin{align}
    \langle W \rangle \sim e^{-\Gcusp[\phi,\as] \log\frac{L}{\eps}}
\,, \label{eq:W_cusp_div}
\end{align}
where $L$ is an infrared cutoff, for example the length of the Wilson loop, and $\eps$ is an ultraviolet cutoff, for example the regulator for the propagator.
$\Gcusp[\phi,\as]$ is the anomalous dimension generated by the cusp, which depends on both the coupling of the theory $\as$ and the cusp angle $\phi$.
We define our cusp angle $\phi$ in this manuscript in Minkowskian signature by $\cosh \phi = \frac{v_1 \cdot v_2}{\sqrt{v_1^2 v_2^2}}$,
where $v_1$ and $v_2$ are unit 4-vectors denoting the directions of the two Wilson lines forming the cusp.
Note that $\phi$ is equivalent to the change in rapidity $\Delta\eta$.
It is important to distinguish that sometimes in literature (e.g. the original Polyakov calculation~\cite{Polyakov:1980ca}), the cusp angle refers to the Euclidean angle $\phi_E$ defined by $\cos\phi_E = \frac{v_1 \cdot v_2}{\sqrt{v_1^2 v_2^2}}$, and one may recover the Minkowskian result by analytic continuation $\phi_E = i \phi$.

\begin{figure}[htbp]
    \centering
    
    \begin{subfigure}[t]{0.49\textwidth}
        \centering
        \includegraphics[width=\textwidth]{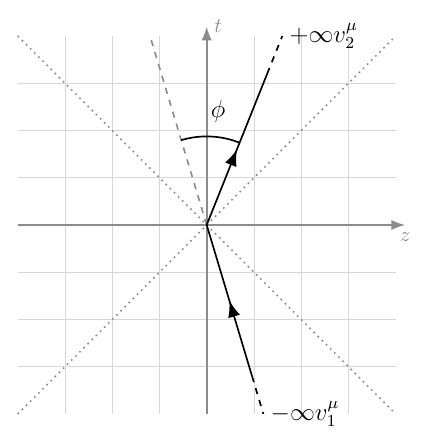}
        \caption{A Wilson line coming from $t=-\infty$ in the direction $v_1^\mu$ forming a cusp with another Wilson line going to $t=+\infty$ in the direction $v_2^\mu$.}
        \label{fig:future_past_lines}
    \end{subfigure}
    \hfill
    \begin{subfigure}[t]{0.49\textwidth}
        \centering
        \includegraphics[width=\textwidth]{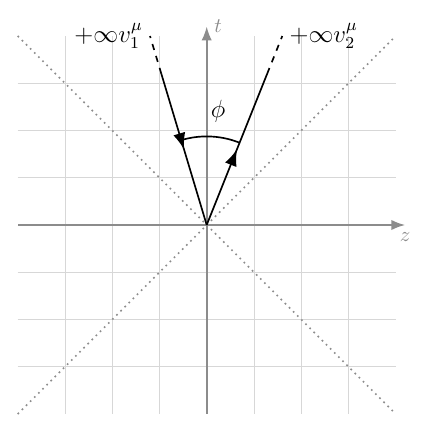}
        \caption{Two Wilson lines going to $t=+\infty$ in the directions $v_1^\mu$ and $v_2^\mu$.}
        \label{fig:future_future_lines}
    \end{subfigure}
    \caption{The convention for the cusp angle $\phi$ for different configurations of Wilson lines. 
    The direction vectors $v_1^\mu$ and $v_2^\mu$ are always future pointing.
    The arrow on the Wilson lines indicate color flow.}
\end{figure}

Another important distinction is between Wilson lines going to future infinity versus those coming from past infinity.
As illustrated in \fig{future_past_lines}, the cusp angle formed by one outgoing Wilson line and one incoming Wilson line is the deviation from a straight line configuration,
and this is the convention in e.g.~\cite{Grozin:2014hna,Grozin:2015kna}.
For two outgoing Wilson lines which is our current case of consideration, as illustrated in \fig{future_future_lines},
the cusp angle is the angle between the two lines.
The two configurations gives the same $\phi$ as the directions $n_1$ and $n_2$ are identical, 
but the cusp anomalous dimension are different,
since these are two different operators.

$\Gcusp[\phi,\as]$ is a universal object appearing in many processes in QCD due to the divergences from gluonic interactions.
Korchemsky and Radyushkin~\cite{Korchemsky:1991zp} were the first to relate the velocity-dependent anomalous dimension of the heavy quark transition current calculated in~\refcite{Falk:1990yz} 
to the vacuum expectation value of cusped Wilson lines.
It was explicitly verified at one-loop that the anomalous dimension $\gamma_w(\as)$ of heavy quark current operator changing velocity $v_1^\mu$ to $v_2^\mu$ is equivalent to the cusp anomalous dimension calculated by Polyakov, provided the Minkowskian angle $\cosh\phi = \frac{v_1 \cdot v_2}{\sqrt{v_1^2 v_2^2}}$.
Explicitly, we have 
\begin{align}
\label{eq:oneloop_cusp}
    \Gamma_\mathrm{cusp}^{1-\mathrm{loop}}[\phi,\as] = \frac{\as}{\pi} C_F 
    (\phi \coth\phi - 1)
\,.\end{align}
Physically, the $-1$ implements the effect of wave function renormalization, so that when the contour is smooth (corresponding to a conserved current), this contribution vanishes. 
The renormalization properties of cusped Wilson loops were subsequently studied extensively~\cite{Dotsenko:1979wb,Brandt:1981kf,Dorn:1986dt,Korchemsky:1987wg,Korchemsky:1992xv}.
When discussing QCD factorization and renormalization group evolution of operators in the fashion of \eqs{chi1_RG1}{chi1_RG2}, such as in the context of transverse-momentum dependent physics,
the cusp anomalous dimension $\bar{\Gamma}_\mathrm{cusp}[\as]$ is defined as the coefficient of the log-enhanced piece of the anomalous dimension,
which only appears in operators (like $\chi_1$) built out of cusped Wilson lines. 
$\bar{\Gamma}_\mathrm{cusp}[\as]$ in this context is the leading coefficient of the lightlike limit, or the large Minkowskian angle limit, of the general, angle-dependent $\Gcusp[\phi,\as]$.

Perturbatively, $\Gcusp$ has been calculated up to four loops in both QCD and $\cN=4$ super Yang-Mills (sYM) theory~\cite{Korchemsky:1987wg,Moch:2004pa,Kidonakis:2009ev,Grozin:2014hna,Grozin:2015kna,Moch:2018wjh,Henn:2019swt,Kidonakis:2023lgc}.
It is also pointed out that up to three loops, the results in QCD and in $\mathcal{N}=4$ sYM have the same structure~\cite{Grozin:2015kna}.
Starting at four loops, a new color factor generates non-planar corrections to the cusp anomalous dimension in $\cN=4$ sYM,
so the results in QCD and its supersymmetric extension differ~\cite{Henn:2019swt}.
In these calculations,
it was shown in both QCD~\cite{Brandt:1981kf,Korchemsky:1987wg,Korchemsky:1991zp,Manohar:2003vb,Bauer:2003pi} and $\cN=4$ sYM~\cite{Drukker:1999zq,Kruczenski:2002fb,Alday:2007mf} that the cusp anomalous dimension scales linearly with the Minkowskian cusp angle for large angles,
\begin{align}
\label{eq:inf_angle_scaling}
    \Gcusp[\phi,\as] \vert_{\phi\to\infty} = \phi \bar{\Gamma}_\mathrm{cusp} (\as) + \ord{\phi^0}
\,.\end{align}
$\bar{\Gamma}_\mathrm{cusp}[\as]$ is the ``cusp anomalous dimension'' in the context of QCD factorization and TMD physics as mentioned above.
In the small Minkowskian cusp angle limit for the configuration in \fig{future_past_lines}, 
the cusp disappears as the two lines become one straight line,
and the cusp anomalous dimension vanishes,
\begin{align}
\label{eq:small_angle_scaling_past}
    \Gcusp[\phi,\as] \vert_{\phi\to 0} = -\phi^2 B (\as) + \ord{\phi^3}
\,,\end{align}
where $B (\as)$ is the so-called bremsstrahlung function.
In the configuration in \fig{future_future_lines},
the small-angle limit corresponds to the zero-distance separation in the rectanglar Wilson loop (two anti-parallel Wilson lines) that defines the heavy quark-antiquark potential~\cite{Grozin:2015kna,Drukker:2011za}.
In this limit, we have
\begin{align}
\label{eq:small_angle_scaling_future}
\Gcusp[\phi',\as] \vert_{\phi'\to 0}
= - \frac{V(\as)}{\phi'} + \ord{\as^4}
\,,\end{align}
where $V(\as)$ is closely related to the heavy quark-antiquark potential.

Properties of $\Gcusp$ at strong coupling can be studied in $\cN=4$ sYM through the AdS/CFT correspondence~\cite{Maldacena:1997re,Witten:1998qj,Maldacena:1998im,Rey:1998ik,Drukker:1999zq}.
In the strong coupling limit, the anomalous dimension exhibits the same scaling behaviors as \eq{inf_angle_scaling}, \eq{small_angle_scaling_past}, and \eq{small_angle_scaling_future}~\cite{Drukker:1999zq,Kruczenski:2002fb,Makeenko:2002qe,Drukker:2011za,Correa:2012at}. In $\cN=4$ sYM, the cusp anomalous dimension also depends on an additional angular parameter, which will become apparent in what follows.
We elaborate on the formalism of using the AdS/CFT correspondence to calculate Wilson loops in the next subsection.

\subsection{Wilson loops in AdS/CFT}
\label{sec:holo_wilson}

One of the many results of the correspondence between type IIB string theory on AdS${}_5 \times S_5$ and $\mathcal{N}=4$ sYM in the large $N_c$ limit~\cite{Maldacena:1997re,Witten:1998qj} is that the expectation values of Wilson loops in the latter theory can be expressed in terms of the action that governs the dynamics of strings in the former theory~\cite{Rey:1998ik,Maldacena:1998im,Drukker:1999zq}.

The Wilson loop that enters this correspondence is given by
\begin{equation}
    W[\mathcal{C}] = \frac{1}{N_c} {\rm Tr} \left[ {\rm P} \exp \left( ig \int_{\mathcal{C}} ds \left[ \dot{x}^\mu A_\mu + \sqrt{\dot{x}^2} \, \hat{n} \cdot \Phi \right] \right) \right] \, , \label{eq:W_def}
\end{equation}
where $x^\mu = x^\mu(s) \in$ Mink${}_4$ together with $\hat{n} = \hat{n}(s) \in S_5$ define the parametrization of the path $\mathcal{C}$, which as is apparent now, requires that one specify both the usual path in 4-dimensional spacetime as well as a path traversed on the $S_5$. The fields $\Phi$ are the six (adjoint-representation) scalar fields of $\mathcal{N}=4$, and enter the Wilson loop above in such a way that i) the Wilson loop is still reparametrization invariant and ii) enjoys more (super)symmetries with respect to the supercharges of the theory than the pure gauge Wilson loop. 

Concretely, the statement of AdS/CFT is that, if $\mathcal{C}$ is the contour over which the Wilson loop is defined, then
\begin{equation}
    \left\langle W[\mathcal{C}] \right\rangle = \int_{\partial \Sigma = \mathcal{C}} D\Sigma \exp \left( - \mathcal{S}[\Sigma] \right) \, , \label{eq:W_PI}
\end{equation}
where the right hand side is, at this point, (mostly) schematic, in the sense that we do not know of a precise definition of the path integral over the degrees of freedom of a string. 
The unambiguous piece of information that this expression contains is that the boundary conditions of the worldsheet $\Sigma$ are determined by the path $\mathcal{C}$ that defines the Wilson loop.

In the strong coupling limit $\lambda = g^2 N_c \to \infty$, fluctuations in the string theory become suppressed and it is sufficient to only consider its bosonic degrees of freedom. One can then write the action in the Nambu-Goto form
\begin{equation}
    \mathcal{S}[\Sigma] = \mathcal{S}_{\rm NG} = \frac{1}{2\pi \alpha'} \int d^2\xi \sqrt{\det \left( \partial_\alpha X^\mu \partial_\beta X^\nu g_{\mu \nu}(X) \right) } \, ,
\end{equation}
where $X^\mu = X^\mu(\xi)$ is a parametrization of the worldsheet $\Sigma$, and $\xi = \xi^a$, $a = 1,2$ are the two world-sheet parameters. Using the fact that the strong coupling limit in the field theory corresponds to $\alpha' \to 0$, the path integral in \eq{W_PI} may be carried out using the saddle point method (i.e., extremizing $\mathcal{S}_{\rm NG}$), and as such the problem of obtaining the leading dependence on $\lambda$ of the Wilson loop is determined by a problem analogous to those of classical mechanics: solving the Euler-Lagrange equations that result from varying the action.

At strong coupling, one can also obtain the usual gauge theory Wilson loop by integrating the right hand side of \eq{W_def} over $\hat{n}$~\cite{Polchinski:2011im}, whereby Neumann boundary conditions are imparted on the string~\cite{Alday:2007he}. In practice, since a constant value of $\hat{n}$ throughout the worldsheet satisfies both Dirichlet and Neumann boundary conditions (because it is a fixed value and its derivative vanishes, respectively),
the distinction between the two types of Wilson loops is inconsequential provided there is no obstruction to having constant $\hat{n}$. That being said, sometimes it is physically warranted to study configurations with nontrivial dependence on $\hat{n}$. We shall see some examples below.

From this starting point, a lot has been learned about the dynamical properties of heavy quarks in gauge theories at strong coupling that can be characterized in terms of Wilson loops. Perhaps the earliest application was the calculation of the quark-antiquark potential~\cite{Maldacena:1998im,Rey:1998ik}. 
It has been particularly insightful in the study of dynamical questions involving heavy quarks in nontrivial gauge theory states. For example, their propagation in a finite temperature $\mathcal{N}=4$ sYM background has received much attention both by studying them as individual particles~\cite{Herzog:2006gh,Gubser:2006bz,Casalderrey-Solana:2006fio,Liu:2006ug,Gubser:2006nz,Casalderrey-Solana:2007ahi,DEramo:2010wup,Rajagopal:2025ukd,Rajagopal:2026urq} as well as by studying the static and dynamical properties of the particle-antiparticle bound states that they form~\cite{Rey:1998bq,Brandhuber:1998bs,Liu:2006he,Liu:2006nn,Chernicoff:2006hi,Ejaz:2007hg,Mateos:2007vn,Faulkner:2008qk,Nijs:2023dks,Nijs:2023dbc}. Gauge theory states with other characteristics, such as angular momentum or a chemical potential have also been considered; see~\cite{Casalderrey-Solana:2011dxg,Giataganas:2018rbq} for reviews.

More central to our present work, Wilson loops with cusps were studied using the AdS/CFT correspondence immediately after its discovery~\cite{Drukker:1999zq}. Subsequent calculations involving lightlike Wilson lines in Minkowski signature~\cite{Kruczenski:2002fb,Makeenko:2002qe} (see also~\cite{Makeenko:2008xr}) established the possibility to study these operators in real time using holographic methods. However, the necessary developments to carry out calculations of matrix elements with nontrivial operator ordering (i.e., not purely time-ordered) were not worked out in full detail until much later~\cite{Skenderis:2008dg,Skenderis:2008dh}. While the methods to handle real-time calculations have developed further since then~\cite{Glorioso:2018mmw,Jana:2020vyx}, they have mostly focused on correlation functions of local operators, rather than extended ones (such as Wilson lines). The calculations of the jet quenching parameter~\cite{DEramo:2010wup} and of the heavy quark momentum change probability~\cite{Rajagopal:2025ukd} in strongly coupled $\mathcal{N}=4$ sYM at finite temperature stand as rare exceptions. The latter calculation pioneered the use of Lagrange multipliers in the stringy description to enforce boundary conditions that follow from the spatial arrangement of the Wilson lines, and showed that these have a natural interpretation as a momentum transfer in the field theory and a momentum flow along the worldsheet in the gravitational theory --- an observation that will prove crucial later in this work.

\subsection{Outline of the paper}

Our purpose in this paper is to study the nonperturbative matrix element of cusped Wilson lines proposed in~\cite{vonKuk:2023jfd,vonKuk:2024uxe} with methods from the AdS/CFT correspondence. Given the large variance in the language that the different physics communities from which we draw input, we will aim to provide an exposition that is as self-contained as possible.

In~\sec{setup}, we make clear the technical details pertaining to our calculation by addressing three key aspects: the computation of Wilson loops in the dual theory, the treatment of non-time-ordered operator correlators, and the role of transverse gauge links.

\begin{figure}
    \centering
    \includegraphics[width=0.49\linewidth]{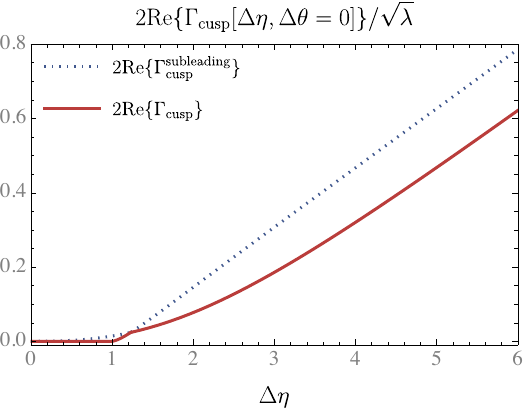}
    \includegraphics[width=0.49\linewidth]{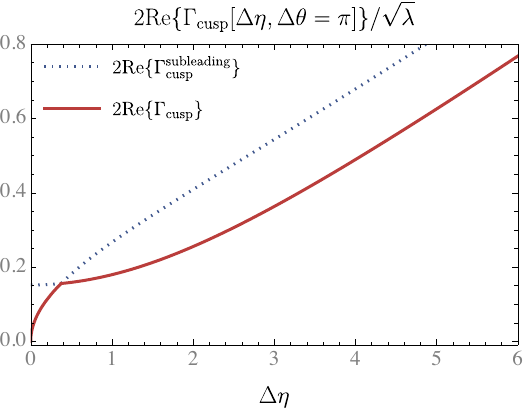}
    \caption{Cusp anomalous dimension for a fundamental Wilson loop in the strongly coupled limit of $\mathcal{N}=4$ sYM theory at large $N_c$ as a function of $\Delta \eta$ for an $S_5$ separation of $\Delta \theta = 0$ (left) and $\Delta \theta = \pi$ (right), obtained from our calculation via the AdS/CFT correspondence in terms of the saddle points of a path integral over string configurations. The solid curves are determined by the configurations with the smallest real part of the action at a given value of $\Delta \eta$ and $\Delta \theta$; the dotted curves show the result of the family of string configurations that are subdominant at said values. Surprisingly (to us), there is an exchange in dominance of the two families of extrema at a relatively small Minkowski angle separation. }
    \label{fig:result_intro}
\end{figure}

In~\sec{onecusp}, we carry out a detailed calculation for the expectation value of a Wilson loop with a single cusp, directly in Minkowski spacetime. We discuss the extremization and regularization of the string theory action, and in particular stress the important insights gained from the approach using Lagrange multipliers. We present the solution for the string action in~\sec{saddle_point_solution}, and extract the cusp anomalous dimension from our computation and relate it to existing literature in~\sec{one_cusp_result}. For the benefit of the reader, we show our result for the anomalous dimension of the operator already in~\fig{result_intro}, which is determined by the largest of two saddles (i.e., the one with smallest value of the action) that contribute to the calculation of the integral over string configurations.

In~\sec{twocusps}, we construct the final calculation of the matrix element $\chi_1(b_\perp)$ in~\eq{chi1_matrix_element} using the techniques discussed in previous parts of the paper. An important realization is the Fourier conjugate of the transverse separation between the two cusps can be related to the momentum flux of the string on the extremal surface, as we discuss in~\sec{trans_momentum_flow}. We then compute the extremal surface for the two-cusp configuration in~\sec{two_cusp_xperpsol}, where we also discuss the large rapidity limit and in particular give the profile of the transverse coordinate in this limit. 
We present the final result for $\chi_1(b_\perp)$ at strong coupling in~\sec{final_chi1_result}, focusing on the cusp anomalous dimension of this operator and extending the result to the lightlike TMD soft function.

We conclude and summarize our results in~\sec{conclusion}, and discuss possible future directions, such as the generalization of our analysis to other holographic theories and potential applications.

\section{Setup}
\label{sec:setup}

This section systematically presents the holographic framework and methodological details necessary for our computation.
We start by providing a more detailed discussion of the calculation of Wilson loops in holography in~\sec{gen_holo}. We then move on to discuss the precise way in which correlation functions of operators with nontrivial operator ordering (i.e., not reducible to purely time-ordered correlators) are described in terms of the dual (gravitational) theory in~\sec{ordering}. This will establish a precise correspondence for the Wilson line segments that are explicitly written in \eq{chi1_matrix_element}. We discuss the often omitted transverse gauge links in more detail in~\sec{trans_gauge_link}.

\subsection{Generalities of the holographic setup} 
\label{sec:gen_holo}

In what follows we use the abstract object $\Sigma$ describing the surface and its parametrization $X^\mu$ interchangeably. Following standard nomenclature, we will also refer to slices of this surface at any fixed time as a ``string,'' which can be thought of as a dynamical degree of freedom dual to heavy quarks by virtue of the holographic correspondence~\cite{Maldacena:1998im}.

The background metric on which the string propagates is given by
\begin{equation}
    ds^2 = \frac{R^2}{z^2} \left[ -dt^2 + d{\bf x}^2 + dz^2 + z^2 d\Omega_5^2 \right] \, ,
\end{equation}
where $z=0$ is called the ``boundary.'' This is the (asymptotic) submanifold where observables of the dual field theory are supported. A point in this boundary is specified by a point in Mink${}_4 \times S_5$. The first space in this product is the spacetime on which the field theory is defined. The variables describing the $S_5$ component of the boundary manifold account for some of the ``internal'' degrees of freedom in $\mathcal{N}=4$ --- concretely, they realize the SO(6) global symmetry defined by ``rotations'' among its scalar fields. $R$ is the AdS curvature scale.

For our purposes, only the combination $R^2/\alpha'$ is relevant, because $R^2$ appears as an overall factor of the metric and can thus be factored out. In terms of field theory parameters, this combination is equal to $\sqrt{\lambda}$. That is to say, if we instead use
\begin{equation}
    ds^2 = \frac{1}{z^2} \left[ -dt^2 + d{\bf x}^2 + dz^2 + z^2 d\Omega_5^2 \right] \, ,
\end{equation}
to define the metric, we can write the Nambu-Goto action as
\begin{equation}
    \mathcal{S}_{\rm NG} = \frac{\sqrt{\lambda}}{2\pi} \int d^2\xi \sqrt{\det \left( \partial_\alpha X^\mu \partial_\beta X^\nu g_{\mu \nu}(X) \right) } \, .
\end{equation}
Then, in the strong coupling limit $\lambda \to \infty$, we can write
\begin{align}
    \left\langle W[\mathcal{C}] \right\rangle &= \int_{\partial \Sigma = \mathcal{C}} DX \exp \left( - \mathcal{S}_{\rm NG}[X] \right) \nonumber \\ 
    &= \exp \left( -  \mathcal{S}_{\rm NG}[\bar{\Sigma}[\mathcal{C}]] + \mathcal{O}(\lambda^0) \right) \, , \label{eq:W_Loop_semiclass_limit}
\end{align}
where we have used that in the large $\lambda$ limit we may evaluate the integral in the saddle point approximation\footnote{To get the complete set of $\mathcal{O}(\lambda^0)$ terms correctly, one needs to calculate a 1-loop determinant that includes the string's fermionic fluctuations~\cite{Drukker:2000ep}, making the calculation more involved, as it is no longer fully described by the Nambu-Goto action.}, by extremizing the action $\mathcal{S}_{\rm NG}$. That is to say, $\bar{\Sigma}[\mathcal{C}]$ is the surface that satisfies
\begin{equation}
    \left. \frac{\delta \mathcal{S}_{\rm NG}}{\delta X^\mu} \right|_{\Sigma = \bar{\Sigma}[\mathcal{C}]} = 0 \, , \label{eq:NG_eqs_formal}
\end{equation}
and satisfies the boundary conditions specified by $\mathcal{C}$, the contour of the Wilson loop which lives at at $z=0$. As such, in principle, all one needs to do is to solve the Nambu-Goto equations of motion~\eqref{eq:NG_eqs_formal} for the given boundary path $\mathcal{C}$, and the answer for the Wilson loop expectation value follows. 

\paragraph{The path on the $S_5$}

As we mentioned earlier, Dirichlet and Neumann boundary conditions on the $S_5$ have Wilson loop expectation values as counterparts with and without the $\mathcal{N}=4$ scalars, respectively. Ideally, we would like to calculate the vacuum matrix element~\eqref{eq:chi1_matrix_element}, where the Wilson lines do not have scalars. However, since this object is defined in QCD, and the theory we discuss here is $\mathcal{N}=4$, we may learn more by considering the different possibilities that are natural in the theory at hand.  Given the discussion in~\cite{Polchinski:2011im}, we anyways expect that the physics of the Wilson loop without scalars may be obtained from the one with scalars by integrating over the configurations on the $S_5$. As such, for definiteness, we will focus on the Dirichlet case, where we fix the position $\hat{n} \in S_5$ by hand to a specific value for each line. We will comment on the Neumann case in the outlook. 

Because the object of interest~\eq{chi1_matrix_element} is obtained as the product of an amplitude with its complex conjugate, it is natural to choose the same\footnote{We discuss why this, and not antipodal $\hat{n}$ directions as one may be tempted to conclude from the discussion in the next paragraph, is the appropriate choice for this operator ordering, in the following section.} $\hat{n}$ for each pair of lines with the same velocity. On the other hand, having in mind the possibility that different physical features may appear, we will consider the possibility of an angular separation $\Delta \theta$ on the $S_5$ between the lines with velocities $v_1$ and $v_2$. 
This will also provide a cross-check of our calculation, as several results for limiting cases of the expectation value of a Wilson loop with a cusp exist in the literature~\cite{Drukker:1999zq,Kruczenski:2002fb,Makeenko:2002qe,Drukker:2011za,Correa:2012at}.

An important reason why we consider the dependence on the $S_5$ separation is to highlight the physical differences that appear in a real-time calculation between the limiting cases $\Delta \theta = 0$ and $\Delta \theta = \pi$. Because of the structure of the Wilson line in \eq{W_def}, it is clear that a backtracking path in its Minkowski components only possesses the so-called ``zig-zag'' symmetry~\cite{Polyakov:1997tj,Drukker:1999zq} if $\hat{n}$ is flipped to its antipodal point at the point where $x^\mu$ begins to backtrack. A Wilson line possesses the zig-zag symmetry if its expectation value is invariant under deformations that deviate from the original path and then retrace back along the same deviations.
More succinctly, the Wilson line with scalars satisfies
\begin{equation}
    \left[ {\rm P} \exp \left( ig \int_{\mathcal{C}} ds \left[ \dot{x}^\mu A_\mu + \sqrt{\dot{x}^2} \, \hat{n} \cdot \Phi \right] \right) \right]^\dagger = {\rm P} \exp \left( ig \int_{\bar{\mathcal{C}}} ds \left[ \dot{x}^\mu A_\mu - \sqrt{\dot{x}^2} \, \hat{n} \cdot \Phi \right] \right) \label{eq:adj_conj} 
\end{equation}
where $\bar{\mathcal{C}}$ is the same set of points as the (open) path $\mathcal{C}$, but traversed in the opposite direction, starting at the endpoint of $\mathcal{C}$ and ending at its starting point, respectively. 
In particular, if the QCD property that the Wilson lines should cancel out in the limit of vanishing cusp angle is to be guaranteed to hold in this setup, then we have to choose $\Delta \theta = \pi$ in this limit.

\subsection{Operator ordering} \label{sec:ordering}

As we have suggested a few times by now, the matrix element~\eqref{eq:chi1_matrix_element} is not a time-ordered correlation function, but rather an expectation value where two of the operators are time-ordered and two are anti-time-ordered in their action on the bra and ket states. Explicitly, 
\begin{align}
\label{eq:chi1_matrix_element_T}
    \chi_1(b_\perp)= \frac{1}{N_c} \Tr
\Mae{0}{ \bar{\mathrm{T}} \{ W^\dagger(b_\perp) \,
   Y_v(b_\perp) \} \, \mathrm{T} \{ Y_v^\dagger(0) \,
W(0) \} }{0} 
\,,\end{align}
which means that, when represented in terms of a path integral, two copies of field degrees of freedom are required in order to enforce the operator ordering as written.\footnote{Conversely, if only Feynman propagators were used, it would amount to imposing a single time-ordering prescription for all operators in the matrix element.}
The path integral containing this doubled field content goes by the name of the Schwinger-Keldysh path integral~\cite{Schwinger:1960qe,Keldysh:1964ud}, or sometimes the closed-time-path (abbreviated as CTP) contour.

The same is true in the dual description in terms of stringy/gravitational physics. That is to say, when the expectation value of interest in the field theory requires multiple copies of the fields to be represented as a path integral, its holographic computation also features a doubled set of fields, with one copy of the bulk (sub)manifold(s) per copy of the corresponding set of degrees of freedom. This was first discussed in detail in the work of Skenderis and van Rees~\cite{Skenderis:2008dg,Skenderis:2008dh}, where they developed the idea that the counterpart of each copy of the field theory path integral is a corresponding holographic path integral on an asymptotically AdS bulk (sub)manifold that has the Schwinger-Keldysh contour as its asymptotic boundary. It generalized the earlier prescription to calculate Schwinger-Keldysh $2$-point correlation functions from the AdS/CFT correspondence, which determines the retarded propagator by imposing an ingoing boundary condition for perturbations~\cite{Son:2002sd,Herzog:2002pc}. More general versions of this prescription, also applicable to non-equilibrium states, have been developed more recently~\cite{Glorioso:2018mmw,Jana:2020vyx}. 
For a pedagogical introduction to this subject, see Section 5 of~\refcite{Haehl:2024pqu}.

Due to the extended nature of the Wilson line operators in the matrix element we study, the Skenderis \& Van Rees prescription is the most natural starting point for our purposes. This is because matching and/or boundary conditions for worldsheet configurations (dual to the Wilson loop) are most naturally imposed in terms of spacetime coordinates (instead of frequency/wavenumber as for $2$-point functions). A previous example of a calculation where this construction was successfully applied is the study of the momentum transfer probability to and from a heavy quark propagating through a finite temperature environment~\cite{Rajagopal:2025ukd}. In our present case, the situation is actually simpler than the scenarios for which the above prescriptions were designed, because we are interested in a vacuum expectation value, rather than a thermal one. In fact, as we will discuss in more detail in the following subsection, all of the cross-talk between the time-ordered and the anti-time-ordered fields happens through the matching conditions at temporal infinity.

\begin{figure}
    \centering
    \includegraphics[width=0.8\linewidth]{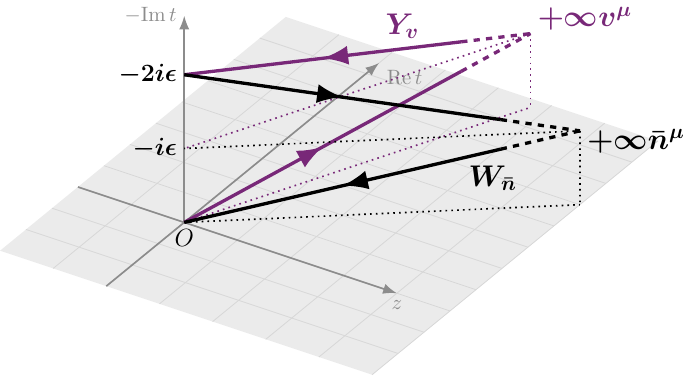}
    \caption{Wilson loop that determines the matrix element~\eqref{eq:chi1_matrix_element} in the complex time plane, where a deformation proportional to $i\epsilon$ enforces the required operator ordering by fixing the contour prescriptions to evaluate time/frequency integrals (fittingly referred to as $i\epsilon$ prescriptions). The lines that have imaginary parts that go between $0$ and $-i\epsilon$ are time-ordered, whereas the ones that have imaginary parts between $-i\epsilon$ and $-2i\epsilon$ are anti-time ordered. The transverse coordinate $x_\perp$ is omitted in this figure.}
    \label{fig:SK_Wilson_arrangement}
\end{figure}

Just like a definite $i\epsilon$ prescription appears in traditional time-ordered field theory calculations, the propagators in the Schwinger-Keldysh path integral feature definite, but different (depending on which copy of the path integral the field corresponds to), $i\epsilon$ prescriptions when correlations between fields are calculated. One way of constructing the Schwinger-Keldysh contour in which these prescriptions appear naturally is by deforming the forward and backward time evolution operators $e^{-iHt}$ and $e^{+iHt}$ to $e^{-iHt (1-i\epsilon) }$ and $e^{+iHt (1+ i\epsilon) }$ (effectively deforming the time evolution contour into the complex plane), formulating the desired correlation functions by inserting operators along each segment, and then taking the $\epsilon \to 0$ limit. Note that the small imaginary time component is such that the time evolution operator is deformed by $e^{-\epsilon t H}$ with a definite sign in the exponent (the opposite sign would be catastrophic because $H$ is not bounded from above). The Wilson loop of interest in our work can be visualized as a function of complex time and longitudinal position as in~\fig{SK_Wilson_arrangement}.

The same considerations carry over for the holographic version of the Schwinger-Keldysh path integral. More concretely for our purposes, because the determinant in the Nambu-Goto action changes sign when going from spacelike to timelike surfaces, a branch cut needs to be specified, and therefore an $i\epsilon$ prescription is necessary for the action $\mathcal{S}_{\rm NG}$ to be uniquely defined for $X$ propagating on a background with Minkowski signature. Fittingly, just like in any quantum (field) theory, we may enforce the corresponding prescription by setting up the path integral by introducing a ``tilt'' in the time evolution contour, deforming it into a complex time path that reproduces the real time evolution in the limit $\epsilon \to 0$. Crucially, this $i\epsilon$ prescription selects the appropriate branch cut to be taken in the square root that enters the definition of the Nambu-Goto action. Effectively, all it does is to ensure that the Nambu-Goto action in real time is defined as the analytic continuation of its Euclidean counterpart while preserving the (crucial) property that, should any off-shell contributions in the string propagation appear, they can do so with the correct sign in the exponential --- very much like a tunneling process in quantum mechanics. 

Defining the area functional of the Nambu-Goto action in this way reflects that we should give a similar treatment to the Wilson line~\eq{W_def}. Consider a Wilson line on an open segment $\mathcal{D}$. In Euclidean space, it is expressed as~\cite{Drukker:1999zq}
\begin{equation}
    W[\mathcal{D}] = \frac{1}{N_c}  {\rm P} \exp \left(  \int_{\mathcal{D}} ds \left[ ig \dot{x}^\mu A_\mu + g \sqrt{\dot{x}^2} \, \hat{n} \cdot \Phi \right] \right) \, , \label{eq:W_def_E}
\end{equation}
whose exponent differs from the expression in~\eq{W_def} by a factor of $i$ in front of the adjoint scalar fields $\Phi$. In Euclidean space, all paths have $\dot{x}^2 > 0$. When analytically continued to Minkowski space, $\dot{x}^2$ becomes negative for timelike paths, and the $i\epsilon$ prescription manifests itself from the different ways of reaching the negative $\dot{x}^2$ real line. When the path is time-ordered, a factor of $(1-i\epsilon)$ multiplies the negative number that determines the sign of the real part of $\dot{x}^2$, meaning that $\sqrt{\dot{x}^2} = i \sqrt{|\dot{x}^2|}$. Conversely, when the path is anti-time-ordered, a factor of $(1+i\epsilon)$ multiplies the negative number that determines the sign of the real part of $\dot{x}^2$, and as such we have $\sqrt{\dot{x}^2} = -i \sqrt{|\dot{x}^2|}$. It then follows that
\begin{align}
    {\rm T} W[\mathcal{D}_{(1-i\epsilon)}] &= {\rm T} \left\{ \frac{1}{N_c} {\rm P} \exp \left(  \int_{\mathcal{D}_{(1-i\epsilon)}} \!\!\!\!\!\! ds \left[ ig \dot{x}^\mu A_\mu + i g \sqrt{\dot{x}^2} \, \hat{n} \cdot \Phi \right] \right) \right\} \, , \label{eq:W_def_T} \\
    \bar{\rm T} W[\mathcal{D}_{(1+i\epsilon)}] &= \bar{\rm T} \left\{ \frac{1}{N_c} {\rm P} \exp \left(  \int_{\mathcal{D}_{(1+i\epsilon)}} \!\!\!\!\!\! ds \left[ ig \dot{x}^\mu A_\mu - i g \sqrt{\dot{x}^2} \, \hat{n} \cdot \Phi \right] \right) \right\} \, , \label{eq:W_def_A} \\
\end{align}
where the time-ordering symbols reflect the operator ordering of the entries of the gauge fields --- i.e., on which copy of the SK path integral they appear. The subscripts on the path symbols $\mathcal{D}$ account for the different analytic continuations of the Euclidean path $\mathcal{D}$.

Note that this means that, as we anticipated earlier, the appropriate setup for the lines on opposite sides of the SK contour with the same velocity (i.e., one on the amplitude and the other on the complex conjugate amplitude) is to have the same value of $\hat{n}$. This is so because the conjugation operation in~\eq{adj_conj} is automatically taken care of by the opposite signs in~\eqs{W_def_T}{W_def_A}: a string with endpoint specified by $\hat{n}$ on the time-ordered manifold of the holographic SK path integral is dual to a Wilson line with the sign in front of the scalars as in~\eq{W_def_T}, whereas a string with the same endpoint for the anti-time-ordered manifold of the holographic SK path integral is dual to a Wilson line with the sign in front of the scalars as in~\eq{W_def_A}.\footnote{A similar discussion can be found in~\cite{Rajagopal:2025ukd}.} Conversely, for Wilson lines to be conjugates of each other on the same SK contour segment, the value of the $\hat{n}$ coordinate for each of the dual strings must be antipodal to each other.

Finally, the construction of Skenderis and van Rees~\cite{Skenderis:2008dg,Skenderis:2008dh} also specifies the matching conditions that the fields --- in our case the worldsheet configurations --- must satisfy at the final time slice, where the two real-time segments of the Schwinger-Keldysh contour meet, and where the two corresponding parts of the bulk manifold meet. Concretely, it specifies that the field variables and their canonical momenta should be continuous and approach the same value at some matching bulk hypersurface (whose boundary is the point on the SK contour where the time-ordered and anti-time-ordered branches meet). In our case, by symmetry, if one of the one-cusp Wilson line configurations lies at $x_\perp = b_\perp$ and the other at $x_\perp = 0$, then at the matching bulk hypersurface both dual worldsheet configurations should approach the midpoint $x_\perp = b_\perp/2$. We note that, when we get to the point where we will need to enforce this condition in~\sec{twocusps}, we will in fact do so by integrating over a conjugate momentum variable, which will itself be closely linked to the canonical momenta of the strings. 

Before moving on to the calculation, it is fitting to point out that this matching procedure, as we just described it, leaves out a subtle point: Given how the Wilson lines are arranged, at the boundary itself the strings hang from $x_\perp = 0$ and $x_\perp = b_\perp$, thus seemingly giving rise to a discontinuity.
The resolution to this apparent discontinuity is that the Wilson lines must be joined by further Wilson line segments --- e.g., transverse gauge links --- that traverse the transverse distance between them in order to define a closed Wilson loop. 
In order to show how they (do not) affect the holographic calculation, it is instructive to discuss the precise arrangement of the transverse gauge links and the field theory considerations that give rise to them.

\subsection{Transverse gauge links} 
\label{sec:trans_gauge_link}

When we first defined our Wilson loop in \eq{chi1_matrix_element}, we omitted the transverse Wilson lines at timelike and lightcone infinity, and only argued for their existence through gauge invariance. 
The complete gauge invariant definition can be written as
\begin{align}
\label{eq:chi1_transverse}
    \chi_1(b_\perp)= \frac{1}{N_c} \Tr
\Mae{0}{W^\dagger(b_\perp) \,
   Y_v(b_\perp) \,  W_T^\dagger (\infty v; 0, b_T) \, Y_v^\dagger(0) \,
W(0) \, W_T (\infty \nbar; 0, b_T)}{0} 
\,,\end{align}
where the transverse Wilson line is defined as
\begin{align}
    W_T(x; 0, b_T)
    = \bar{\mathrm P} \Bigl[ \exp \Bigl(
   -i g \int_0^{b_T} \! \df s \, \hat{b}_\perp \cdot A(x + \hat{b}_\perp s)\Bigr)\Bigr]
\,.\end{align}
This configuration is represented by the closed Wilson loop in~\fig{Wilson_chi1}.

\Refcite{Belitsky:2002sm} established the importance of including transverse Wilson lines at lightcone infinity to ensure gauge invariance of TMD parton distribution functions (PDFs). While the fact that the loop is closed ensures gauge invariance, calculations in different gauges will give different importance to different segments.
In covariant, physical gauges, such as the Feynman gauge, the vector potential vanishes asymptotically, therefore one doesn't expect gauge fields at infinity to contribute.
In singular gauges such as the lightcone gauge $A_+=0$, however, 
the vector potential along the lightcone formally vanishes.
Without additional transverse Wilson lines at infinity, the effects of final state interactions are missing and the definition of TMD PDFs are not gauge-invariant under different choices of the gauge boundary conditions~\cite{Brodsky:2002ue,Brodsky:2002cx,Ji:2002aa}.
These transverse Wilson lines are also important for constructing lattice analogues of TMDs~\cite{Ebert:2019okf}.
Inclusion of transverse Wilson lines at the effective field theory Lagrangian level are discussed in~\refscite{Idilbi:2010im,Garcia-Echevarria:2011ewi}.

Similarly, in the definition of TMD fragmentation functions (FFs), from which the Wilson loop $\chi_1(b_\perp)$ defined in \eq{chi1_matrix_element} originates, we must also include transverse Wilson lines at lightcone infinity. This is manifested as the transverse Wilson line at lightcone infinity $W_T(\infty \nbar; 0, b_T)$ in the definition \eq{chi1_transverse}.
For $\chi_1(b_\perp)$, we also have a connecting gauge link at timelike ($\propto v$) infinity for the two heavy quark Wilson lines, resulting from summing over all final state heavy hadrons~\cite{vonKuk:2023jfd,vonKuk:2024uxe}.

As noted above, in covariant, physical gauges the vector potential vanishes asymptotically, and the transverse gauge links do not contribute.
We can interpret this intuitively as the Wilson line at infinite
distance not impacting the physics at finite distance; the paths that the transverse gauge links take may be arbitrary in the transverse plane.
Perturbatively, it is instructive to discuss how this emerges in different gauges.
In covariant gauges such as the Feynman gauge, the transverse sections $W_T$ and $W_T^\dagger$
\eq{chi1_transverse} effectively reduce to the identity as the gauge fields in them decouple from the rest of the Wilson line contributions due to being infinitely far away, so they cannot enter the renormalization of the operator at any order in perturbation theory --- consistent with the fact that the Wilson lines may be joined at infinity with any path. Because the anomalous dimension of an operator is gauge-invariant, this all-orders statement demands the configuration at infinity carries no anomalous dimension for our operator of interest.
In singular gauges such as $A^+=0$,
the transverse Wilson lines are nontrivial and do contribute, but only to restore the same gauge-invariant total.\footnote{The order of limits may be subtle, see, for example, the discussion in~\cite{Scheihing-Hitschfeld:2022xqx}.}

On the holographic side, the fact that the precise configuration of the transverse gauge links at infinity does not contribute to the result is apparent in the fact that, by symmetry of the equations that determine the saddle point worldsheet configurations dual to the Wilson loop, the strings have to meet at the intermediate value of the $x_\perp$ coordinate in the bulk, regardless of how the transition from $x_\perp = 0$ to $x_\perp =  b_\perp$ is arranged at the boundary. And, much like in non-singular-gauge perturbation theory, the contributions to the anomalous dimension of the operator emerge naturally from the spacetime region that surrounds either of the two cusps --- the contributions from lightcone or temporal infinity amount to sums over final states that match the configurations of the amplitude and the complex conjugate amplitude to each other.

\section{One cusp}
\label{sec:onecusp}

As a necessary intermediate step, we now turn to examining the extremal surfaces that describe the contributions to the expectation value of a Wilson loop coming from the neighborhood of a single cusp, where both of the lines going out of the cusp are timelike and going towards the future, as illustrated by~\fig{future_future_lines}, separated by a spacetime rapidity $\Delta \eta = \phi = {\rm arccosh}\Big(- \frac{v_1 \cdot v_2}{\sqrt{ v_1^2 v_2^2 } } \Big)$.\footnote{To the (likely) annoyance of the readers, note that at this point we switched to a ``mostly plus'' metric convention.}
With those results in hand, we then discuss the comparison of the rapidity dependence of the corresponding cusp anomalous dimensions with existing results in literature.

\subsection{Extremal surfaces for one cusp}

In addition to the rapidity separation, we consider an arbitrary angular separation $\Delta \theta$ on the $S^5$ between the two lines coming out of the cusp. That is to say, the surfaces we seek to characterize will be specified by two numbers, $\Delta \eta$ and $\Delta \theta$.

Crucially, neither of these parameters breaks the conformal symmetry of the theory. Therefore, because both i) the equations of motion that need to be solved in order to extremize the action $\mathcal{S}_{\rm NG}$ and ii) the boundary conditions of these equations are invariant under scale transformations, it follows that the solution to these equations are also scale invariant.

Explicitly, letting $(t,x)$ be timelike and spatial coordinates describing the same plane that the vectors $v_1$, $v_2$ span, and writing down the metric in terms of $\eta = {\rm arctanh}(x/t)$, $a = \sqrt{t^2 - x^2}$ and a great circle angle $\theta$ on the $S^5$
\begin{align}
    ds^2 = \frac{1}{z^2} \left[ -da^2 + a^2 d\eta^2 + dz^2 + z^2 d\theta^2 \right] \, ,
\end{align}
it follow that we can parametrize the surface coordinates as 
\begin{align}
    X^\mu = \left(a, \eta\left(\frac{z}{a}\right), z, \theta\left(\frac{z}{a}\right)\right) \, , \label{eq:one_cusp_parameter}
\end{align}
where $z/a$ describes the one non-trivial scale-invariant combination of the independent coordinates $z,a$ that we have chosen to describe the surface. Other coordinate choices are, of course, possible, but these have the advantage of making the consequences of conformal symmetry manifest.

With that in mind, it is fruitful to define a scale-invariant coordinate $u \equiv z/a$ to parametrize the surface and study the equations of motion that describe its extremal configuration. For completeness, we note that, in terms of the new coordinate pair $(a,u)$ to describe the surface, the metric reads
\begin{equation}
    ds^2 = \frac{1}{u^2} \left[ - (1 - u^2) \frac{da^2}{a^2} + \frac{2 da \, du}{a} + d\eta^2 + du^2 + u^2 d\theta^2 \right] \, .
\end{equation}

The area element in the Nambu-Goto action is then specified by the determinant 
\begin{align}
    \det (\partial_\alpha X^\mu \partial_\beta X^\nu g_{\mu\nu})
    = -\frac{1}{z^4} \left[ 1 + (1-u^2) (\eta')^2 + u^2(1-u^2) (\theta')^2 \right]
\,,\end{align}
where prime denotes derivatives with respect to $u$,
and this determinant determines
\begin{align}
    \mathcal{S}_{\rm NG}[\Sigma] &= \frac{\sqrt{\lambda}}{2\pi} \int d^2\xi \sqrt{\det \left( \partial_\alpha X^\mu \partial_\beta X^\nu g_{\mu \nu}(X) \right) } \nonumber \\ &= 2 \times \frac{\sqrt{\lambda}}{2 \pi} \int \frac{da}{a} \frac{du}{u^2} \sqrt{ - \left[ 1 + (1-u^2) (\eta')^2 + u^2(1-u^2)(\theta')^2 \right]} \, . \label{eq:one_cusp_SNG_Euclidean}
\end{align}
Note that we have introduced a factor of $2$ to account for the fact that the parametrization~\eqref{eq:one_cusp_parameter} only describes half of the surface that we seek to characterize --- provided that, as we will see momentarily, the solutions reach a maximum value of $u$ as $\eta,\theta$ vary from their corresponding values at either boundary line, with this maximum being reached at exactly the midpoint value for the dependent coordinates.

Note that in \eq{one_cusp_SNG_Euclidean} is the first time we explicitly encounter the ambiguity we anticipated earlier regarding how to choose the branch cut in the area element of the Nambu-Goto action. We emphasize that this apparent ambiguity is resolved by the fact that there is a definite $i\epsilon$ prescription resulting from obtaining the definition of the area functional by analytically continuing it from its Euclidean formulation. For the time-ordered branch of the Schwinger-Keldysh contour, this results in an overall $(1-i\epsilon)$ in the argument of the square root, which in turn implies that, as long as all of the surface being described lies in the corresponding bulk submanifold, we can write
\begin{align} \label{eq:action_real}
    \mathcal{S}_{\rm NG}[\Sigma] &= i \frac{\sqrt{\lambda}}{\pi} \int \frac{da}{a} \frac{du}{u^2} \sqrt{1 + (1-u^2) (\eta')^2 + u^2(1-u^2)(\theta')^2} \, ,
\end{align}
where the integrand is now real and positive for all configurations being integrated over in the first line of \eq{W_Loop_semiclass_limit}, because those configurations correspond to real (bosonic) degrees of freedom, and thus $\eta'$ and $\theta'$ are real in said integral. 

\subsection{First attempt: direct extremization via Euler-Lagrange equations}

To extremize this action, we look for solutions to the corresponding equations of motion that satisfy 
\begin{align}
    \frac12 \Delta \eta &= \int_0^{u_{\rm max}} \eta'(u) \, du  \, , \qquad 
    \frac12 \Delta \theta = \int_0^{u_{\rm max}} \theta'(u) \, du  \, , \label{eq:deta_dtheta_constraint}
\end{align}
where $u_{\rm max}$ is the ``turning point'' of the surface --- at which $(\eta')^{-1} = \frac{du}{d\eta}$ and $(\theta')^{-1} = \frac{du}{d\theta}$ vanish.
Note that this is required by symmetry and continuity of the surface.

In principle, it is possible to simply take a variation of the action and obtain equations of motion. In fact, since the action density does not depend explicitly on the value of $\eta$ or $\theta$ --- only on their derivatives --- and  is completely independent of $a$, it follows that there exist conserved quantities $E_\eta$, $E_\theta$ such that
\begin{align}
    \frac{  (1-u^2) \eta'}{u^2 \sqrt{1 + (1-u^2)\eta'^2 + u^2 (1 - u^2) \theta'^2 }} &= E_\eta \, ,\\
    \frac{  u^2 (1-u^2) \theta'}{u^2 \sqrt{1 + (1-u^2)\eta'^2 + u^2 (1 - u^2) \theta'^2 }} &= E_\theta  \, ,
\end{align}
and therefore one may solve for $\eta',\theta'$ algebraically. The solutions to these equations are
\begin{align} \label{eq:eta_theta_sol_first_attempt}
    \eta' &= \frac{E_\eta u^2}{\sqrt{(1-u^2)(1-u^2(1+E_\theta^2) - E_\eta^2 u^4)}}
     \\
    \theta' &= \frac{E_\theta}{\sqrt{(1-u^2)(1-u^2(1+E_\theta^2) - E_\eta^2 u^4)}} \, ,
\end{align}
which can be integrated directly to yield
\begin{align} \label{eq:sol_eta_E_theta_eta}
        \Delta \eta &= \frac{\pi}{2}  E_\eta u_{+}^3 F_1 \! \left( \frac32 , \frac12 , \frac12, 2 ; u_{+}^2 , \frac{u_{+}^2 }{u_-^2} \right) \, ,  \\ \label{eq:sol_theta_E_theta_eta}
        \Delta \theta &= \frac{2 E_\theta u_{+}}{\sqrt{1 - u_{+}^2 } } K \! \left( \frac{ u_{+}^2 }{u_-^2} \frac{1 - u_{-}^2  }{ 1- u_{+}^2  }  \right) \, ,
\end{align}
where we have introduced functions of the integration constants $u_{\pm}$ (whose arguments we just omitted for brevity) as
\begin{align}
    u_{\pm}^2(E_\eta,E_\theta) = \frac{1}{2 E_\eta^2} \left( \pm \sqrt{4 E_\eta^2 + (1+E_\theta^2)^2} -(1+E_\theta^2)\right) \, ,
\end{align}
where $F_1(a,b_1,b_2,c;x,y)$ and $K(m)$ are the Appell Hypergeometric function of two variables and the Elliptic $K$ function\footnote{The Elliptic $K$ function is defined in terms of the parameter $m$ as \begin{equation} K(m) = \int_0^1 \frac{dt}{\sqrt{(1-t^2) (1- m t^2)}  } \, .
    \nonumber
\end{equation} }, respectively. Furthermore, $u_+(E_\eta,E_\theta) = u_{\rm max}$ because it is the only turning point that appears in the $u>0$ domain.

\begin{figure}
    \centering
    \includegraphics[width=0.49\linewidth]{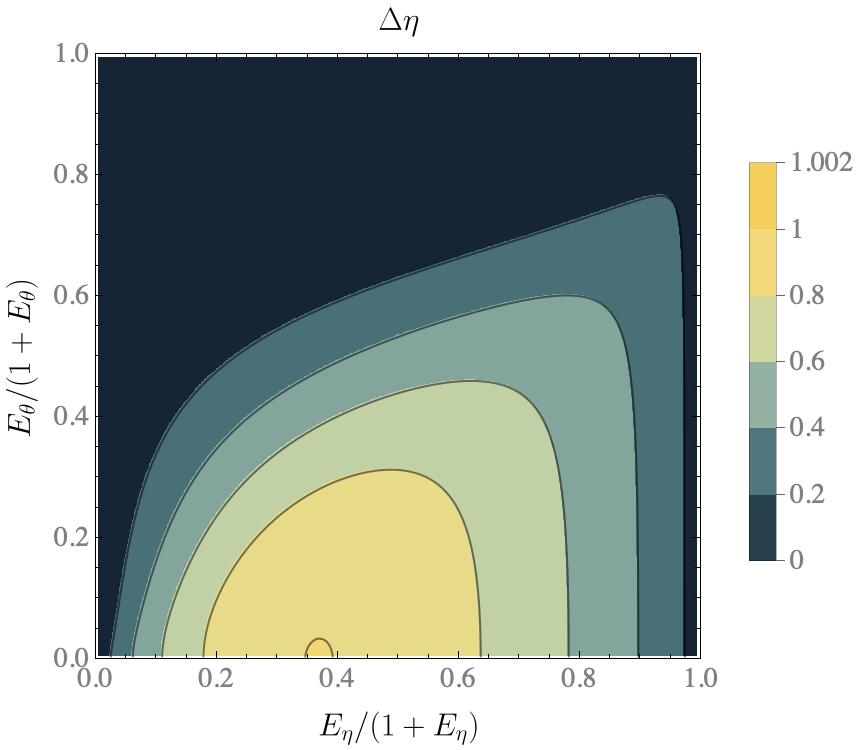}
    \includegraphics[width=0.49\linewidth]{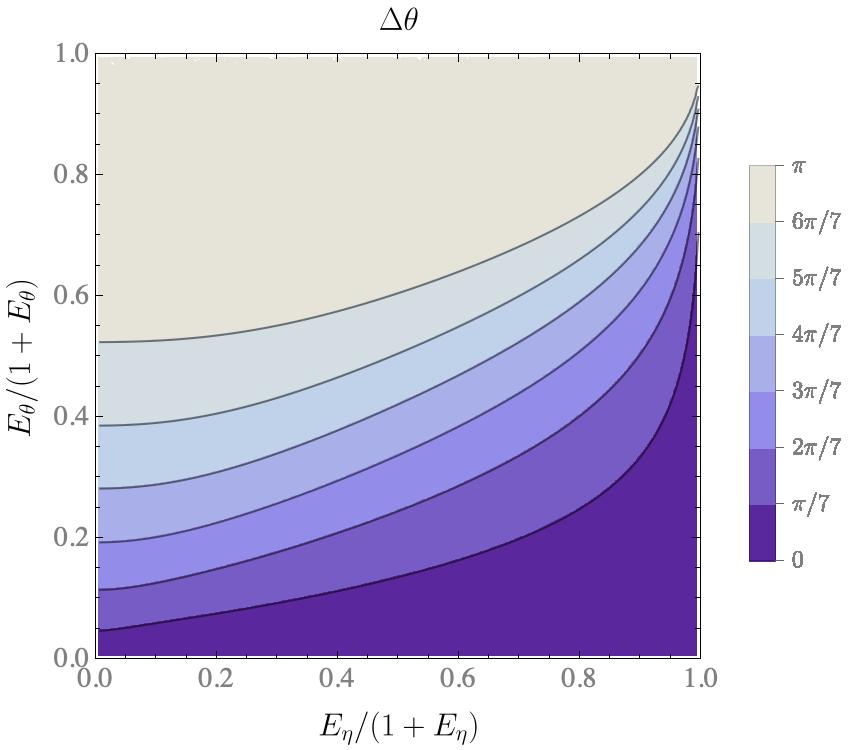}
    \caption{Contour plots with the values of $\Delta \eta$ (left) and $\Delta \theta$ (right) as a function of the integration constants $E_\theta$, $E_\eta$, corresponding to the total rapidity and angle differences spanned by the string solutions from the boundary $u = 0$ to the corresponding turning point $u_{\rm max}$. The corners of the domain correspond to the four pairs $(E_\eta,E_\theta)$ made up from combining $E_\eta \in \{ 0, \infty \}$ and $E_\theta \in \{0,\infty \}$. The result for integration constants with opposite signs can be obtained via appropriate reflections of these contours.}
    \label{fig:dtheta_deta_plot_real_constants}
\end{figure}

\begin{figure}
    \centering
    \includegraphics[width=0.49\linewidth]{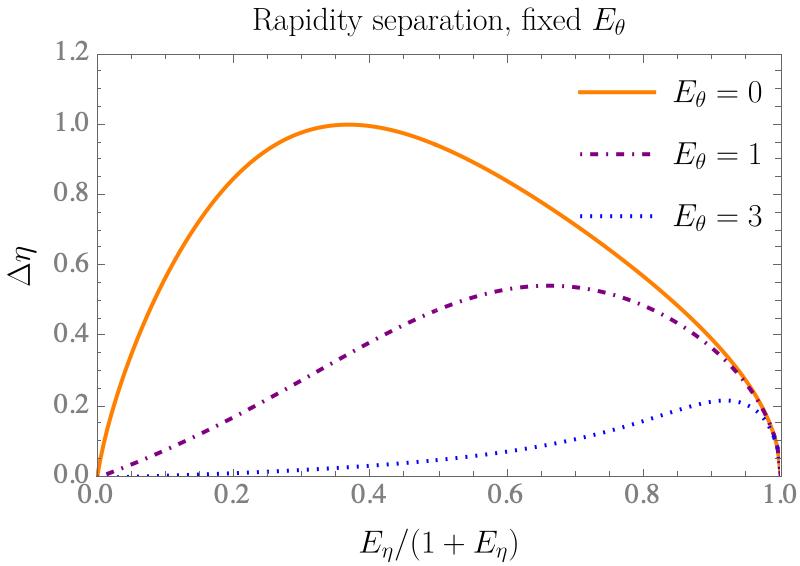}
    \includegraphics[width=0.49\linewidth]{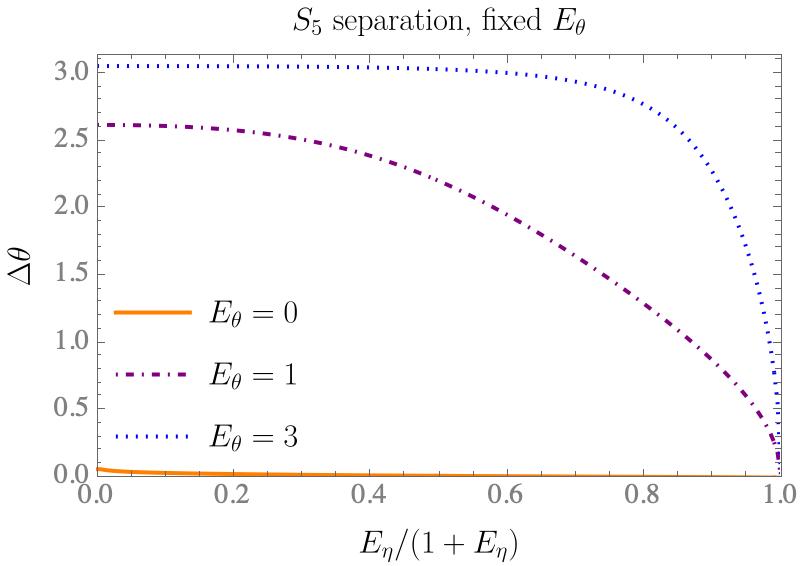}
    \caption{Values of $\Delta \eta$ (left) and $\Delta \theta$ (right) in \fig{dtheta_deta_plot_real_constants} along the $E_\theta = 0, 1, 3$ lines. The magnitude of $\Delta \eta$ reaches a global maximum along $E_\theta = 0$, with a value slightly above $\Delta \eta = 1$.}
    \label{fig:deta_plot_real_constant_Etheta_zero}
\end{figure}

It is clear that having purely real solutions $\eta(u)$, $\theta(u)$ corresponds to real integration constants $E_\theta$, $E_\eta$. However, one can determine by simple inspection that not all values of $(\Delta \eta,\Delta \theta)$ are reached in this way. In fact, as we can see from~\fig{dtheta_deta_plot_real_constants}, there exists no pair of values $(E_\eta,E_\theta)$ that yields a value of $\Delta \eta$ above a calculable bound $\Delta \eta_c \approx 1.002$ (see also ~\fig{deta_plot_real_constant_Etheta_zero}). This should be contrasted with the Euclidean calculation~\cite{Drukker:1999zq}, where the extremal surface for all conceivable values of the angle between the Wilson lines can be described in terms of real-valued integration constants.

On the other hand, having Wilson lines in a field theory with a rapidity separation $\Delta \eta$ is a well-defined quantity to calculate for any value of $\Delta \eta$, not just below $\Delta \eta_c$. At the very least, this means that the analysis we just sketched is insufficient to complete the calculation. Later on in this section, and in hindsight, we shall see that the missing ingredient in our discussion is to consider complex-valued solutions for $\eta$ and $\theta$. However, in order to justify this in a way that can be directly related to the original formulation of the problem, it is better to take a step back and redo the analysis, (re)starting from all the way from the first line of \eq{W_Loop_semiclass_limit}.

\subsection{Second attempt: extremization with Lagrange multipliers}
\label{sec:extremize_lagrange}

Given that the setup of the problem is invariant under scale transformations, we evaluate the path integral that determines the Wilson loop (with one cusp, denoted as $\mathcal{C}_1$) as
\begin{align}
    \left\langle W[\mathcal{C}_1] \right\rangle &= \int_{\partial \Sigma = \mathcal{C}_1} DX \exp \left( - \mathcal{S}_{\rm NG}[X] \right) \nonumber \\ 
    &= \int_0^\infty du_{\rm max} \int_{\substack{ \eta(u_{\rm max}) \, - \, \eta(0) \, = \, \Delta \eta/2 \\ \theta(u_{\rm max}) \, - \, \theta(0) \, = \, \Delta \theta/2 \\ \left.\frac{du}{d\eta}\right|_{u = u_{\rm max} } =\, 0 \, = \left.\frac{du}{d\theta}\right|_{u = u_{\rm max} } }} D\theta D\eta \exp \left( -  \mathcal{S}_{\rm NG}[X^\mu = (a,\eta(u),z,\theta(u) )] \right) \, , \label{eq:W_Loop_step_1}
\end{align}
where this equality only holds for the exponentially enhanced/suppressed dependence on the coupling. Note that this equality follows from the fact that the second line contains all saddle point candidates that enjoy all of the symmetry in the action and boundary conditions. We have introduced the turning point variable $u_{\rm max}$ explicitly, which is a necessary parameter in order to specify boundary conditions, and also enters the action~\eqref{eq:one_cusp_SNG_Euclidean} as an upper integration boundary for the $u$ integral.

Because the Nambu-Goto action only depends on $\eta$ and $\theta$ through their derivatives with respect to $u$, it proves useful to impose the total angle difference constraints via Lagrange multipliers at the level of the action,
\begin{align} \label{eq:effective_action_real}
    \tilde{\mathcal S}[\eta',\theta';C_\eta,C_\theta,u_{\rm max}] &=  \frac{i \sqrt{\lambda}}{ \pi} \int \frac{da}{a}\left\{ \int_0^{u_{\rm max}} \frac{du}{u^2} \,  \sqrt{1 + (1-u^2) (\eta')^2 + u^2(1-u^2)(\theta')^2} \right.
    \nn \\ 
    &\left. \quad \quad \quad \quad \quad \quad + C_\eta \left[ \frac{\Delta \eta}{2} - \int_0^{u_{\rm max}} \eta' \, du  \right] 
    + C_\theta \left[ \frac{\Delta \theta}{2} - \int_0^{u_{\rm max}} \theta' \, du \right] 
    \right\} \, ,
\end{align}
where the Wilson loop expectation value now reads 
\begin{align}
    \left\langle W[\mathcal{C}_1] \right\rangle 
    &= \int  dC_\eta dC_\theta du_{\rm max} \int_{\substack{ \left.\frac{du}{d\eta}\right|_{u = u_{\rm max} } =\, 0 \\ \left.\frac{du}{d\theta}\right|_{u = u_{\rm max} } =\, 0  }} D\theta D\eta \exp \left( -  \tilde{\mathcal S}[\eta',\theta';C_\eta,C_\theta,u_{\rm max}] \right) \, . \label{eq:W_Loop_step_2}
\end{align}

Crucially, the integrand on the right hand side of \eq{W_Loop_step_2} does not directly depend on the value of $\eta$ or $\theta$ at all --- it only depends on the derivatives $\eta',\theta'$. Then, via a linear change of variables, we can write (to leading order in $\sqrt{\lambda}$) exactly the same expression
\begin{align}
    \left\langle W[\mathcal{C}_1] \right\rangle 
    &= \int  dC_\eta dC_\theta du_{\rm max} \int_{\substack{ \left.\frac{du}{d\eta}\right|_{u = u_{\rm max} } =\, 0 \\ \left.\frac{du}{d\theta}\right|_{u = u_{\rm max} } =\, 0  }} D\theta' D\eta' \exp \left( -  \tilde{\mathcal S}[\eta',\theta';C_\eta,C_\theta,u_{\rm max}] \right) \, , \label{eq:W_Loop_step_3}
\end{align}
but with the path measure over $\eta,\theta$ being instead given in terms of their derivatives.

With these new integration variables, one can carry out the integrals over $\eta'$ and $\theta'$ analytically in the saddle point approximation. Taking functional derivatives of $\tilde{\mathcal{S}}$ with respect to $\eta'$ and $\theta'$ at any $u \neq u_{\rm max}$, one finds
\begin{align}  \label{eq:eta_sol_Ceta_Ctheta}
    \eta' &= \frac{C_\eta u^2}{\sqrt{(1-u^2)(1-u^2(1+C_\theta^2) - C_\eta^2 u^4)}} \, ,
     \\
    \theta' &= \frac{C_\theta}{\sqrt{(1-u^2)(1-u^2(1+C_\theta^2) - C_\eta^2 u^4)}} \, , \label{eq:theta_sol_Ceta_Ctheta}
\end{align}
and we see that the Lagrange multipliers $C_\eta,C_\theta$ are exactly the integration constants $E_\eta,E_\theta$ we had earlier, but now promoted to integration variables as part of the determination of the saddle point configurations. 

The derivative boundary conditions at $u = u_{\rm max}$ are satisfied if
\begin{align}
    u_{\rm max} = 1  & & {\rm or} & & u_{\rm max} = u_+(C_\eta,C_\theta) \, ,
\end{align}
where the first possibility implies that the string configuration becomes complex between $u = u_+$ and $u = 1$, with $\eta'$ and $\theta'$ taking imaginary values. Because of this, the action associated with the saddle point with $u_{\rm max} = 1$ will be complex-valued, and thus will feature an exponential suppression in addition to a phase.

In the end, it will turn out that both possibilities are relevant, exchanging dominance depending on the values of $\Delta \eta$ and $\Delta \theta$. At first glance, one would be inclined to think that only $u_{\rm max} = u_+$ contributes to the final result, because the integrals over $C_\eta$ and $C_{\theta}$ are defined over the real line, with which, after using~\eqs{eta_sol_Ceta_Ctheta}{theta_sol_Ceta_Ctheta},  $\tilde{\mathcal{S}}$ is a purely imaginary number for $u_{\rm max} = u_+$ and a complex number with positive real part for $u_{\rm max} = 1$. This means that across the entire domain of integration of $C_\eta$ and $C_\theta$ the saddle for $(\eta',\theta',u_{\rm max})$ with $u_{\rm max} = 1$ is exponentially suppressed relative to the one with $u_{\rm max} = u_+$. However, the saddle points for the integral over the Lagrange multipliers $C_\eta$, $C_\theta$ need not lie on the original integration domain. In fact, one can  conclude from~\fig{dtheta_deta_plot_real_constants} that $\Delta \eta > \Delta \eta_c$ requires going outside of the original integration domain. In this situation, both $u_{\rm max} = 1$ and $u_{\rm max} = u_+$ will have complex-valued actions, and there is no a priori guarantee that one will always dominate over the other.

In the case with $u_{\rm max} = 1$, one needs to specify how to evaluate the square roots in \eqs{eta_sol_Ceta_Ctheta}{theta_sol_Ceta_Ctheta} when their argument becomes negative (i.e., what is the correct contour prescription for the integral over $u$). We give a detailed account of the complex analysis considerations needed to fix the sign of the imaginary part of $\eta'$ and $\theta'$, as well as that of the real part of $\tilde{\mathcal{S}}$ in Appendix~\ref{app:complex_analysis_eta_theta}.

Then, at leading order in $\sqrt{\lambda}$ (omitting all $\mathcal{O}(\lambda^0)$ terms),
one arrives at
\begin{align}
    \left\langle W[\mathcal{C}_1] \right\rangle 
    &= \int_{-\infty}^\infty dC_\eta dC_\theta  \left[ \exp \left( -  \tilde{\mathcal S}^{(+)}(C_\eta,C_\theta) \right) +  \exp \left( -  \tilde{\mathcal S}^{(1)}(C_\eta,C_\theta) \right) \right] \, , \label{eq:W_Loop_step_4}
\end{align}
where
\begin{align}
    \tilde{\mathcal S}^{(+)}(C_\eta,C_\theta) &= \frac{i \sqrt{\lambda}}{\pi} 
    \int \frac{da}{a} \left\{ \frac{C_\eta \Delta \eta}{2} + \frac{C_\theta \Delta \theta}{2} + \int_0^{u_+(C_\eta,C_\theta)}\frac{du}{u^2} \sqrt{\frac{1-(1+C_\theta^2)u^2 - C_\eta^2 u^4}{1-u^2}} \right\}  \, , \\
    \tilde{\mathcal S}^{(1)}(C_\eta,C_\theta) &= \frac{i \sqrt{\lambda}}{\pi} 
    \int \frac{da}{a} \left\{ \frac{C_\eta \Delta \eta}{2} + \frac{C_\theta \Delta \theta}{2} + \int_0^{1}\frac{du}{u^2} \sqrt{\frac{1-(1+C_\theta^2)u^2 - C_\eta^2 u^4 - i \epsilon}{1-u^2}} \right\} \, \, ,
\end{align}
where the $i\epsilon$ inside the square root specifies that the $u$ integration contour has to pass \textit{above} $u_+(C_\eta,C_\theta)$ in the complex plane. In what follows, we will use superscripts $(+)$ and $(1)$ to distinguish quantities corresponding to the saddles with $u_{\rm max} = u_+$ and $u_{\rm max} = 1$, respectively.

As such, we have reduced the problem of obtaining the leading $\sqrt{\lambda}$ dependence to calculating a 2-dimensional integral over real numbers. After having obtained the result, one can re-interpret these integrals as contour integrals over complex $C_\eta$ and $C_\theta$, and the integration paths may be then deformed into the complex plane to evaluate the integrals using the steepest descent method. To rigorously substantiate this last observation (which we shall make use of later on in section~\ref{sec:saddle_point_solution}), in Appendix~\ref{app:complex_analysis_complex_C} we discuss how the functions $\tilde{\mathcal{S}}_1$ and $\tilde{\mathcal{S}}_+$ are naturally extended into the complex $C_\eta$ and $C_\theta$ planes at the level of the Wilson loop~\eqref{eq:W_Loop_step_3} from its definitions in terms of a path integral over $\eta'$ and $\theta'$.

\subsection{Divergences and Regularization} 
\label{sec:divergences_regularization}

So far, we have discussed the extremization of the Nambu-Goto action without paying close attention to the fact that, as written, all of its expressions are infinite quantities. This is something that can be observed as early as Eqs.~\eqref{eq:one_cusp_SNG_Euclidean} or~\eqref{eq:action_real}, where
\begin{enumerate}
    \item There is a small-$u$ divergence coming from the integral $\int \tfrac{du}{u^2}$, which contains in it the (infinite) mass of the probe particle that acts as a color source and generates the Wilson line. The contribution from said mass is not a genuine contribution to the Wilson loop; all it does is to shift the overall energy of the state. 
    \item As a consequence of considering scale-invariant worldsheet configurations, the integral over proper time $\int \tfrac{da}{a}$ is logarithmically divergent, and features both UV and IR divergences.
\end{enumerate}
Because of the variables we have chosen to parametrize the surface, these two divergences are apparently multiplying each other. 

First, to isolate the genuine contributions to the Wilson loop, which come solely from the light degrees of freedom of $\mathcal{N}=4$ sYM, we subtract from the action the corresponding result for two straight Wilson lines, given in Minkowski coordinates by $\tfrac{i\sqrt{\lambda}}{2\pi } \left( \tfrac{1}{v_1^0} + \frac{1}{v_2^0} \right) \int dt \int_0^\infty \frac{dz}{z^2}$, where $v_1^0 = \gamma(v_1)$ and $v_2^0 = \gamma(v_2)$ are the corresponding boost factors of each line in the frame defined by the $t$ coordinate. After rewriting this in terms of $(a,u)$ coordinates, and doing the sum in equal proper time steps along each line, it reads $\tfrac{i\sqrt{\lambda}}{\pi} \int \frac{da}{a} \int_0^\infty \frac{du}{u^2}$. The result of doing this yields
\begin{align}
    \tilde{\mathcal S}_{\rm eff}(C_\eta,C_\theta) = \frac{i \sqrt{\lambda}}{\pi} 
    \int \frac{da}{a} \Bigg\{ & \frac{C_\eta \Delta \eta}{2} + \frac{C_\theta \Delta \theta}{2} - \frac{1}{u_{\rm max}} \nonumber \\ & + \int_0^{u_{\rm max}}\frac{du}{u^2} \left[\sqrt{\frac{1-(1+C_\theta^2)u^2 - C_\eta^2 u^4 - i\epsilon}{1-u^2}} - 1 \right]  \Bigg\}  \, . 
    \label{eq:tilde_S_eff}
\end{align}
where $u_{\rm max}$ can be either $1$ or $u_+(C_\eta,C_\theta)$, and we have included the $i\epsilon$ that specifies the contour prescription for $u_{\rm max} = 1$. 

Subtracting $\tfrac{i\sqrt{\lambda}}{\pi} \int \frac{da}{a} \int_0^\infty \frac{du}{u^2}$ is independent of $\Delta \eta$ and $\Delta \theta$, and also independent of $C_\eta$ and $C_\theta$. Even if one were concerned that the change of variables in the divergent integral from $(t,z)$ to $(a,u)$ might change the actual value being subtracted from the result, both are explicitly independent of the physical parameters on which the Wilson loop depends, making it a manifestly uniform change in the energy of the state being evolved. For our purposes, we could have simply stated this as
\begin{align}
    \frac{\partial \tilde{\mathcal S}}{\partial C_\eta} = \frac{\partial \tilde{\mathcal S}_{\rm eff}}{\partial C_\eta} \, , & & \frac{\partial \tilde{\mathcal S}}{\partial C_\theta} = \frac{\partial \tilde{\mathcal S}_{\rm eff}}{\partial C_\theta} \, .
\end{align}

This means that, after doing the subtraction we just discussed, all that remains is to understand and discuss the physics of the logarithmic divergence in the integral over proper time, which is unambiguously isolated as the only remaining divergent contribution. To make contact with the discussion in the introduction, it is helpful to introduce UV and IR regulators $(\Lambda^{-1},L)$ in the calculation, in the form of step functions $\Theta(z-\Lambda^{-1})$ and $\Theta(L-a)$, respectively, as part of the integration measure. The scale $\Lambda$ is the inverse of the length scale that defines the usual UV regulator in holographic calculations, denoted as $\epsilon$ in \eq{W_cusp_div}. 
For simplicity, we have chosen an IR regulator in terms of the proper time coordinate; however, in practice, this IR regulator should be set by the scale at which the Wilson loop is not well-described by a single cusp, and can have a physical effect.\footnote{For the Wilson loop to be closed (and thus a well-defined gauge-invariant quantity), there must be a distance scale beyond which the geometry of the problem cannot be described only in terms of the scale-invariant geometry we have discussed so far. For example, for a loop with a cusp at the origin in a $2D$ plane, described by the radial distance to the cusp $r$, in terms of an angular parameter $\phi \in (0,\pi/2)$ as $(x,y) = (r(\phi) \cos(\phi) , r(\phi) \sin \phi )$ and $r(\phi) = \sin(2\phi)$, $L$ will be an $\mathcal{O}(1)$ number.}
In $(a,u)$ coordinates, these step functions amount to $\Theta(u - 1/(a\Lambda)) \Theta(L- a)$. Carrying out the integral over $a$, one gets 
\begin{equation}
    \int \frac{da}{a} \Theta(u - 1/(a\Lambda)) \Theta(L- a) = \ln (\Lambda L u) \, \Theta(u - (\Lambda L)^{-1} )  \approx \ln(L/\epsilon) + \ln(u) \, ,
\end{equation}
where the first term on the right hand side is exactly the logarithm appearing in the exponent of \eq{W_cusp_div}, and the second term is a $u$-dependent piece that is dependent on the choice of IR regulator (e.g., an IR cutoff on Minkowski time $t$ would lead to a different residual $u$-dependent contribution). The step function $\Theta(u - (\Lambda L)^{-1} )$ may be set to unity because the $u$ integral in \eq{tilde_S_eff} is finite as $u \to 0$ (even with the extra $\ln(u)$ multiplicative factor).

Therefore, up to a UV and IR finite term, we can write
\begin{align}
    \tilde{\mathcal S}_{\rm eff}(C_\eta,C_\theta) &= \frac{i \sqrt{\lambda}}{\pi} 
    \ln(\Lambda L) \Bigg\{ \frac{C_\eta \Delta \eta}{2} + \frac{C_\theta \Delta \theta}{2} - \frac{1}{u_{\rm max}} \nonumber \\ & \quad\quad\quad\quad\quad\quad + \int_0^{u_{\rm max}}\frac{du}{u^2} \bigg[\sqrt{\frac{1-(1+C_\theta^2)u^2 - C_\eta^2 u^4 - i\epsilon}{1-u^2}} - 1 \bigg]  \Bigg\} \, ,
      \label{eq:tilde_S_eff_log}
\end{align}
where the mass scale $\Lambda$ is a UV regulator, and $L$ is an IR scale characteristic of the temporal extent of the cusp geometry. The logarithmic divergence has the same origin as in~\cite{Drukker:1999zq}, where the UV cutoff is given in terms of a length $\epsilon$ (a symbol we already used for a different purpose). 
Note that we may already anticipate from the structure of this expression (by comapring with~\eq{W_cusp_div}) that the cusp anomalous dimension will emerge as the numerical coefficient in front of the logarithm, once the remaining integrals are carried out.

With these considerations, everything is explicitly finite. In hindsight, we see that, if anything, the largeness of $\Lambda L$ only enhances the validity of our saddle point approximation for the path integrals over $\eta(u),\theta(u)$.

\subsection{Finding the saddle points for the Lagrange multipliers} 
\label{sec:saddle_point_solution}

Finally, we can evaluate the Wilson loop in terms of the regularized actions as
\begin{align}
    \left\langle W[\mathcal{C}_1] \right\rangle 
    &= \int_{-\infty}^\infty dC_\eta dC_\theta  \left[ \exp \left( -  \tilde{\mathcal S}^{(+)}_{{\rm eff}}(C_\eta,C_\theta ) \right) + \exp \left( -  \tilde{\mathcal S}^{(1)}_{{\rm eff}}(C_\eta,C_\theta ) \right) \right] \, , \label{eq:W_Loop_step_5}
\end{align}
where each instance of $\tilde{\mathcal S}_{\rm eff}$ is given by \eq{tilde_S_eff_log} with the corresponding value of $u_{\rm max}$.

Using the multiplication of large parameters $\sqrt{\lambda}\ln(\Lambda L)$ as a large parameter, we can now evaluate the integral over $C_\eta,C_\theta$ of each term with yet another application of the steepest descent method. Note that, while the saddle point analysis we present in what follows is convenient and fitting for the limiting case of interest here, if all one wants is to obtain a number, one could simply carry out the integral in~\eq{W_Loop_step_5} numerically. We discuss this approach for selected examples in Appendix~\ref{app:numbers}. 

As one may anticipate from our earlier analysis, it will not always be possible to find real saddles for $C_\eta,C_\theta$. However, with the present approach in our calculation, we can directly inspect the integrand for singularities and unambiguously decide whether the integration contour can be deformed into the complex plane to go over a given candidate saddle point. While our analysis in the main body of this work only concerns itself with a thorough characterization of the saddle points, there could in principle be other contributions to the integral (e.g., if a singularity or branch cut prevents one from deforming the integration contour in a way that only the saddle points contribute). See Appendix~\ref{app:complex_analysis_C} for more details on all of these points.

The saddle point conditions that result from varying $\tilde{\mathcal{S}}_{\rm eff}$ with respect to $C_\eta$ and $C_\theta$ are simply~\eq{deta_dtheta_constraint}, with $\eta', \theta'$ given by~\eqs{eta_sol_Ceta_Ctheta} {theta_sol_Ceta_Ctheta}. Since the explicit expressions depend on the value of $u_{\rm max}$, the relation that determines $C_\eta$ and $C_\theta$ as a function of $\Delta \eta$ and $\Delta \theta$ is different in each case. Explicitly, 
\begin{align}
    \Delta \eta &= 2C_\eta \int_{\mathcal{L}(u_{\rm max} ) } \frac{u^2 d u}{\sqrt{(1-u^2) (1 - (1 + C_\theta^2) - C_\eta^2 u^4 ) } } \, , \label{eq:contour_generic_eta} \\
    \Delta \theta &= 2C_\theta \int_{\mathcal{L}(u_{\rm max} ) } \frac{d u}{\sqrt{(1-u^2) (1 - (1 + C_\theta^2) - C_\eta^2 u^4 ) } } \, , \label{eq:contour_generic_theta}
\end{align}
where, for real $C_\eta$ and $C_\theta$, $\mathcal{L}(u_{\rm max} )$ is any path in the complex $u$ plane from $0$ to $u_{\rm max}$ that only goes through the upper right quadrant. For complex values of the Lagrange multipliers, the integration contour $\mathcal{L}(u_{\rm max} )$ should be arranged in such a way that the singularities do not cross the integration path as the integrand is analytically continued. The starting configuration of the path (i.e., when the Lagrange multipliers are real) is fixed by the $i\epsilon$ prescription discussed in Appendix~\ref{app:complex_analysis_eta_theta}. Effectively, this prescription is equivalent to saying that the real part of the action is positive.

\paragraph{Finding the saddle points for $u_{\rm max} = u_+$.}

\begin{figure}
    \centering
    \includegraphics[width=0.7\linewidth]{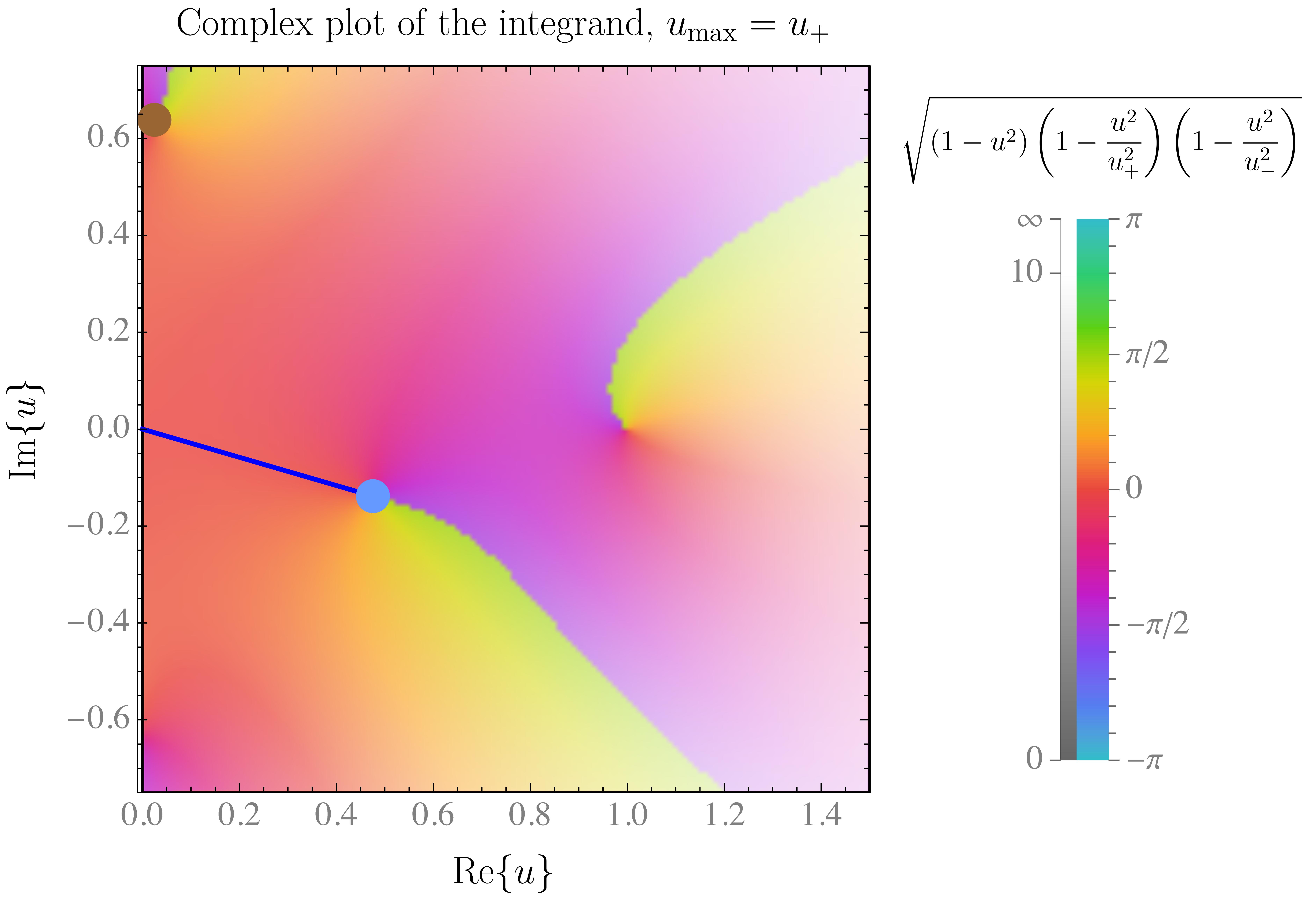}
    \caption{Plot of the denominator $\sqrt{(1-u^2) (1 - (1 + C_\theta^2) - C_\eta^2 u^4 )}$ in~\eq{contour_generic_eta} and~\eq{contour_generic_theta} in the complex $u$ plane. The blue line is the integration contour that defines the analytic continuation of the expressions for $\Delta \eta$ and $\Delta \theta$ as functions of complex $C_\eta$, $C_\theta$ for the $u_{\rm max} = u_+$ saddle. Here we chose $C_\eta = 3+i$ and $C_\theta = 1 + i$. The light blue point is the position of $u_+$; the brown point is the position of $u_-$. There are branch cuts starting from both of these points as well as from $u = 1$.}
    \label{fig:complex_umax_uplus}
\end{figure}

When $u_{\rm max} = u_+$, the saddle point condition takes the same form as when we discussed the real saddles, given in~\eqs{sol_eta_E_theta_eta}{sol_theta_E_theta_eta}. In terms of $C_\eta$ and $C_\theta$, they read
\begin{align} 
        \Delta \eta &= \frac{\pi}{2}  C_\eta u_{+}^3(C_\eta,C_\theta) F_1 \! \left( \frac32 , \frac12 , \frac12, 2 ; u_{+}^2(C_\eta,C_\theta), \frac{u_{+}^2(C_\eta,C_\theta) }{u_{-}^2(C_\eta,C_\theta)} \right) \, ,  
    \label{eq:sol_C_eta}
        \\
        \Delta \theta &= \frac{2 C_\theta u_{+}(C_\eta,C_\theta)}{\sqrt{1 - u_{+}^2(C_\eta,C_\theta) } } K \! \left( \frac{ u_{+}^2(C_\eta,C_\theta) }{u_{-}^2(C_\eta,C_\theta)} \,\frac{1 - u_{-}^2(C_\eta,C_\theta)  }{ 1- u_{+}^2(C_\eta,C_\theta)  }  \right) \,,
    \label{eq:sol_C_theta}
\end{align}
where, crucially, the functions on the right hand side are analytic functions --- i.e., they are well-defined into the complex plane, and, up to branch cuts, define the unique analytic continuation of the result on the real line. In terms of a complex contour integral, $\mathcal{L}(u_+)$ may simply be understood as the integral from $0$ to $u_+$. See~\fig{complex_umax_uplus} for a typical configuration. 

Therefore, the task at hand is to solve for $C_\eta,C_\theta$ at fixed, real-valued $\Delta \eta,
\Delta \theta$. Some algebraic manipulations are helpful for this purpose. First, it is convenient to introduce the parameter of the Elliptic $K$ function
\begin{equation}
    \tilde{x} \equiv \frac{ u_{+}^2 }{u_{-}^2} \,\frac{1 - u_{-}^2  }{ 1- u_{+}^2  } = \frac{u_+^2 C_\theta^2 - 1}{1 - u_+^2} \, ,
\end{equation}
(where we have omitted the arguments of $u_\pm(C_\eta,C_\theta)$) and use it to solve for $u_+$ in terms of $\Delta \theta$. One obtains
\begin{equation}
    u_+^2 = 1 - \left[ \left(\frac{\Delta \theta}{2K(\tilde{x})} \right)^2 - \tilde{x} \right]^{-1} \, ,
\end{equation}
and as a byproduct one arrives at explicit expressions for $C_\eta,C_\theta$ in terms of $\tilde{x}$ that parametrize the subset of Lagrange multiplier values that solve \eq{sol_C_theta}:
\begin{align}
    C_\theta^2 = \frac{\left(\frac{\Delta \theta}{2K(\tilde{x})} \right)^2 }{\left(\frac{\Delta \theta}{2K(\tilde{x})} \right)^2 - \tilde{x} - 1} \, , \qquad
    C_\eta^2 = \frac{\left[1- \left(\frac{\Delta \theta}{2K(\tilde{x})} \right)^2 \right] \left[\left(\frac{\Delta \theta}{2K(\tilde{x})} \right)^2 - \tilde{x} \right]}{\left[\left(\frac{\Delta \theta}{2K(\tilde{x})} \right)^2 - \tilde{x} - 1\right]^2} \, .
\end{align}

The remaining step is to insert these expressions into \eq{sol_C_eta} and find $\tilde{x}$ that satisfies the equation for a given $(\Delta \eta,\Delta \theta)$. Operationally, it is easier to first fix $\Delta \theta$ and then plot the imaginary and real parts of what would be $\Delta \eta$, i.e. the right hand side of \eq{sol_C_eta}, as a function of complex $\tilde{x}$. Candidate saddle points are then obtained by finding the values of $\tilde{x}$ such that $\Delta\eta$ is real. 
Whether the candidate saddle points can be reached or not by deforming the integration contours starting from the real line into the complex plane is a task to be ascertained case-by-case. We discuss this analysis in more detail in Appendix~\ref{app:complex_analysis_C}; here we simply focus on describing the position of the saddle points. 

\begin{figure}
    \centering
    \hspace{0.01\linewidth} 
    \includegraphics[width=0.475\linewidth]{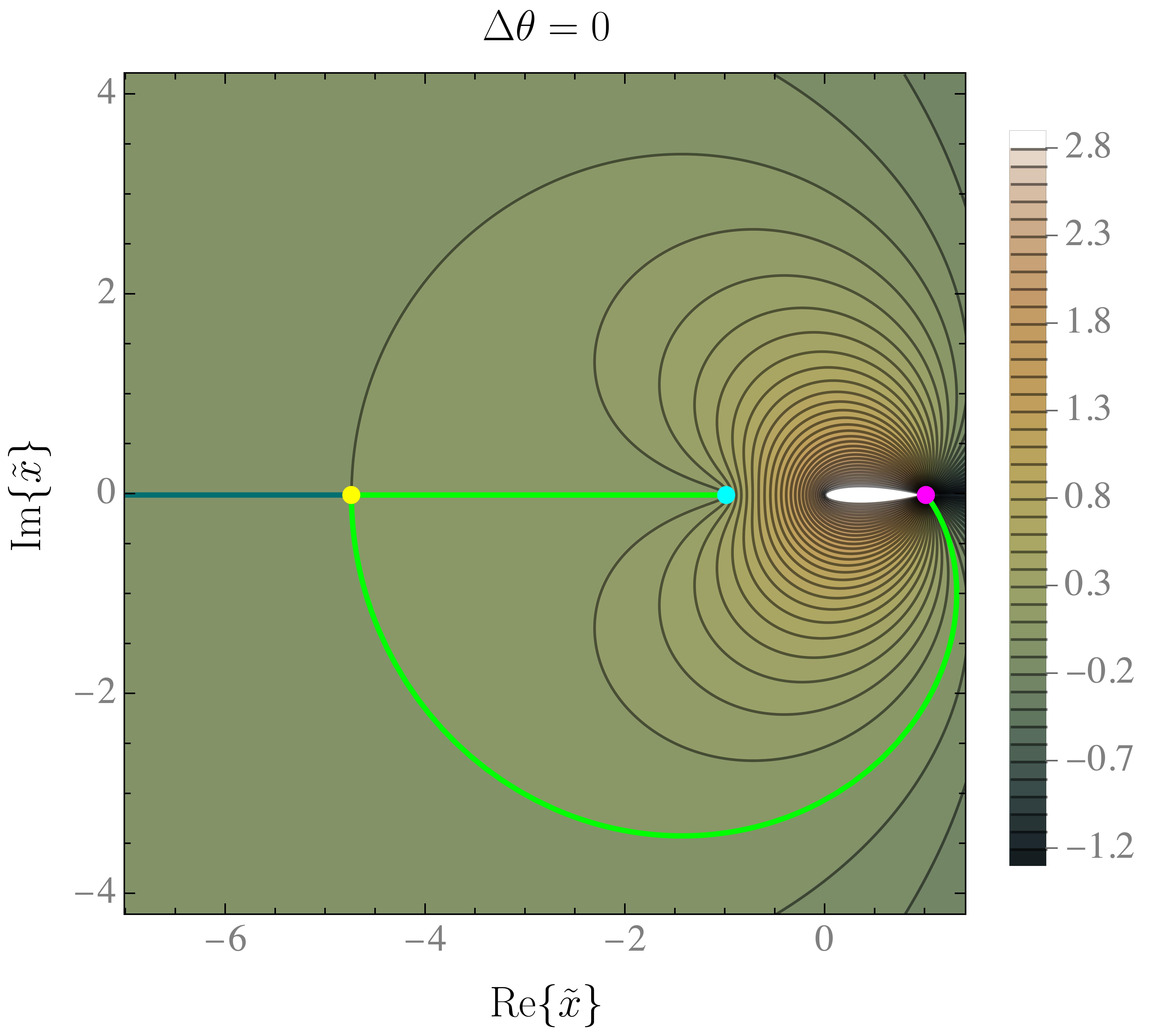}
    \hspace{0.01\linewidth}
    \includegraphics[width=0.475\linewidth]{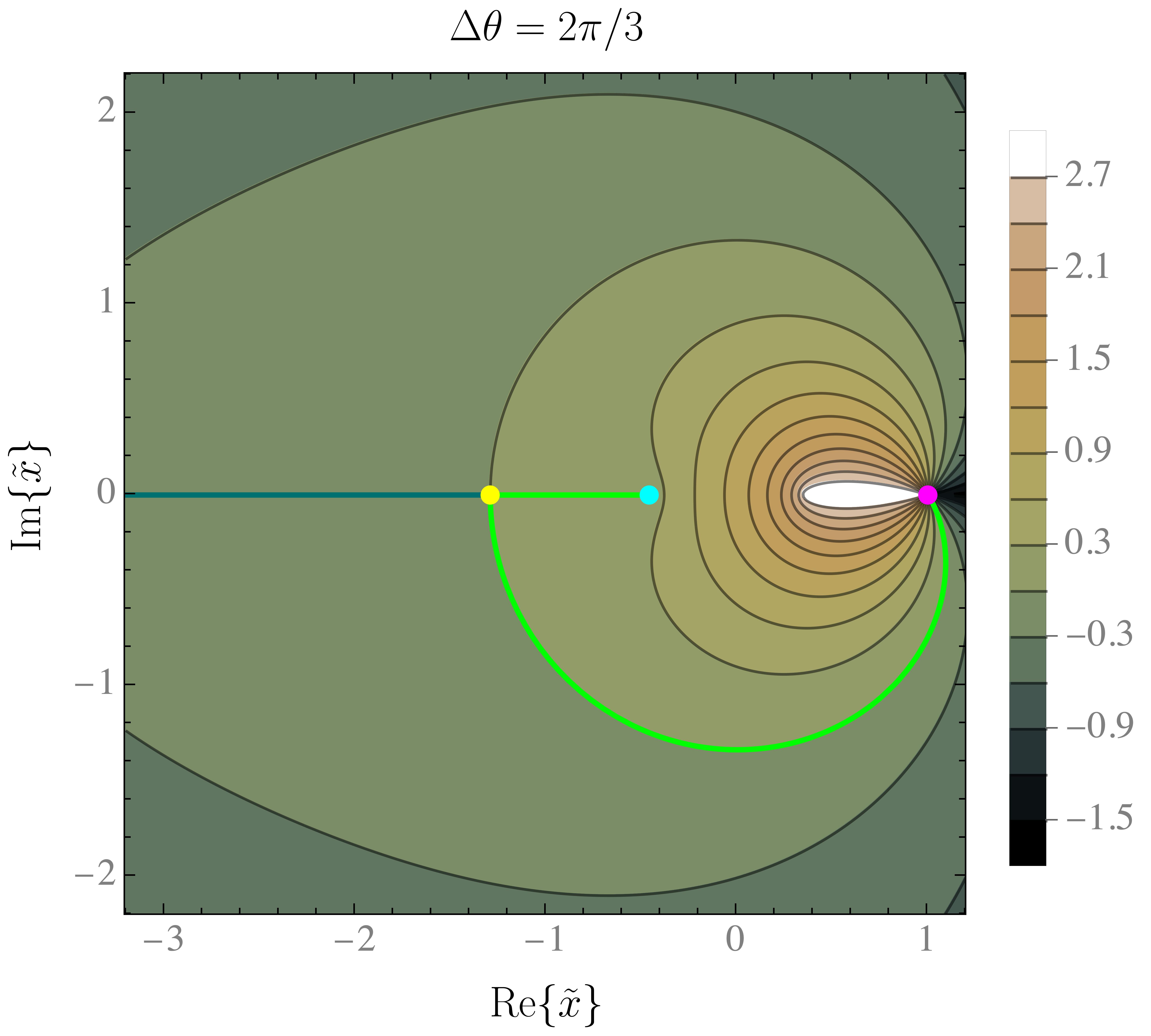}
    
    \vspace{0.2cm} \includegraphics[width=0.49\linewidth]{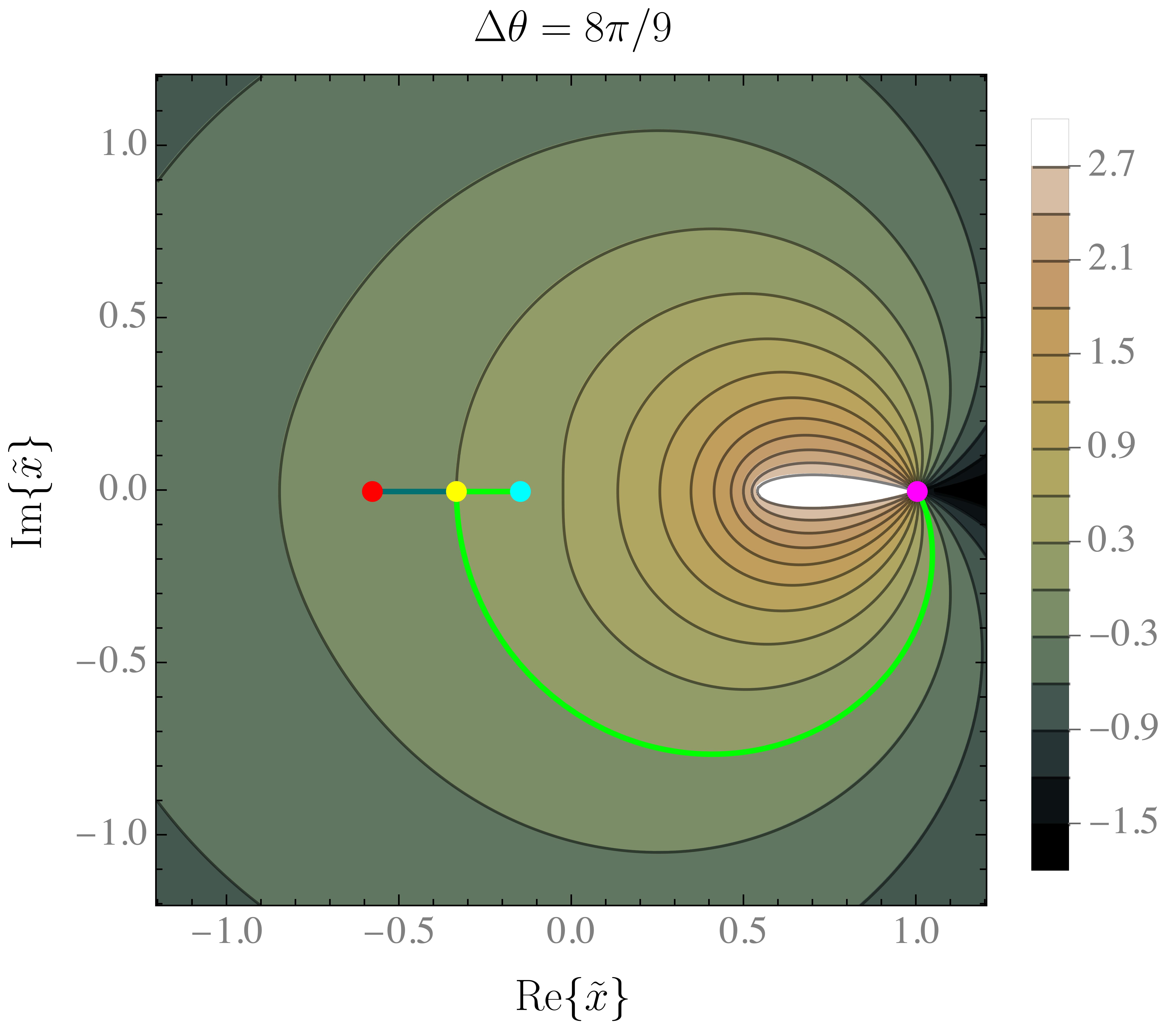}
    \includegraphics[width=0.49\linewidth]{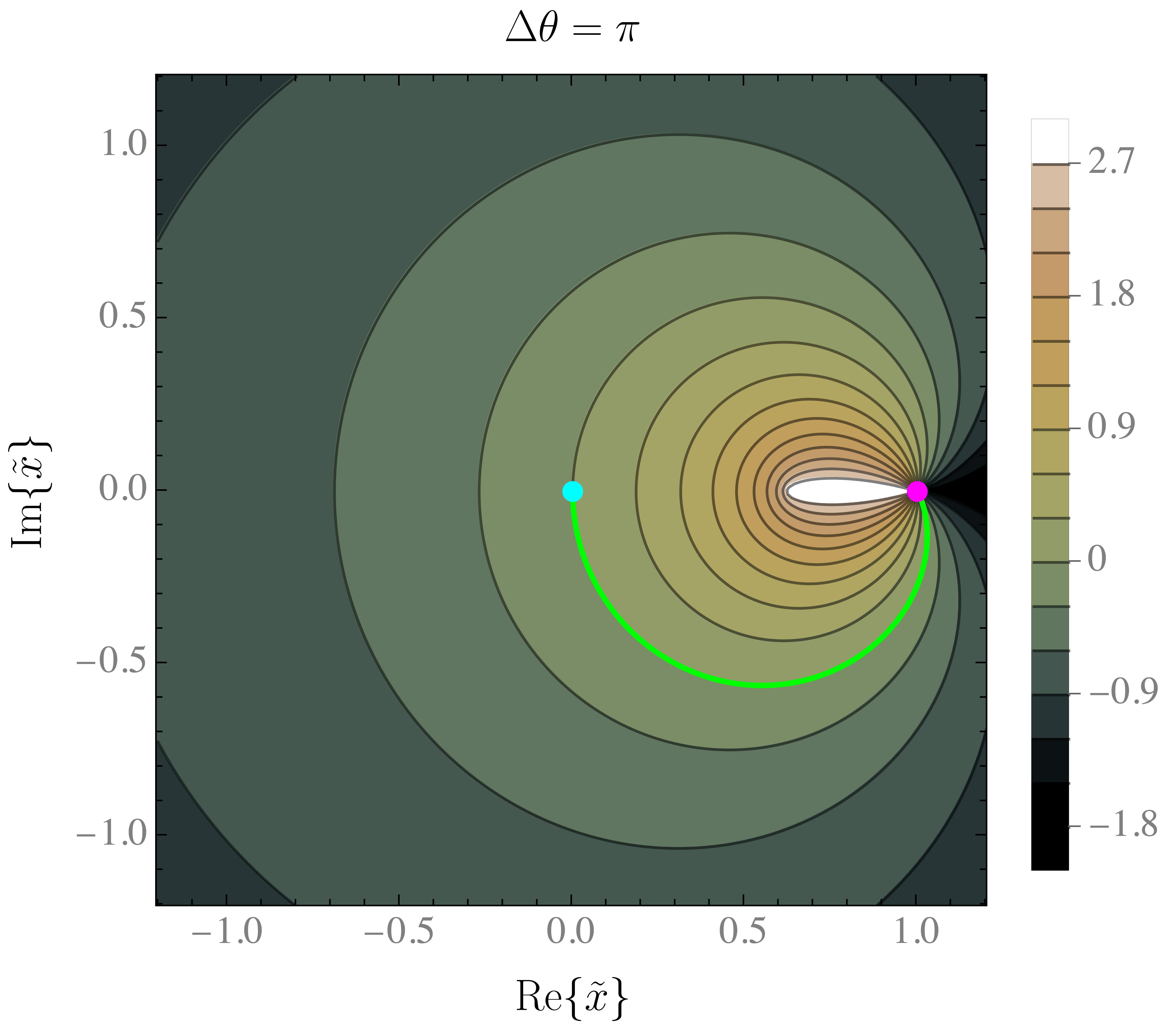}
    \caption{Values of the imaginary part of $\Delta \eta$ as a function of $\tilde x$, computed through~\eq{sol_C_eta}.
    The highlighted green contour marks vanishing $\mathrm{Im}\{\Delta\eta\}$. The cyan and red points label solutions $\tilde x$ for which $\Delta\eta=0$, the magenta point labels a solution $\tilde x$ for which $\Delta\eta\to \infty$, and the yellow point labels a solution $\tilde x$ for which $\Delta \eta$ reaches its maximum along the real line.
    See main text for more a detailed discussion.}
    \label{fig:latte}
\end{figure}

In~\fig{latte} we show contour plots of the imaginary part of $\Delta \eta$ as specified by \eq{sol_C_eta}, for four values of $\Delta \theta \in \{0, 2\pi/3, 8\pi/9, \pi \}$. Even though they will not be part of our subsequent discussion, we examine intermediate values of $\Delta \theta$ in addition to $0$ and $\pi$ because the contour plots for these extreme cases exhibit a qualitative difference in the number of segments that provide candidate saddles. We highlight the contours with vanishing imaginary parts, where the contributing saddle points (i.e., the ones that can be reached via contour deformations) lie, with a green line. As we will explain below, we draw one of them with a dark cyan line. The endpoints of these segments are given by:
\begin{itemize}
    \item The solution for $\tilde{x}$ in the equation 
    \begin{equation}
        1 + \tilde{x} = \left( \frac{\Delta \theta}{2 K(\tilde{x})} \right)^2 \, ,
    \end{equation}
    which corresponds to $u_+^2 = 0$ and $C_\eta^2, C_\theta^2 \to \infty$. This point is colored cyan in~\fig{latte}. Geometrically, this corresponds to a worldsheet profile that does not go into the bulk at all, and has $\Delta \eta = 0$. The value of $\tilde{x}$ that satisfies this equation is a real number between $-1$ (for $\Delta \theta = \pi$) and $0$ (for $\Delta \theta = 0$).
    \item The solution for $\tilde{x}$ in the equation
    \begin{equation}
        \left( \frac{\Delta \theta}{2 K(\tilde{x})} \right)^2 = 1 \, ,
    \end{equation}
    which implies that $C_\eta^2 = 0$ and $u_+^2 = 1/(1+C_\theta^2)$. This point is colored red in~\fig{latte} (only visible for $\Delta \theta = 8\pi/9$). This corresponds to a surface that does go into the bulk and has $\Delta \eta = 0$. For all $0<\Delta \theta < \pi$, the value of $\tilde{x}$ that satisfies this equation is a real number between $-\infty$ and $-1$ being equal to the former if $\Delta \theta = 0$ and to the latter if $\Delta \theta = \pi$. 
    \item $\tilde{x} = 1$. This point corresponds to $u_+^2 = 2$, $C_\eta^2 = -1/4$, and $C_\theta^2 = 0$, because $|K(\tilde{x})| \to \infty$ there. This point is colored magenta in~\fig{latte}. This corresponds to the configuration with $\Delta \eta \to \infty$, and coincides with the ones found by~\cite{Kruczenski:2002fb,Makeenko:2002qe} for lightlike Wilson lines.
    \item There is an additional point of interest, which is the point along the real $\tilde{x}$ axis when the green contours in~\fig{latte} take a 90 degree turn into the complex plane, colored yellow in the figure and that we shall refer to as $\tilde{x}^*$. For all $\Delta \theta < \pi$, this point is given by the value of $\tilde{x}$ for which $\Delta \eta$ as written in~\eq{sol_C_eta} reaches a maximum along the real line (we have not found analytic expressions for $\tilde{x}^*$ or the value of $\Delta \eta$ at this point; we only display numerical determinations as shown in the figures).
\end{itemize}

We display the candidate saddle points that are further down the negative $\tilde{x}$ real axis in a different color (dark cyan) because (as we will see in the next subsection) these correspond to configurations with higher energy (imaginary part of the action) than the other points on the real axis (which span the same $\Delta \eta$ interval). As such, they are unstable configurations.
The case $\Delta \theta = \pi$ is special in that $\tilde{x}^* = 0$ and the contour immediately heads into the complex plane starting from $\Delta \eta = 0$.
In all cases, the interval with real values of $\tilde{x}$ corresponds to saddle points with real $C_\eta, C_\theta$, and when $\tilde{x}$ becomes complex so do the Lagrange multipliers.

\paragraph{Finding the saddle points for $u_{\rm max} = 1$.}

\begin{figure}
    \centering
    \includegraphics[width=0.7\linewidth]{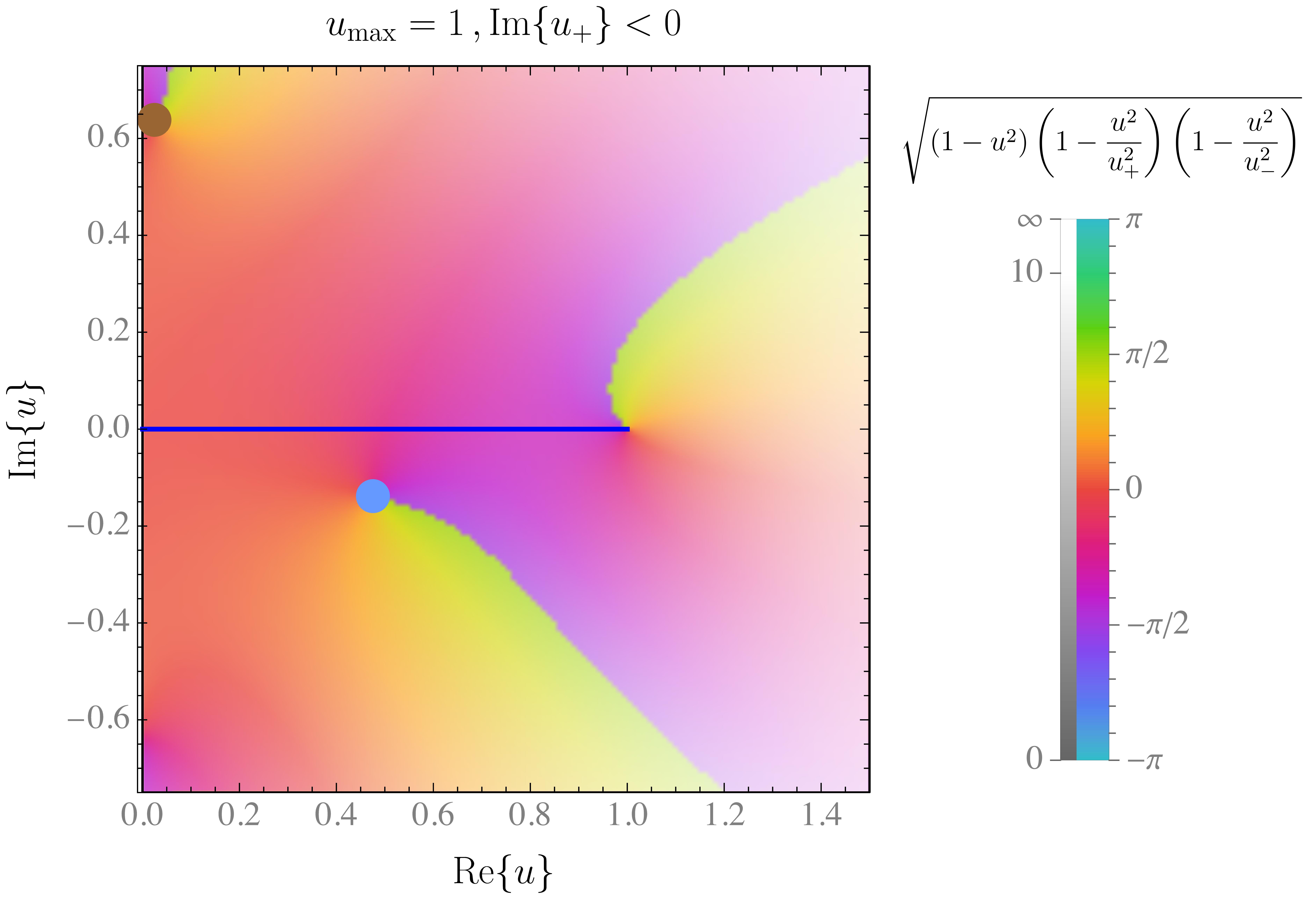}
    \includegraphics[width=0.7\linewidth]{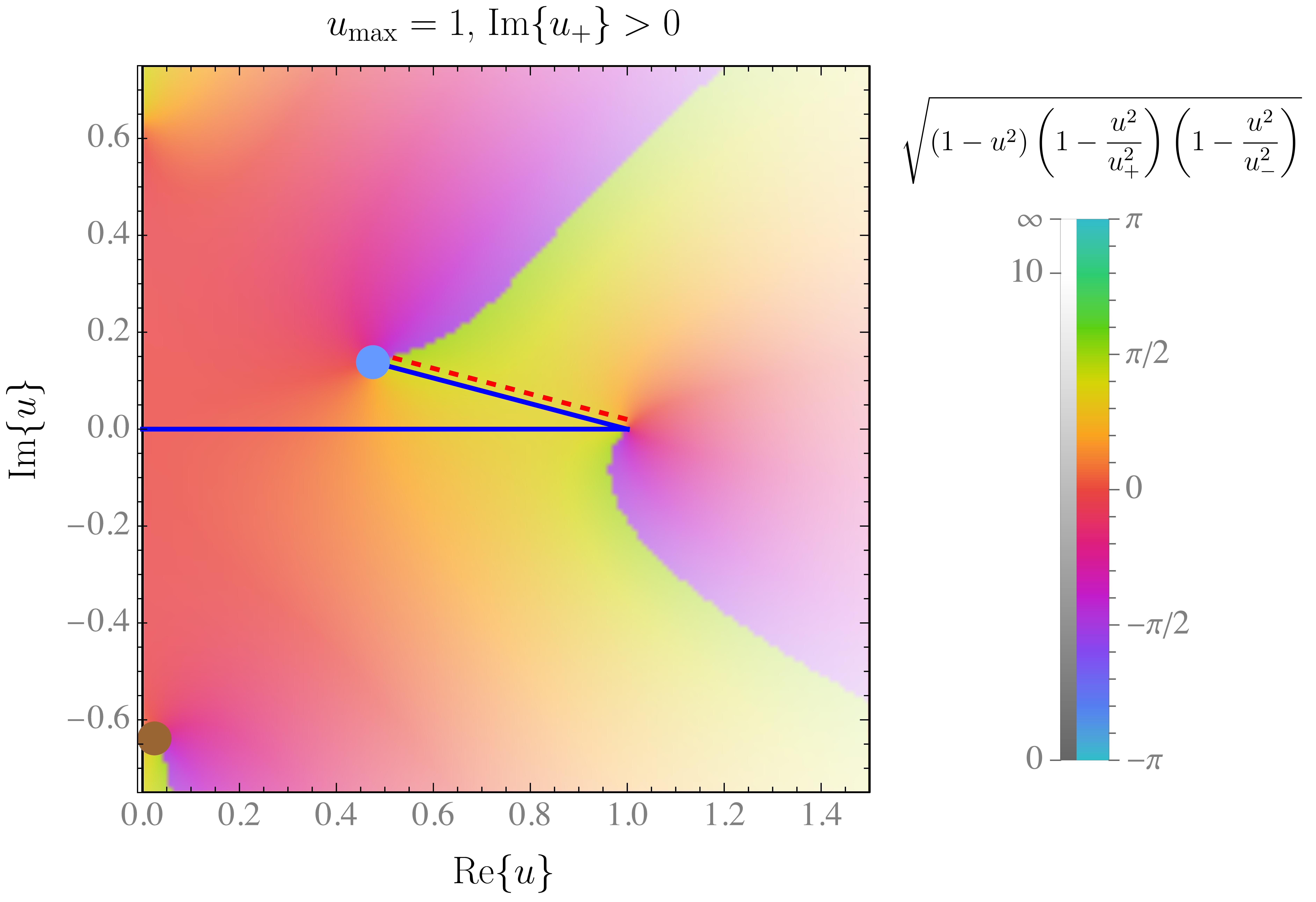}
    \caption{Plot of the denominator $\sqrt{(1-u^2) (1 - (1 + C_\theta^2) - C_\eta^2 u^4 )}$ in~\eq{contour_generic_eta} and~\eq{contour_generic_theta} in the complex $u$ plane. In each panel, the light blue point is the position of $u_+$; the brown point is the position of $u_-$. There are branch cuts starting from both of these points as well as from $u = 1$. Top panel: Typical configuration with ${\rm Im}\{u_+\} < 0$. The blue line is the integration contour that defines the analytic continuation of the expressions for $\Delta \eta$ and $\Delta \theta$ as functions of complex $C_\eta$, $C_\theta$ for the $u_{\rm max} = 1$ saddle. In the top panel we chose $C_\eta = 3+i$ and $C_\theta = 1 + i$. Bottom panel: Typical configuration with ${\rm Im}\{u_+\} > 0$. The blue lines together with the red dashed line (on the second Riemann sheet behind the cut starting at $u = u_+$) is the integration contour that defines the analytic continuation of the expressions for $\Delta \eta$ and $\Delta \theta$ as functions of complex $C_\eta$, $C_\theta$ for the $u_{\rm max} = 1$ saddle.  In the bottom panel we chose $C_\eta = 3-i$ and $C_\theta = 1 - i$.}
    \label{fig:complex_umax_one}
\end{figure}

In this case, the saddle point conditions are given by the integrals of $\eta'$ and $\theta'$ up to $u = 1$, with a contour integration path that passes above the singularity at $u = u_+$ (see~\fig{complex_umax_one}). When ${\rm Im}\{u_+\} < 0$, this path can be taken to be a straight line to $u=1$. On the other hand, when ${\rm Im}\{u_+\} > 0$, a deformation is required. In this situation, the integration contour over $u$ can be deformed into three straight segments: 
\begin{enumerate}[label=(\roman*)]
    \item A segment going from $u = 0$ to $u = 1$,
    \item a segment going from $u = 1$ to $u = u_+$,
    \item and a segment going from $u = u_+$ to $u = 1$, on the other side of the branch cut that starts at $u = u_+$.
\end{enumerate}
Because the last two segments (the blue segment and the dashed red segment in the lower panel of~\fig{complex_umax_one}) go on opposite sides of a branch cut introduced by a square root, they contribute the same amount and with the same sign (the minus sign from the different branch compensates the sign from the direction of integration). Explicitly, we have
\begin{align}
    \Delta \eta &= 2C_\eta \Bigg[ \int_{0}^1 \!\! \frac{u^2 d u}{\sqrt{(1-u^2) (1 - (1 + C_\theta^2)u^2 - C_\eta^2 u^4 ) } } + 2\int_{1}^{u_+} \!\!\!\!\! \frac{u^2 d u}{\sqrt{(1-u^2) (1 - (1 + C_\theta^2)u^2 - C_\eta^2 u^4 ) } } \Bigg] \, , \label{eq:Delta_Eta_As_Contour_umax1} \\
    \Delta \theta &= 2C_\theta \Bigg[ \int_{0}^1 \!\! \frac{d u}{\sqrt{(1-u^2) (1 - (1 + C_\theta^2)u^2 - C_\eta^2 u^4 ) } } +  2 \int_{1}^{u_+} \!\!\!\!\! \frac{d u}{\sqrt{(1-u^2) (1 - (1 + C_\theta^2)u^2 - C_\eta^2 u^4 ) } } \Bigg] \, , \label{eq:Delta_Theta_As_Contour_umax1}
\end{align}
where we emphasize that some care needs to be taken in that the branch cut prescriptions for the square root should be continuously connected to the one for real values of the Lagrange multipliers. It turns out that only this second situation (with ${\rm Im}\{u_+\}>0$; illustrated in the lower panel of~\fig{complex_umax_one}) can lead to real values of $\Delta \eta$; we therefore focus on this one in what follows.

While it is possible to rewrite the expression for $\Delta \eta$ in terms of Appell Hypergeometric functions of the first kind, in practice we find it more convenient to evaluate the integral expression directly. On the other hand, The expression for $\Delta \theta$ can be developed further in a way that, as in the $u_{\rm max} = u_+$ case, greatly simplifies the problem of finding candidate saddle points by reducing the problem from two complex variables to one. Starting from~\eq{Delta_Theta_As_Contour_umax1}, direct integration and some algebraic manipulations yield
\begin{equation}
    \Delta \theta = 2 C_\theta \sqrt{\frac{u_+^2}{u_+^2 - 1}} \tilde{K}(\tilde{y}) \, , \label{eq:Delta_Theta_Elliptic_K_umax1_1}
\end{equation}
where we have introduced
\begin{align}
    \tilde{K}(\tilde{y}) &\equiv \left[ K(\tilde{y}) - 2 \sqrt{1/{\tilde{y}}} \, K(1/\tilde{y} ) \right] \, , \\
    \tilde{y} &\equiv \frac{1}{u_-^2} \frac{u_+^2 - u_-^2}{u_+^2 - 1} \, .
\end{align}
Using that $u_+^2/u_-^2 = (1 + C_\theta^2) u_+^2 - 1$, we may rewrite $C_\theta$ in terms of $\tilde{y}$ and $u_+$, which then by replacing it in~\eq{Delta_Theta_Elliptic_K_umax1_1} returns an equation that determines $u_+$ in terms of $\Delta \theta$ and $\tilde{y}$. One obtains
\begin{equation}
    u_+^2 = \frac{2 - \tilde{y} + \left( \frac{\Delta \theta}{2 \tilde{K}(\tilde{y}) } \right)^2 }{1 - \tilde{y} + \left( \frac{\Delta \theta}{2 \tilde{K}(\tilde{y}) } \right)^2} \, ,
\end{equation}
which can then be used to parametrize the values of $C_\eta^2$ and $C_\theta^2$ that correspond to configurations with a given value of $\Delta \theta$
\begin{align}
    C_\theta^2 &= \frac{\left( \frac{\Delta \theta}{2 \tilde{K}(\tilde{y}) } \right)^2 }{2 - \tilde{y} + \left( \frac{\Delta \theta}{2 \tilde{K}(\tilde{y}) } \right)^2} \, , \\
    C_\eta^2 &= - \frac{\left[1 - \tilde{y} + \left( \frac{\Delta \theta}{2 \tilde{K}(\tilde{y}) } \right)^2 \right] \left[1  + \left( \frac{\Delta \theta}{2 \tilde{K}(\tilde{y}) } \right)^2 \right] }{\left[2 - \tilde{y} + \left( \frac{\Delta \theta}{2 \tilde{K}(\tilde{y}) } \right)^2\right]^2} \, .
\end{align}

\begin{figure}
    \centering
    \includegraphics[width=0.49\linewidth]{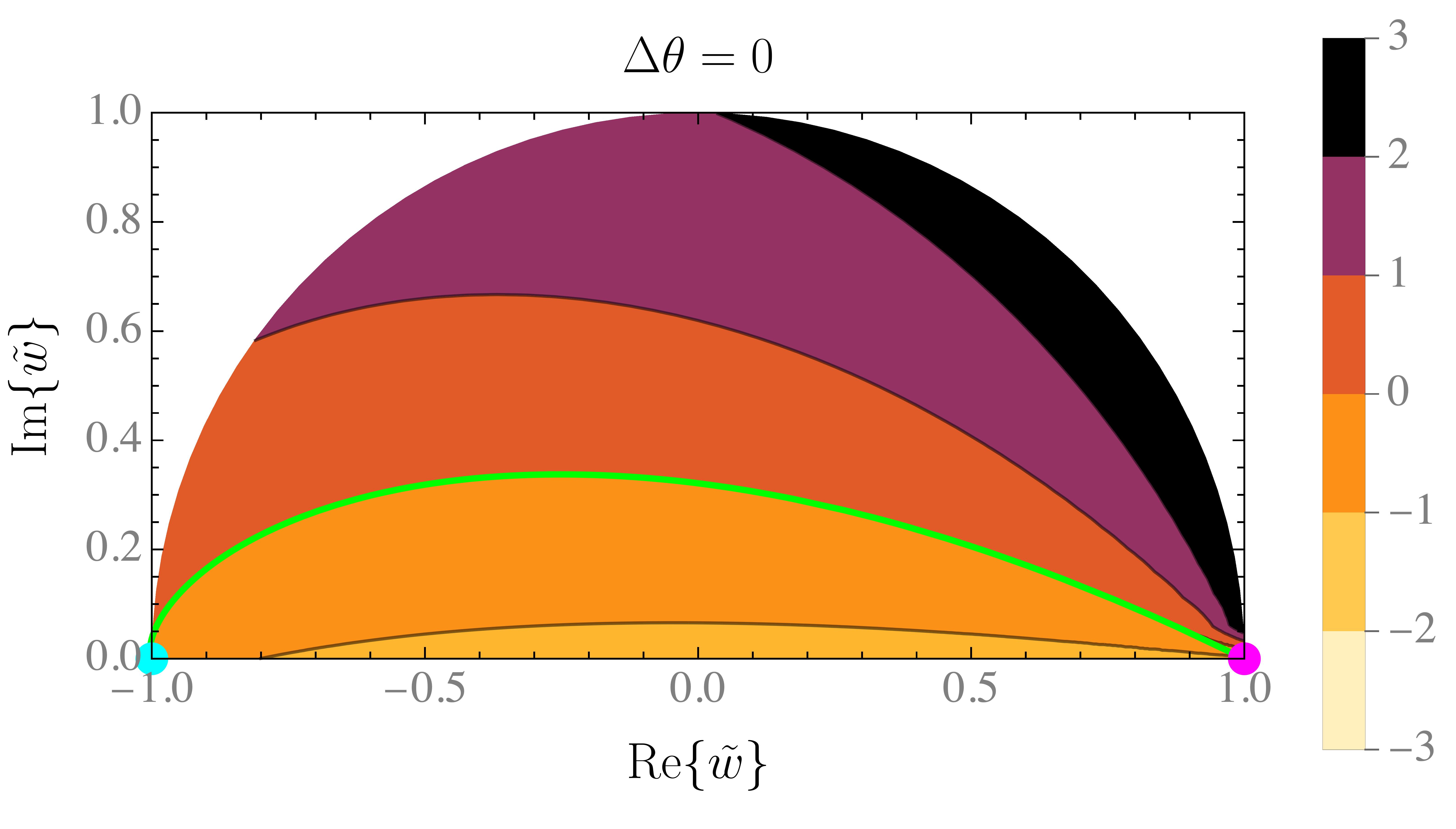}
    \includegraphics[width=0.49\linewidth]{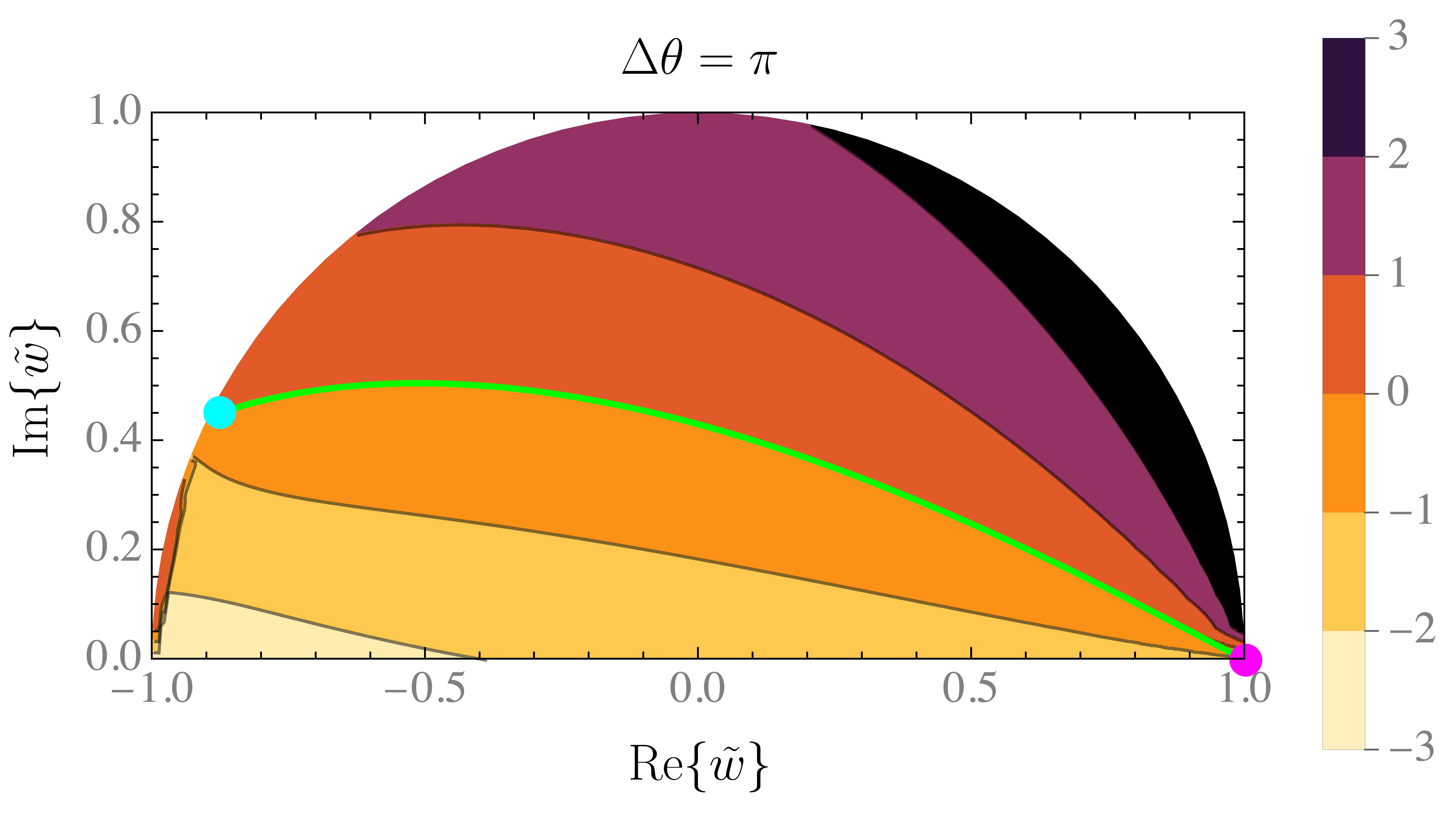}
    \caption{Values of the imaginary part of $\Delta \eta$ as a function of $\tilde{w}$, computed through~\eq{Delta_Eta_As_Contour_umax1}. The highlighted green contour marks vanishing ${\rm Im}\{\Delta \eta\}$. The cyan point labels the $\Delta \eta = 0$ solution, and the magenta point labels the solution for which $\Delta \eta \to \infty$.}
    \label{fig:sunset}
\end{figure}

Again, as in the $u_{\rm max} = u_+$ case, the remaining step is to insert these expressions into~\eq{Delta_Eta_As_Contour_umax1} and find $\tilde{y}$ that satisfies the equation for a given $(\Delta \eta, \Delta \theta)$. For this purpose, we have found it useful to further define $\tilde{w} = \frac{1+\tilde{y} }{1 - \tilde{y} }$, which is a conformal transformation that maps the left half of the complex plane --- where we have found the saddle point candidates to lie --- to the unit circle. In~\fig{sunset} we show contour plots of the imaginary part as specified by~\eq{Delta_Eta_As_Contour_umax1} for $\Delta \theta \in \{ 0, \pi \}$. Because there is no qualitative change in the fact that there is a unique segment that describes ${\rm Im}\{\Delta \eta\} = 0$ spanning all values of ${\rm Re}\{\Delta \eta\}$, and since our main focus is on these two extreme values, we do not display the intermediate cases. We highlight the contour with vanishing imaginary part in green. Its endpoints are given by: 
\begin{itemize}
    \item The solution for $\tilde{y}$ in the equation
    \begin{equation}
        \left(\frac{\Delta \theta}{2\tilde{K}(\tilde{y}) }\right)^2 + 1 = 0 \, ,
    \end{equation}
    which corresponds to $\Delta \eta = 0$ by setting $C_\eta = 0$. This point is colored cyan in~\fig{sunset}. For the case with $\Delta \theta = 0$, it corresponds to $\tilde{y} \to \infty$, which gets mapped to $\tilde{w} = -1$.
    \item $\tilde{y} = 1$. In the same way as $\tilde{x} = 1$ in the $u_{\rm max} = 1$ case, this point corresponds to $u_+^2 = 2$, $C_\eta^2 = -1/4$ and $C_\theta^2 = 0$, because $|\tilde{K}(\tilde{y})| \to \infty$ as $\tilde{y} \to 1$. This corresponds to the configuration with $\Delta \eta \to \infty$.
\end{itemize}

In all cases (and for both values of $u_{\rm max}$), we found the green contour segments described above, i.e., the solutions with real $\Delta \eta, \Delta \theta$, by numerically finding the zeroes of the imaginary part of $\Delta \eta$.

\subsection{Extract the cusp anomalous dimension at strong coupling}
\label{sec:one_cusp_result}

\begin{figure}
    \centering
    \includegraphics[width=0.49\linewidth]{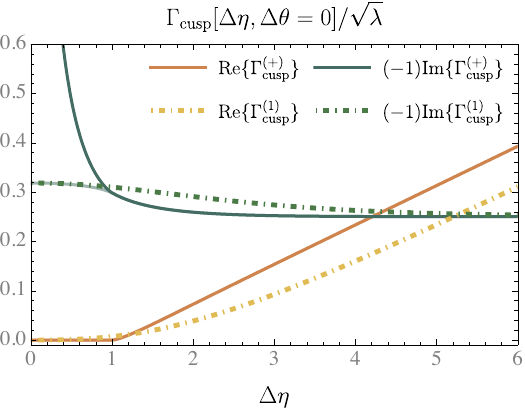}
    \includegraphics[width=0.49\linewidth]{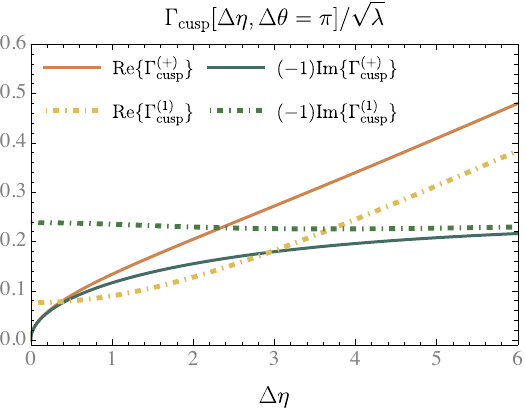}
    \caption{Candidate values of the cusp anomalous dimension, calculated from the expectation value of the Wilson loop in terms of $\tilde{S}_{\rm eff}$ evaluated at its saddle points, as a function of $\Delta \eta$ for $\Delta \theta = 0$ (left) and $\Delta \theta = \pi$ (right). For each value of $\Delta \eta$, the curve with the lowest value of the real part of $\Gamma_{\rm cusp}$ is the one that gives the dominant contribution to the Wilson loop, and therefore determines the behavior of the cusp anomalous dimension. We show the resulting value from the action with $u_{\rm max} = u_+$ (solid curves, lighter green for the unstable branch of the real configurations) and with $u_{\rm max} = 1$ (dot-dashed curves).}
    \label{fig:cuspy}
\end{figure}

With the values of $\tilde{x}$ and $\tilde{y}$ determined (respectively, for $u_{\rm max} = u_+$ and $u_{\rm max} = 1$) for each desired value of the pair $(\Delta \eta, \Delta \theta)$, the expectation value of the Wilson loop~\eqref{eq:W_Loop_step_5} in terms of the extrema of the integral becomes
\begin{align}
    \langle W[\mathcal{C}_1]\rangle &= \exp \left( - \tilde{\mathcal{S}}_{\rm eff}^{(+)}(\tilde{C}_\eta^{(+)}(\Delta \eta, \Delta \theta ), \tilde{C}^{(+)}_\theta(\Delta \eta, \Delta \theta )) \right) \nonumber \\
    &\quad  + \exp \left( - \tilde{\mathcal{S}}_{\rm eff}^{(1)}(\tilde{C}^{(1)}_\eta(\Delta \eta, \Delta \theta ), \tilde{C}^{(1)}_\theta(\Delta \eta, \Delta \theta )) \right) \, , \label{eq:W_Loop_step_6}
\end{align}
where $\tilde C^{(+,1)}_\eta(\Delta \eta, \Delta \theta ), \tilde C^{(+,1)}_\theta(\Delta \eta, \Delta \theta )$ are the values of the Lagrange multipliers at the contributing saddle points determined in the way we described in the previous subsection.\footnote{Strictly speaking, this sum can have three terms if we consider the two saddles with $u = u_+$ at small $\Delta \eta$. Because one represents a state with higher energy than the other, we will only keep the one with smaller (more negative) imaginary part in this discussion --- we anyways show its value when we present our results in~\fig{cuspy}.}\textsuperscript{,}\footnote{Furthermore, if obstructions to deform the original integration contour into a steepest descent contour exist, there could in principle be additional terms in this sum coming from regions of the complex plane that are not saddle points. One example of where such an obstruction could in principle take place is given by the magenta dot-dashed lines in~\fig{contours_umax_uplus_one} in~\app{complex_analysis_C}. While these segments are not part of the steepest descent contour, we have checked by direct numerical integration over the original integration contour that they do not contribute (see~\app{numbers}) in the instances where we have good numerical control. That being said, at a formal level, the situation would be more satisfactory if one could find an argument that systematically shows that these segments do not modify the asymptotic behavior of the integral (a counterexample would be even more interesting).}
It follows that the
anomalous dimension for the one-cusp configuration is simply given by the coefficient of $\ln (\Lambda L)$ in \eq{tilde_S_eff_log}, for the action $\tilde{\mathcal{S}}_{\rm eff}$ with the smallest real part, as that one will give the dominant contribution to the right hand side of~\eq{W_Loop_step_6}. That is to say, the result is given by calculating
\begin{equation}
\label{eq:gcusp_onecusp_sol}
    \Gamma_{\rm cusp}^{(s)}[\Delta \eta, \Delta \theta] = \frac{i \sqrt{\lambda}}{\pi} 
    \Bigg\{   \int_{\mathcal{L}(u_{\rm max}(s)) } \frac{du}{u^2} \left[ \sqrt{\frac{1-u^2}{1 - (1 + C_\theta^2) u^2  - C_\eta^2 u^4 }} - 1 \right] - \frac{1}{u_{\rm max}(s) }  \Bigg\} \, ,
\end{equation}
where $s \in \{+,1\}$ as before, and when $s = +$ we evaluate $u_+ = u_+(\tilde C^{(+)}_\eta, \tilde C^{(+)}_\theta)$, and $\tilde C^{(+)}_\eta, \tilde C^{(+)}_\theta$ are defined as the solutions to \eqs{sol_C_eta}{sol_C_theta} as above. Once~\eq{gcusp_onecusp_sol} is calculated, all one has to do is compare the results coming from the two saddles.

We plot this quantity in~\fig{cuspy} for the two classes of saddle points as a function of $\Delta \eta$ for $\Delta \theta \in \{0 , \pi \}$.
We observe the following features, all consistent with previous results in the literature~\cite{Kruczenski:2002fb,Makeenko:2002qe,Makeenko:2008xr,Drukker:2011za}, but which also include new results:
\begin{itemize}
    \item At small $\Delta \eta$, the saddle that determines the cusp anomalous dimension is the one with $u_{\rm max} = u_+$ (i.e., $s = +$). At sufficiently large $\Delta \eta$, the cusp anomalous dimension is governed by the saddle with $u_{\rm max} = 1$ (i.e., $s=1$) in both cases, as the real part of the action becomes the smallest. This saddle dominance exchange is a surprising result to us. See~\fig{result_intro} for plots with only the real part and highlighting the results from the dominant configurations.
    \item At large $\Delta \eta$, the leading behavior of $\Gamma_{\rm cusp}$ is linear in $\Delta \eta$ (as was known even before an AdS/CFT calculation was carried out~\cite{Korchemsky:1985xj,Korchemsky:1987wg,Ivanov:1985np,Korchemsky:1988hd,Alday:2007mf}), and takes the form
    \begin{equation}\label{eq:Gcusp_lrap_limit}
        \Gamma_{\rm cusp} \sim \frac{\sqrt{\lambda}}{4\pi} \Delta \eta \, ,
    \end{equation}
    independent of the value of $\Delta \theta$, the separation on the $S_5$, and is the same for both saddles that entered our analysis. This linear behavior agrees with the established result~\cite{Kruczenski:2002fb,Makeenko:2002qe} in \eq{inf_angle_scaling}. The imaginary part of $\Gamma_{\rm cusp}$ approaches a finite value, given by
    \begin{equation}
        \lim_{\Delta \eta \to \infty} {\rm Im} \{\Gamma_{\rm cusp}\} = \frac{\sqrt{\lambda}}{\pi} \left\{ \int_0^1 \frac{du}{u^2} \left[ \sqrt{\frac{(1-u^2/2)^2}{1-u^2}} -1 \right] -1   \right\} = -\frac{\sqrt{\lambda}}{4} \, ,
    \end{equation}
    also independent of $\Delta\theta$ and the saddle under consideration. This result can be most directly obtained from the expression in~\eq{tilde_S_eff_log} and using that $\lim_{\Delta \eta \to \infty} {\rm Re} \{ \tilde{C}_\eta^{(s)} \Delta \eta \} = 0 $, which is a consequence of the fact that the integrands that define $\Delta \eta$ diverge only logarithmically with the difference between $C_\eta^2$ and $-1/4$. On the other hand, the subleading term in the real part (a constant offset) at large $\Delta \eta$ does depend on these specific considerations.
    \item For $\Delta \theta = 0$ (left plot in \fig{cuspy}), the real part of $\Gamma_{\rm cusp}$ vanishes for $\Delta \eta < \Delta \eta_c$ and the imaginary part diverges $ \propto 1/\Delta \eta$. The divergence of the imaginary part can be understood via the relation of a Wilson loop with a cusp and the (generalized) heavy quark-antiquark potential~\cite{Drukker:2011za}, which diverges with the inverse of the spatial separation between the lines, as in \eq{small_angle_scaling_future}. In fact, this qualitative behavior is present for all $\Delta \theta < \pi$, as one may verify by direct calculation from the $s=+$ saddle (although the value of $\Delta \eta_c$ decreases to zero as $\Delta \theta$ is taken to $\pi$).
    \item The point $\Delta \eta = 0$ in the $\Delta \theta = \pi$ case (right plot in \fig{cuspy}) has a vanishing anomalous dimension, because the path is a backtracking contour that satisfies the ``zig-zag'' symmetry. In particular, this configuration preserves the supersymmetry needed for the loop to be BPS~\cite{Zarembo:2002an}. There is no inconsistency with the fact that the near-BPS expansion~\cite{Correa:2012at,Drukker:2011za} is proportional to $\phi_E^2 - \theta^2$, because in the near-BPS expansion the coefficient of $(\phi_E^2 - \theta^2)$ diverges for a backtracking contour. 
        \item The $\Delta \eta$ dependence of the cusp anomalous dimension coming from the $u_{\rm max} = 1$ saddle is qualitatively very similar to its result at weak coupling: int is a convex function which starts off quadratically from zero and smoothly transitions to a linear growth at large $\Delta \eta$. 
\end{itemize}

The matrix element~\eq{chi1_matrix_element} is defined 
such that each cusp has a timelike Wilson line along $v^\mu$ and a lightlike Wilson line along $\nbar^\mu$. In the one-cusp situation we have discussed in this section, the configuration of each cusp corresponds to the limit $\Delta \eta \to \infty$, in which case the anomalous dimension diverges with the rapidity separation, at a rate determined by $\bar{\Gamma}_{\rm cusp} = \sqrt{\lambda}/(4\pi)$, with the same meaning as in the perturbative counterpart \eq{chi1_RG2}. 
Note that because we start with two timelike directions $v_1$ and $v_2$, this approach is equivalent to regularizing the rapidity divergence by taking the $\nbar-$Wilson line off lightcone in the timelike direction, which is also the regularization scheme adopted in~\refcite{Ji:2019sxk} to enable lattice calculation of the TMD soft function.
We point out that this limit $\Delta \eta \to \infty$ is insensitive to the value of the separation $\Delta \theta$ on $S_5$, as one could have expected based on the fact that the adjoint scalars $\Phi$ decouple from the Wilson loop~\eq{W_def} if $\dot{x}^2 = 0$ (see also the discussion in~\cite{Alday:2007he}).
It is also insensitive to whether only one of the directions approaches the lightcone, or both. Therefore $\bar\Gamma_\mathrm{cusp} = \sqrt{\lambda}/(4\pi)$ holds for both timelike-lightlike and lightlike-lightlike cusp configurations.

It may appear curious at first that we obtain a nontrivial imaginary part for $\Gcusp$, but this is actually an established phenomenon in perturbative QCD. 
Take the physical example where two timelike Wilson lines represent two partons, and the cusp angle is determined by the parton masses and the momentum transfer. For the cases where both partons are incoming or outgoing, the momentum transfer invariant develops an imaginary phase, which originate from Glauber gluon exchanges and are thus known as Glauber phases~\cite{Rothstein:2016bsq,Jouttenus:2011wh,Banerjee:2025kkq}. 
Note that these contributions are absent if one parton is incoming and the other outgoing, 
as the case illustrated in \fig{future_past_lines}.
For the cusp configuration like \fig{future_future_lines} which show up in $\chi_1$ or the TMD soft function, 
we do have both partons outgoing,
and therefore expect an imaginary Glauber phase for each cusp. 
However, the full operator involves the one-cusp Wilson line and its complex conjugate,
and the imaginary part cancels and we expect the cusp anomalous dimension of the operator to only depend on $2\,\mathrm{Re}\{\Gcusp\}$.

\section{Two cusps with transverse separation }
\label{sec:twocusps}

In the previous section we examined the calculation of the action of the string dual to a Wilson line with a cusp, discussed various subtle points, and recovered known results in the corresponding limits. While the techniques to do this calculation have been available for a long time, we are not aware of any systematic analysis of the cusp anomalous dimension that results from the whole range of values of $\Delta \eta$ and $\Delta \theta$, as we have now done.

Having completed this, we now turn to our main interest in this work: to evaluate \eq{chi1_matrix_element} via a direct calculation in a strongly coupled gauge theory. This problem now involves two cusps in a Wilson loop, one that is time-ordered and another that is anti-time-ordered as operators acting on the gauge theory Hilbert space, which can be identified as the product of an amplitude and a complex conjugate amplitude summed over final states.

It turns out that with what we have done so far it is already possible to guess what the result should be: other than the added segments, the main addition in this problem is the appearance of a new distance scale $b_\perp$, which should now enter the result explicitly. However, there is no other scale in the problem --- with the exception of the regulators that we discussed in~\sec{divergences_regularization}. As such, it can only enter the result of a practical calculation in combination with the UV or IR regulators $\Lambda, L$. However, as discussed in~\sec{divergences_regularization}, the IR regulator is determined exactly by the distance at which the Wilson loop stops being described by the contribution from the cusp alone. It therefore follows that $b_\perp$ is exactly this kind of IR scale. As such, a natural guess is
\begin{align}
    \chi_1(b_\perp;\Delta \eta, \Delta \theta) &\propto \exp \left( - 2 {\rm Re}\{\Gamma_{\rm cusp}[\Delta \eta,\Delta \theta] \} \ln ( \Lambda b_\perp ) \right) \nonumber \\
    &= |b_\perp \Lambda|^{- 2 {\rm Re}\{\Gamma_{\rm cusp}[\Delta \eta,\Delta \theta] \} } \, , \label{eq:chi1_result_guess}
\end{align}
where the imaginary part of the cusp anomalous dimension provided by the AdS/CFT calculation cancels between the contribution of the amplitude and complex conjugate amplitude. 

We shall see in what follows how \eq{chi1_result_guess} can be justified further with a holographic calculation. This also opens the possibility of examining the effects of non-conformal holographic backgrounds on top of which the string propagates, and investigate the behavior of $\chi_1$ in non-conformal quantum field theories --- a prospect we shall return to in our conclusions.

\subsection{From transverse separation to momentum flow}
\label{sec:trans_momentum_flow}

Generalizing our starting point for the problem with one cusp, for a path with two cusps denoted $\mathcal{C}_2$ we write
\begin{align}
    \left\langle W[\mathcal{C}_2] \right\rangle &= \int_{\partial \Sigma = \mathcal{C}_2} DX \exp \left( - \mathcal{S}_{\rm NG}[X] \right) \nonumber \\ 
    &= \int_{ \substack{\{X^T(z=0)\} \, = \, \mathcal{C}_1(0) \\ \{X^A(z=0)\} \, = \, \mathcal{C}_1^*(b_\perp) \\ X^T(a \to \infty) \, = \,  X^A(a \to \infty) } } DX^T DX^A \exp \left( - \mathcal{S}^T_{\rm NG}[(X^T)^\mu] - \mathcal{S}^A_{\rm NG}[(X^A)^\mu] \right) \, ,
\end{align}
where we have doubled the number of fields in the path integral, labeled with superscripts $T$ and $A$, representing the part of the Wilson loop that is time-ordered and the part that is anti-time-ordered, respectively. We have denoted the path for the time-ordered Wilson loop as $\mathcal{C}_1(0)$, and $\mathcal{C}_1^*(b_\perp)$ for the anti-time-ordered case, with their arguments indicating their position along the $x_\perp$ direction. The last boundary condition states that the surfaces hanging from $\mathcal{C}_1$ and $\mathcal{C}_1^*$ connect at $a\to\infty$.
It is convenient to write the parametrization of the worldsheet in either branch (be it time-ordered or anti-time-ordered) as
\begin{equation}
    X^\mu = (a,\eta,x_\perp(a,\eta),z(a,\eta),\theta(a,\eta)) \, ,
\end{equation}
where, in addition to choosing $\eta$ as the worldsheet parameter instead of $u$, we have also introduced a new field $x_\perp(a,u)$ which describes the position of the worldsheet in a direction orthogonal to the plane defined by the cusps, along the direction defined by $b_\perp$ (see~\fig{Wilson_chi1}). Because of the introduction of $b_\perp$, there is now a scale in the problem that allows for nontrivial $a$ dependence in the configurations that dominate the path integral. As such, all of the fields are parametrized in terms of both $a$ and $\eta$. 

Since $b_\perp$ is the only ingredient that breaks the scale symmetry of the problem, it proves convenient to analyze its effects through a Lagrange multiplier, similar to 
our one cusp analysis in the previous section. 
Concretely, we introduce a Lagrange multiplier $C_\perp$ which enforces the net displacement between $\mathcal{C}_1$ and $\mathcal{C}_1^*$ along $x_\perp$ to equal $b_\perp$. This yields
\begin{align}
    \left\langle W[\mathcal{C}_2] \right\rangle(b_\perp)
    &= \int_{ \substack{\{X^T(z=0)\} \, = \, \mathcal{C}_1(0) \\ \{X^A(z=0)\} \, = \, \mathcal{C}_1^*(0) \\ X^T_{{\rm no} \, x_\perp}(a \to \infty) \, = \,  X^A_{{\rm no} \, x_\perp}(a \to \infty) } } DX^T DX^A \exp \left( - \mathcal{S}_{\rm NG}^T[(X^T)^\mu] - \mathcal{S}_{\rm NG}^A[(X^A)^\mu] \right) \nonumber \\
    & \quad \times \int D{C}_\perp e^{i \int d\eta \,{C}_\perp(\eta) \cdot \left[ \int_0^\infty \dot{x}^T_\perp(a,\eta) \, da \, - \, \int_0^\infty \dot{x}^A_\perp(a,\eta) \, da \, - \, b_\perp \right]  }
    \, . \label{eq:W_C2_step_2}
\end{align}
We have moved the boundary condition regarding the transverse separation of the cusp completely into the integral over the Lagrange multiplier. The boundary conditions in the path integral now impose
$x_\perp(z=0) = 0$ for both contours $\mathcal{C}_1$ and $\mathcal{C}_1^*$.
This is justified because after inserting the Lagrange multiplier, each copy of the action becomes separately translationally invariant under uniform shifts in $x_\perp$, meaning their exact transverse coordinates do not carry any physical significance, and the only physically important condition, i.e. the net transverse separation, is enforced by the Lagrange multiplier.
It is important to point out the surfaces still only connect at $a\to \infty$.

It is instructive to then consider the Fourier transform of this object,
\begin{equation}
    \chi_1(q) = \int db_\perp e^{i q \, b_\perp} \chi_1(b_\perp) = \int db_\perp e^{i q \, b_\perp} \left\langle W[\mathcal{C}_2] \right\rangle(b_\perp) \, ,
\end{equation}
which results in the insertion of a Delta function
\begin{equation}
    \delta
    \Big( q - \int d\eta \, C_\perp(\eta) \Big)
\end{equation}
in the path integral. That is to say, the Lagrange multiplier that enforces the spatial separation between the two cusps carries information about the momentum transfer conjugate to $b_\perp$.

This has a natural interpretation in terms of the string equations of motion: if one writes $\mathcal{S}_{\rm NG}^{T/A} = (\pm i) \int da d\eta \, \mathcal{L}$, where the sign choice reflects which set of fields enter the action (being $+i$ for time-ordered $(T)$ fields and $-i$ for anti-time-ordered $(A)$ fields), then the Euler-Lagrange equations resulting from varying the exponent of the integrand of the path integral with respect to $x_\perp$ is, for any $a < \infty$
\begin{align}
    \frac{\partial}{\partial a} \left( \frac{\partial \mathcal{L}}{\partial \dot{x}_\perp^T } \right) + \frac{\partial}{\partial \eta} \left( \frac{\partial \mathcal{L}}{ \partial( \partial_\eta {x}_\perp^T) } \right) 
    &= 0 \, , \\
    -\frac{\partial}{\partial a} \left( \frac{\partial \mathcal{L}}{\partial \dot{x}_\perp^A } \right) - \frac{\partial}{\partial \eta} \left( \frac{\partial \mathcal{L}}{ \partial(\partial_\eta {x}_\perp^A) } \right) 
    &= 0 \, ,
\end{align}
which is one of the Nambu-Goto equations of motion. 
The first term in parenthesis in each of these equations
\begin{equation}
    \Pi_a^{T/A} = \frac{\partial \mathcal{L}}{\partial \dot{x}_\perp^{T/A} } \, ,
\end{equation}
has a natural interpretation as a momentum flux. One can easily convince oneself from the equation of motion that the integrated flux $\int d\eta \, \Pi_a$ is a conserved quantity with respect to $a$.

Note that the term with the Lagrange multiplier $C_\perp$ does not affect the Nambu-Goto equations of motion as it is a total derivative. 
However, this term does affect the variation with respect to $x_\perp(a=\infty,\eta)$, which is not constrained by boundary conditions anymore because the Lagrange multiplier $C_\perp$ lifted this restriction. As such, the variation of the terms in the exponent of the second line of~\eq{W_C2_step_2} yields
\begin{equation}
    \delta \left[ i \int d\eta \int_0^\infty \dot{x}_\perp^T da \right] = \delta \left[ i \int d\eta \,  (x_\perp^T(a = \infty) -  x_\perp^T(a = 0) ) \right] = i \int d\eta \,  \delta x_\perp^T(a = \infty) \, ,
\end{equation}
where $x_\perp^T(a = 0)$ is still fixed by the boundary condition.
Furthermore, the variation with respect to $x_\perp^T(a = \infty)$ also generates a contribution from the action itself --- concretely, the upper limit of integration over $a$ that appears in the integration by parts
\begin{align}
    \delta \mathcal{S}_{\rm NG} &= \int da d\eta \, \left( \frac{\partial \mathcal{L}}{\partial  \dot{x}_\perp} \partial_a(\delta x_\perp) +\frac{\partial \mathcal{L}}{\partial (\partial_\eta x_\perp)}\partial_\eta(\delta x_\perp) \right) \nonumber \\
    &= \int d\eta \left[ \delta x_\perp \frac{\partial \mathcal{L}}{\partial \dot{x}_\perp } \right]_{a=0}^{a=\infty} + \int da \left[ \delta x_\perp \frac{\partial \mathcal{L}}{\partial (\partial_\eta \dot{x}_\perp) } \right]_{\eta=-\Delta \eta/2}^{\eta=+\Delta \eta/2} \nonumber \\ 
    & \quad - \int da d\eta \, \left[ \frac{\partial}{\partial a} \left( \frac{\partial \mathcal{L}}{\partial \dot{x}_\perp } \right) + \frac{\partial}{\partial \eta} \left( \frac{\partial \mathcal{L}}{ \partial(\partial_\eta {x}_\perp) } \right) \right] \delta x_\perp \nonumber \\
    &= \int d\eta \,  \delta x_\perp(a=\infty,\eta) \left. \frac{\partial \mathcal{L}}{\partial \dot{x}_\perp } \right|_{a=\infty}  - \int da d\eta \, \left[ \frac{\partial}{\partial a} \left( \frac{\partial \mathcal{L}}{\partial \dot{x}_\perp } \right) + \frac{\partial}{\partial \eta} \left( \frac{\partial \mathcal{L}}{ \partial(\partial_\eta {x}_\perp) } \right) \right] \delta x_\perp \, .
\end{align}
Therefore, requiring that the variations of the exponent in the path integral with respect to $x_\perp^T(a=\infty,\eta)$ and $x_\perp^A(a=\infty,\eta)$ be stationary implies that
\begin{align}
    - i  \Pi_a^T (a=\infty,\eta) + i C_\perp(\eta) &= 0 \, , \\
    i  \Pi_a^A(a=\infty,\eta) - i C_\perp(\eta) &= 0 \, ,
\end{align}
i.e., $C_\perp$ is exactly the momentum flux $\Pi_a^{T/A}$ at infinity,
and $\Pi_a^{T} (a = \infty) = \Pi_a^{A} (a = \infty)$.

Then, since the integrated flux $\int d\eta \, \Pi_a^{T/A}$ is a conserved quantity with respect to $a$, the value of the total flux is set by the source term proportional to the delta function. Explicitly, we have
\begin{equation}
    \int d\eta \, \Pi_a^{T/A}(a,\eta) = \int d\eta \, C_\perp(\eta) = q \, , \label{eq:momentum_conservation}
\end{equation}
for \textit{all} values of $a$. Note that in this subsection we have chosen to write $x_\perp$ as a function of $\eta$ instead of $u = z/a$ as we had done before. While this parametrization has the advantage that the integration boundaries can be specified in a straightforward manner ($\eta$ spans all numbers between $0$ and $\Delta \eta$, or $\pm \Delta \eta/2$), we could choose any coordinate that to carry out the integrals in~\eq{momentum_conservation}, as long as such coordinate spans any 1D string  with constant $a$ on the extremal surface.

Because the flux is conserved, one can use it at any value of $a$ along the worldsheet to relate its properties in different regions of the geometry determined by the Wilson loop. In particular, to study how the matching condition at infinity affects the behavior of the surface near the cusp. This is our next task.

\subsection{The transverse profile of the two-cusp configuration}
\label{sec:two_cusp_xperpsol}

Once we have established that the integral of the momentum flux $\Pi_a$ along the worldsheet is a conserved quantity that matches the integral of the Lagrange multiplier $C_\perp$, which in turn is fixed by the conjugate momentum variable $q$ to $b_\perp$, we can proceed to study solutions to the string equations of motion.

Our main purpose is to verify \eq{chi1_result_guess}. That is to say, we have to verify that
\begin{enumerate}
    \item The introduction of $b_\perp$ (equivalently, the momentum flow along the worldsheet) does not change the UV divergence that gives rise to the cusp anomalous dimension, and
    \item that this new scale acts as an IR regulator in the action.
\end{enumerate}
We will do so by inspecting the region near each cusp, by looking for perturbative (i.e., linearized) solutions  around the worldsheet configurations defined by each cusp that carry a fixed momentum flux.

We begin by considering the Nambu-Goto action for a worldsheet parametrized as
\begin{equation}
    X^\mu = (a,\eta(u),x_\perp(a,u),z=au, \theta(u)) \, ,
\end{equation}
where we have only kept a nontrivial dependence on $a$ in the argument of $x_\perp$, because (as long as $x_\perp$ is a small perturbation) the $a$-dependent back-reaction on the other fields will be higher order in the transverse displacement. Formally, our expansion in what follows will be in powers of $(x_\perp/a)$.

Note that we again choose to parametrize the worldsheet with coordinates $(a,u)$. We emphasize two key advantages of this parametrization that connect to our earlier analysis. 
First, since $u$ is a coordinate that spans the strings on this worldsheet, our analysis in~\sec{trans_momentum_flow} regarding conserved momentum flux will carry through with respect to $u$ instead of $\eta$.
Second, since we consider $x_\perp$ as a small perturbation on the one-cusp configuration, we can utilize the extremization for the other coordinates discussed in detail in~\sec{onecusp} as a background.

Keeping only terms up to quadratic order in $x_\perp$, the action (in either the time-ordered or the anti-time-ordered branch) is
\begin{align}
    \mathcal{S}_{\rm NG}[\Sigma] &= \frac{\sqrt{\lambda} }{\pi} \int \frac{da}{a} \int_0^{u_{\rm max} } \! \frac{du}{u^2} \sqrt{-\left[ 1 + (1 - u^2) h \right]} \nonumber \\
+& \frac{\sqrt{\lambda} }{2\pi} \int \frac{da}{a^3} \int_0^{u_{\rm max} } \! \frac{du}{u^2} \frac{ (-1 + u^2 h) x_\perp'^2 + (1+h)(a \dot{x}_\perp - u x_\perp' )^2 - 2 u^2 h (a \dot{x}_\perp - u x_\perp') x_\perp'}{\sqrt{-\left[ 1 + (1 - u^2) h \right]} } \, , \label{eq:action_xperp_expanded}
\end{align}
where, for brevity, we have introduced $h \equiv \eta'^2 + u^2 \theta'^2$.
Again, a dot (prime) denotes the derivative with respect to $a (u)$. 
The first term is the zeroth-order action~\eq{action_real} that we have extremized to extract the cusp anomalous dimension in~\sec{onecusp}.

In the limit where we neglect the back-reaction from $x_\perp$, we can fix $\eta$, $\theta$ as before, by extremizing the zeroth order action. Perturbatively in $(x_\perp/a)$, the support of the worldsheet along the $u$ coordinate is still determined by $u_{\rm max}$, which is fixed in terms of the Lagrange multipliers of the one-cusp calculation. However,  
recall that when evaluating the final integral in \eq{W_Loop_step_5}, an analytic continuation for the Lagrange multipliers was necessary, by means of which $u_+(C_\eta,C_\theta)$ becomes complex. This means that, in order to obtain the equations of motion for $x_\perp$ in a controlled way, for the case $u_{\rm max} = u_+$ it is better to rescale $u = w \, u_{\rm max} $ before evaluating the integral over the ``background'' fields and thus fixing the values of the Lagrange multipliers, and solve for $x_\perp(a,w)$ instead. For the case where $u_{\rm max} = 1$ the situation is subtle in a different way, as the integration contour over $u$ may have to be deformed to properly define the analytic continuation. Nonetheless, since the solutions are analytic functions of $u$, in practice it is possible to analyze them directly in terms of the $u$ coordinate.

The equation of motion can be obtained straightforwardly:
\begin{align}
    & \frac{\partial}{\partial u} \left[ \frac{2(-1+u^2 h) x_\perp' - 2(1+h) u (a \dot{x}_\perp - u x_\perp') + 2 u^2 h (2u x_\perp' - a \dot{x}_\perp) }{a^3 u^2 \sqrt{-\left[ 1 + (1 - u^2) h \right]} } \right] \nonumber \\
    + & \frac{\partial}{\partial a} \left[ \frac{ 2(1 + h) a (a \dot{x}_\perp - u x_\perp' ) - 2 u^2 h a x_\perp'  }{a^3 u^2 \sqrt{-\left[ 1 + (1 - u^2) h \right]} } \right] = 0 \, , \label{eq:xperp_eom}
\end{align}
from which one immediately identifies the term in square brackets in the second line as the momentum flow density $\Pi_a$ along the worldsheet. 
In general, one expects to obtain solutions of \eq{xperp_eom} of the form
\begin{equation}
    x_\perp(a,u) = \sum_\lambda c_{\lambda} a^{\lambda} f_{\lambda}(u) \, ,
\end{equation}
where the functions $f_\lambda(u)$ are the ``quasinormal modes'' of the cusp, each characterized by a scaling dimension $\lambda$.
In order for the integrated flux with respect to $u$ (equivalent to $\eta$) of a given mode of the solution to be conserved, 
we need
\begin{align}
\label{eq:momflux_mode}
    a^{\lambda-2} c_\lambda \int du \, 
    \frac{2(1+h) \left[\lambda f_\lambda(u) - u f_\lambda'(u) - 2u^2 h f_\lambda'(u) \right]}{u^2 \sqrt{-\left[ 1 + (1 - u^2) h \right]}}
    = \text{constant in } a
\,.\end{align}
We see that there is a solution that carries a fixed momentum flow, given by choosing $\lambda = 2$:
\begin{equation}
\label{eq:xperp_c2f2}
    x_\perp(a,u) = c_2 \, a^2\, f_2(u) \, .
\end{equation}
The solution $f_2$ is special for our purposes because it is the only nontrivial solution that carries momentum flux. 
For $\lambda \neq 2$, the integral over $u$ in \eq{momflux_mode} needs to vanish, therefore the mode $f_\lambda(u)$ does not contribute to the momentum flux which is defined by this integral.
Extremizing the action, all of such contributions will therefore be set to zero.

The equation that determines $f_2$ can be cast as a first-order equation
where only the first term in \eq{xperp_eom} survives,
\begin{equation}
    \frac{(-1+u^2 h) f_2' - (1+h) u (2f_2 - u f_2') + u^2 h (2u f_2' -2f_2) }{u^2 \sqrt{-\left[ 1 + (1 - u^2) h \right]} } = A \, ,
\end{equation}
for some constant $A$. This constant is arbitrary, as it sets the normalization of $f_2$. In this equation, $h$ is determined in terms of the values of the Lagrange multipliers that determine the extremal surface of the corresponding one-cusp geometry
\begin{equation}
    h = h(u) = \frac{u^4 C_\eta^2 + u^2 C_\theta^2}{(1-u^2) \left( 1 - (1 + C_\theta^2) u^2 - u^4 C_\eta^2 \right) } \, . \label{eq:h_of_u}
\end{equation}

One can then solve for $f_2$ by writing this equation as a standard first-order ordinary differential equation
\begin{equation}
    f_2' + 2u \frac{1+h+uh}{1-u^2-2(1+u)u^2h} f_2 + A u^2 \frac{\sqrt{-\left[ 1 + (1 - u^2) h \right]}}{1-u^2-2(1+u)u^2h} = 0 \, ,
\end{equation}
which one may integrate directly. Imposing that $f_2(u=0) = 0$, one obtains
\begin{align}
    f_2(u) = - A \int_0^u d t &\,  t^2 \frac{\sqrt{-\left[ 1 + (1 - t ^2) h(t) \right]}}{1-t^2-2(1+t)t^2h(t)} \nonumber \\ \times & \exp \left( -2 \int_{t}^u ds \, s \frac{1+h(s)+s h(s)}{1-s^2-2(1+s)s^2h(s)}  \right) \, , \label{eq:f2_result}
\end{align}
with $A$ an arbitrary normalization constant. The matching condition at the turning point (where the surface hanging from one side of a single cusp meet that from the other) is automatically satisfied provided $f'_2$ is finite at $u = u_{\rm max}$. This is so because (geometrically) the tangent vector to the surface at the turning point is a linear combination of $\partial_\eta$ and $\partial_\theta$, and as such $x_\perp$ smoothly reaches an extremum if $\tfrac{dx_\perp}{d\eta} = \tfrac{du}{d\eta} \tfrac{dx_\perp}{du} \sim \sqrt{u_{\rm max} - u}  \tfrac{dx_\perp}{du} = 0$. Therefore, as long as $\left|\tfrac{dx_\perp}{du} \right| < \infty$, the surface with the $x_\perp$ deformation is smooth simply by having the same $x_\perp$ profile on each side of the surface that hangs from a single cusp. Note that when the Lagrange multipliers are analytically continued into the complex plane, the expression for $f_2$ in~\eq{f2_result} determines via the position of its singularities what contours of integration along the complexified $u$ variable are suitable to define its analytic continuation.

For the specific case at hand, $c_2$ in~\eq{xperp_c2f2} is fixed by requiring that
\begin{align}
    q = \int du \, \Pi_a(a,u) &= -i \frac{\sqrt{\lambda}}{2\pi} \int du \frac{ 2(1 + h) a (a \dot{x}_\perp - u x_\perp' ) - 2 u^2 h a x_\perp'  }{a^3 u^2 \sqrt{-\left[ 1 + (1 - u^2) h \right]} } \nonumber \\
    &= -i \frac{\sqrt{\lambda}}{\pi} \int du \frac{ (1 + h)  (2 c_2 f_2 - u c_2 f_2' ) -  u^2 h c_2 f_2'  }{ u^2 \sqrt{-\left[ 1 + (1 - u^2) h \right]} } \, ,
\end{align}
meaning that
\begin{equation}
    c_2 = \frac{i \pi q}{\sqrt{\lambda} } \left( \int_0^{u_{\rm max}} du \frac{ (1 + h)  (2 f_2 - u f_2' ) -  u^2 h f_2'  }{ u^2 \sqrt{-\left[ 1 + (1 - u^2) h \right]} } \right)^{-1} \equiv \frac{q}{\sqrt{\lambda}} \tilde{c}_2 \, . \label{eq:c2_tilde_def}
\end{equation}
Whenever $u_+$ is complex, the  integration should be understood as the analytic continuation of the function, and the integration contour along $u$ be suitably deformed to define said analytic continuation.

In conclusion, at small $x_\perp/a$, one has
\begin{equation}
    x_\perp(u) = \frac{q}{\sqrt{\lambda}} a^2 \tilde{c}_2 f_2(u) \, , \label{eq:xperp_small_a}
\end{equation}
where the arbitrariness in choosing $A$, i.e., in the normalization of $f_2$, cancels between the expression for $c_2$ in~\eq{c2_tilde_def} and the explicit $f_2$ in~\eq{xperp_small_a}. 
Parametrically, \eq{xperp_small_a} is reliable as long as
\begin{equation}
    \frac{q}{\sqrt{\lambda}} a^2 \ll a \implies a \ll \frac{\sqrt{\lambda}}{q} \, , \label{eq:a_small}
\end{equation}
because in this regime we may neglect the back-reaction onto the one-cusp background geometry.
It follows that each copy of the action will no longer be well-described by the one-cusp geometry when $a \sim \sqrt{\lambda}/q$. For larger values of $a$, having a nontrivial $x_\perp$ solution necessarily modifies the worldsheet profile along the other coordinates. 
Therefore, in calculating the contribution to the action with the logarithmically divergent piece --- i.e., the contribution proportional to the cusp anomalous dimension --- the scale $ \sqrt{\lambda}/q$ acts as an IR regulator, as it sets the scale for $a$ beyond which the zeroth-order term in \eq{action_xperp_expanded} is no longer dominant.

\begin{figure}
    \centering
    \includegraphics[width=0.6\linewidth]{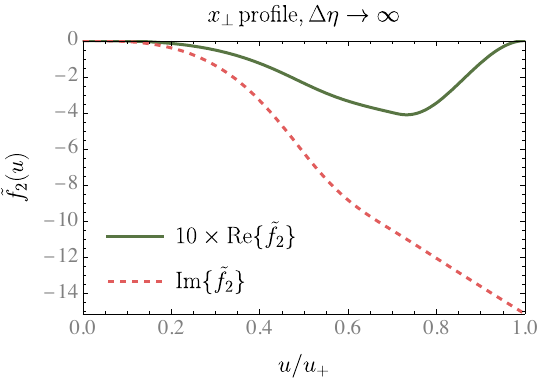}
    \caption{Profile of $x_\perp$ at a fixed (but small) $a$ as $\Delta \eta \to \infty$, with $\tilde{f}_2$ defined via the rescaling $x_\perp = \frac{q \, a^2}{\sqrt{\lambda} \Delta \eta } \tilde{f}_2(u)$. For concreteness (and simplicity) we plot $f_2$ between $0$ and $u_+ = \sqrt{2}$ along the real axis. In the case when $u_{\rm max} = 1$, the solution has to be extended back to $u = 1$, along the same contour that defines the one cusp action.}  
    \label{fig:f2}
\end{figure}

\paragraph{The $\Delta\eta\to\infty$ limit}

Because of our initial motivation to study the matrix element~\eq{chi1_matrix_element}, it is interesting to study the behavior of $x_\perp$ in the lightlike limit $\Delta \eta \to \infty$. This limit corresponds to $C_\eta^2 \to -1/4$ and $C_\theta^2 \to 0$, meaning that $h(u)$ in~\eq{h_of_u} is rational function of $u$ with explicit coefficients, and thus fully specifies the integrand on the right hand side of~\eq{f2_result}. One may check that the integrand that defines $f_2(u)$ has no singularities all the way up to (and including) $u_+(\tilde{C}_\eta(\Delta \eta \to \infty),\tilde{C}_\theta(\Delta \eta \to \infty) ) = \sqrt{2}$. On the other hand, while $f_2(u)$ is without singularities, the momentum flow integral in~\eq{momentum_conservation}, which translates to the integral in the middle expression of~\eq{c2_tilde_def}, develops a singularity in the limit $(C_\eta^2,C_\theta^2) \to (-1/4,0)$. This can be seen because in this limit $h(u)$ itself develops a double pole at $u = \sqrt{2}$, and as such the integral defining the momentum flow develops a logarithmic divergence near the turning point $u_+$. Fittingly, this is the same divergence generated near the turning point when one integrates~\eq{eta_sol_Ceta_Ctheta} to obtain the value of $\Delta \eta$ corresponding to a given pair of Lagrange multipliers.\footnote{This is true for both $u_{\rm max} = 1$ and $u_{\rm max} = u_+$, where in the latter case the integration contours for the momentum flow and for $\Delta \eta$ only approach $u_+$ once, while in the former they each do so twice, meaning that the proportionality constant between $\tilde{c}_2$ and $\Delta \eta$ is the same.} As such, $\tilde{c}_2$ can be easily evaluated in this limit in terms of $\Delta \eta$. We obtain (up to an overall sign depending on whether the solution is for a time-ordered or anti-time-ordered field) 
\begin{equation}
    \lim_{\Delta \eta \to \infty} \Delta \eta \, \tilde{c}_2 = \frac{-2\pi i (1+ \sqrt{2}) }{  \, f_2(u = \sqrt{2})} \, .
\end{equation}

Then, introducing $\tilde{f}_2(u) = \Delta \eta \, \tilde{c}_2 f_2(u)$, which is now independent of the (arbitrary) normalization constant $A$, we arrive at
\begin{equation}
    x_\perp(a,u) = \frac{q \, a^2}{\sqrt{\lambda} \Delta \eta} \tilde{f}_2(u) \, ,
\end{equation}
where $\tilde{f}_2(u)$ approaches finite values in the limit $\Delta \eta \to \infty$. We show this limiting profile in~\fig{f2}. Physically, the reason why $1/\Delta \eta$ appears as a normalization factor can be understood because the momentum flow is distributed across the whole range of spacetime rapidity covered by the one-cusp geometry, here described via the $u$ coordinate. As such, when the momentum flow $\Pi_a$ is integrated over the whole $u$ (or $\eta$) range it yields a finite answer equal to the prescribed momentum transfer $q$.

Because of the $1/\Delta \eta$ factor, one can in principle refine the estimate for the regime of validity of the small $x_\perp$ approximation in~\eq{a_small}. It may seem then that our calculation becomes arbitrarily good in the lightlike limit. However, as we will discuss in the next section, the natural scale for $q$ (for a given, fixed $b_\perp$) is actually proportional to $\sqrt{\lambda} \Delta \eta / b_\perp$, meaning that the criterion in~\eq{a_small} is (parametrically) simply $a \ll b_\perp$.

\subsection{The matrix element $\chi_1$ at strong coupling}
\label{sec:final_chi1_result}

Finally, we are ready to evaluate the two-cusp Wilson loop in Fourier space. Starting from~\eq{W_C2_step_2},
using the results from the two previous subsections as well as the one-cusp zeroth-order action from~\sec{onecusp}, we have
\begin{align}
    \chi_1(q) &= \int db_\perp e^{i q b_\perp} \langle W[\mathcal{C}_2] \rangle(b_\perp) \nonumber \\
    &=\int_{ \substack{\{X^T(z=0)\} \, = \, \mathcal{C}_1(0) \\ \{X^A(z=0)\} \, = \, \mathcal{C}_1^*(0) \\ X^T_{{\rm no} \, x_\perp}(a \to \infty) \, = \,  X^A_{{\rm no} \, x_\perp}(a \to \infty) } } DX^T DX^A DC_\perp \, \delta \! \left(q - \int du C_\perp(u) \right) \nonumber \\ 
    & \quad \quad \quad \quad \quad \quad \times \exp \left( - \mathcal{S}_{\rm NG}^T[(X^T)^\mu] + i \int du \,{C}_\perp(u) \int_0^\infty \dot{x}^T_\perp(a,u) \, da \right)  \nonumber \\
    & \quad \quad \quad \quad \quad \quad \times \exp \left(- \mathcal{S}_{\rm NG}^A[(X^A)^\mu] - i \int du \,{C}_\perp(u) \, \int_0^\infty \dot{x}^A_\perp(a,u) \, da \right)
    \, ,  \nonumber \\
    &= \Big[ \Lambda {\rm -independent} \, {\rm prefactor} \Big] \times \exp \left( - \ln(\Lambda \sqrt{\lambda}/q ) \Gamma_{\rm cusp}[\Delta \eta,\Delta \theta] \right) \nonumber \\
    & \quad\quad\quad\quad\quad\quad\quad\quad\quad\quad\quad\quad\quad\, \times \left[ \exp \left( - \ln(\Lambda \sqrt{\lambda}/q ) \Gamma_{\rm cusp}[\Delta \eta,\Delta \theta] \right) \right]^* \, , \label{eq:W_C2_step_3}
\end{align}
where in the last equality we keep only the contribution from the small $a$ domain defined by~\eq{a_small}.
From the form of $x_\perp(a,u)$ given in~\eq{xperp_small_a},
we see that the first-order action contributes no additional divergence at small $a$, therefore the only UV divergence is the same logarithmic divergence as in the zeroth-order action regulated by $\Lambda$. 
In particular, $\Gcusp[\Delta\eta,\Delta\theta]$ is given by the one-cusp result defined by~\eq{gcusp_onecusp_sol}.
The IR cutoff is now provided by $a = \sqrt{\lambda}/q$, below which the solution for $x_\perp$ discussed in~\sec{two_cusp_xperpsol} is valid,
and therefore enters the logarithm together with $\Lambda$.

We may extract the anomalous dimensions of the operator as
\begin{align}
    \gamma_{\chi_1} &= \frac{d \ln \chi_1(q)}{d \ln \Lambda} = - 2 {\rm Re} \left\{ \Gamma_{\rm cusp}[\Delta \eta, \Delta \theta] \right\} \, .
    \label{eq:anom_dim_result}
\end{align} 
Given the large rapidity limit of $\Gcusp[\Delta\eta,\Delta\theta] \sim \sqrt{\lambda}/ (4 \pi) \, \Delta\eta$ as in~\eq{Gcusp_lrap_limit}, we see that our result here has the scaling expected from the evolution equations for $\chi_1$ as in ~\eqs{chi1_RG1}{chi1_RG2}.

Crucially, since $q$ and $\Lambda$ are the only dimensionful quantities in the problem, they have to appear together to form a dimensionless ratio. Together with the anomalous dimension~\eq{anom_dim_result}, it follows from here that
\begin{equation}\label{eq:chi1_q_cusp}
    \chi_1(q; \Delta \eta, \Delta \theta) = \left(\frac{q^2}{\Lambda^2} \right)^{{\rm Re} \left\{ \Gamma_{\rm cusp}[\Delta \eta, \Delta \theta] \right\}} \, .
\end{equation}
Taking a Fourier transform back to position space, we obtain
\begin{equation}\label{eq:chi1_bperp_cusp}
    \chi_1(b_\perp; \Delta \eta, \Delta \theta) = \left( b_\perp^2 \Lambda^2 \right)^{-{\rm Re} \left\{ \Gamma_{\rm cusp}[\Delta \eta, \Delta \theta] \right\}} \, ,
\end{equation}
where we neglected $\mathcal{O}(\lambda^0)$ terms in the exponent when carrying out the Fourier transform, consistent with the fact that our calculation is only accurate at leading order in the strong coupling expansion.

We note that if one carries out the Fourier transform in the saddle point approximation, using the large $\sqrt{\lambda}$ limit, one obtains a value of $q$ at the extremum given by 
\begin{equation}
    \frac{\partial}{\partial q} \left[ - i q b_\perp - 2 \ln( \Lambda \sqrt{\lambda}/q ) {\rm Re}\{ \Gcusp[\Delta \eta, \Delta \theta] \} \right] = 0 \implies q = - 2 i \, {\rm Re}\{ \Gcusp[\Delta \eta, \Delta \theta] \}/b_\perp \, ,
\end{equation}
which means that the typical scale of $q$ in the integral is proportional to $\sqrt{\lambda}$, and in particular, proportional to $\sqrt{\lambda} \Delta \eta$ in the large $\Delta \eta$ limit.\footnote{Reassuringly, this is also self-consistent with the fact that we are evaluating the integral with a saddle point method, as all terms in the exponent acquire the same overall scaling with a large prefactor.}

We note that one could have arrived at the same conclusion without introducing the Lagrange multiplier $C_\perp$, or equivalently, the conserved flux~\eq{momentum_conservation}. Using the same logic as when we first discussed regularization in~\sec{divergences_regularization}, one can argue that $b_\perp$ plays the role of an IR regulator, because when $a \sim b_\perp$ the worldsheet configuration should be directly affected by the modified boundary condition at infinity, in contrast to the one-cusp case which has no obstacle to maintain the scale-invariant profile. It then follows that the expression for the action can only be given by the single-cusp configuration at small enough $a$, thus retaining the logarithmic dependence on the UV cutoff, and replacing the IR cutoff with $b_\perp$.

The advantage of the analysis we carried out here, in terms of the conjugate momentum variable $q$, is that we obtain explicit access to the string profile at small $a$, and thus can be more precise regarding what constitutes ``small enough'' $a$ for the one-cusp worldsheet configurations to be a good approximation.
We foresee that these solutions --- for which we showed examples in \fig{f2} --- may be most useful as inital conditions to numerically solve for string configurations with fixed momentum flow, integrating along $a$ from the cusp outwards towards infinity.

\paragraph{Generalization to lightlike-lightlike soft function}

In our setup, it is clear that the $\Delta\eta\to \infty$ limit does not distinguish between rapidity divergences from 1) $v_1$ timelike, $v_2$ lightlike and 2) $v_1, \, v_2$ both lightlike. This observation was already made in~\sec{one_cusp_result} for the one-cusp case, i.e. the result $\Gcusp(\Delta\eta\to\infty,\Delta\theta) = \bar\Gamma_\mathrm{cusp} \,\Delta \eta = \sqrt{\lambda}/(4\pi) \, \Delta \eta$ holds for both timelike-lightlike and lightlike-lightlike cusp configurations.
This correspondence is also present in perturbation theory. For example, it was shown in~\refcite{Vladimirov:2017ksc} that the rapidity divergence of the one-loop soft factor is generated by a single lightlike Wilson line with a coefficient that is independent of the nature of the other Wilson line. Consequently, the timelike-lightlike and lightlike-lightlike configurations share the same leading rapidity divergence.
This remains the case for the two-cusp case we consider in this section.
Perturbatively, this was already stated in~\eq{chi1_RG2}, where it was calculated~\cite{vonKuk:2024uxe} that the cusp anomalous dimension of $\chi_1(b_\perp)$ is the same, universal lightlike cusp anomalous dimension $\bar\Gamma_\mathrm{cusp}[\as(\mu)]$ for both configurations in QCD.
Furthermore, in the lightlike limit, we obtain the rapidity evolution governed by the Collins-Soper kernel%
\footnote{For readers familiar with the Collins-Soper kernel from perturbative QCD, taking a derivative with respect to the rapidity (separation) is consistent with Collins' definition in~\refcite{Collins:2011zzd} of the Collins-Soper kernel $\tilde K$ directly through the soft function. The modern logarithmic derivative with respect to the Collins-Soper scale $\zeta$ is easily recovered by realizing $\sqrt{\zeta}\sim e^{-\eta}$.}:
\begin{align}
    \gamma_{\zeta} = \frac{d \ln \chi_1}{d \Delta \eta }\Bigg|_{\Delta\eta\to\infty} 
    = - 2 \ln (\Lambda b_\perp) \frac{d \,\mathrm{Re}\{\Gamma_{\rm cusp} \}}{d \Delta \eta}\Bigg|_{\Delta\eta\to\infty}
    = - 2 \ln (\Lambda b_\perp) \frac{\sqrt{\lambda}}{4\pi}
    \label{eq:CSkernel_largeeta}
\,.\end{align}
Note that~\eq{CSkernel_largeeta} is schematically the same as~\eq{chi1_RGrho},
and together with~\eq{anom_dim_result} satisfies the integrability constraint~\eq{chi1_RG2}.

Our analysis and result in this section regarding $\Delta\eta \to \infty$ applies directly to the $\mathcal{N}=4$ sYM counterpart of the TMD soft function, which has the configuration similar to $\chi_1(b_\perp)$ in~\fig{Wilson_chi1} but with lightlike Wilson lines $W_n, \, W_n^\dagger$ (different from $W_{\nbar}$) instead of timelike Wilson lines $Y_v, \, Y_v^\dagger$.
Explicitly, we anticipate~\eqs{chi1_q_cusp}{chi1_bperp_cusp} to be true also for the Wilson loop configuration provided by the TMD soft function,
and the large rapidity string profile shown in~\fig{f2} to hold as well.

\section{Conclusions and outlook}
\label{sec:conclusion}

We have presented a thorough analysis of the expectation value of cusped Wilson loops with a timelike cusp angle characterized by a spacetime rapidity separation $\Delta \eta$ (sometimes referred to as the Minkowski angle $\phi$) between the Wilson lines. 
We study both the configuration of a single cusp and a novel configuration of two cusps separated by a transverse distance, directly in Minkowski signature.
In the large rapidity separation limit, the vacuum matrix element of the Wilson loop with two cusps becomes the $\mathcal{N}=4$ sYM analog of the QCD TMD soft function. 
Our calculation relies on taking the large $N_c \to \infty$ and large $\lambda = g^2 N_c \to \infty$ limits of $\mathcal{N}=4$ sYM, so that the expectation value of the Wilson loop in this field theory becomes equivalent to a calculation of (semi)classical dynamics of a string in an AdS${}_5 \times S_5$ background. Mathematically, this corresponds to finding the extremal surfaces/string configurations that satisfy the boundary conditions imposed by the Wilson loop, with the area functional given by the Nambu-Goto action.
In the course of our calculation, we have observed several structural features of physical interest, on which we remark in what follows.

We first studied the contributions to the expectation value of a Wilson loop with a timelike cusp angle coming from the immediate vicinity of the cusp. In this region, the problem enjoys a scale symmetry that allows one to simplify the problem of finding the extremal string configurations from equations on a 2D worldsheet to an effective 1D extremization problem. Even though it is simpler than the original, this problem is still nontrivial due to the fact that, except for a bounded interval at small rapidity separation, the saddle points for the path integral over the string configurations are not real: they are complex. The main technical ingredient that allowed us to make progress is the introduction of Lagrange multipliers to enforce the desired separation along the Minkowski and internal $(S_5)$ angles. Once the boundary conditions are enforced by multipliers that can be grouped together with the rest of the path integral, carrying out the integration over the string degrees of freedom becomes straightforward. After solving the equations of motion of the string (i.e., extremizing the action), these Lagrange multipliers become equal to (conserved) momentum fluxes along the string. Their conservation is due to boost invariance and rotational invariance on the $S_5$, respectively. Thus, the expectation value of the Wilson loop becomes an integral over two real numbers; and while the saddle points of this integral are unavoidably complex, the fact that it is a finite-dimensional integral makes it tractable relative to having to find saddles in the infinite-dimensional path integral over the string degrees of freedom.

An important subtlety emerges in that there is not only one family of saddles that can contribute to the expectation value of the Wilson loop. In fact, we showed that not only  are there two families of saddle points in this problem, but that they actually exchange dominance as $\Delta \eta$ is increased from $0$ at a finite value of the rapidity separation. The family of saddles that dominates at small $\Delta \eta$ can be cross-checked with many results in the literature, and contrasted with analog calculations in Euclidean signature. On the other hand, the family of saddles that dominates at large $\Delta \eta$ is quantitatively and qualitatively novel; there is no natural Euclidean counterpart of them because the turning point of the surface rests along a null line into the bulk of AdS${}_5$ (in our notation, $u = z/a = 1$). This family of surfaces is complex (with nonzero real and imaginary parts), and requires a somewhat more involved complex analysis calculation for its determination. It is worth mentioning that both families of saddles, even if used separately, would imply the same large $\Delta \eta$ asymptotics for the cusp anomalous dimension found by~\refcite{Kruczenski:2002fb}. Put together, the results from both families determine the cusp anomalous dimension at all values of the rapidity separation. In~\fig{result_intro} and~\fig{cuspy} we showed them for $\Delta \theta = 0$ and $\Delta \theta = \pi$; the same calculation can be directly carried out from our setup for intermediate values of the $S_5$ separation $\Delta \theta$ if so desired. The case with Neumann boundary conditions can be obtained by integrating over $\Delta \theta$, which, comparing with the intermediate steps in our calculation, sets $C_\theta = 0$ and as such reduces to the $\Delta \theta = 0$ case, as one could have anticipated based on the fact that this configuration automatically satisfies Neumann boundary conditions on the $S_5$.

At a technical level for the one-cusp configurations, an interesting direction that we did not explore is to carry out a systematic analysis of the quadratic fluctuations around each saddle, which, in addition to allowing for an in-depth analysis of the stability properties of each of them, also contains physics information about local operator insertions along either line coming out of the cusp. It would also be interesting to explore whether the individual saddles we found could be isolated in their contribution to observables --- and thus provide further cross-checks of the configurations we found.

Second, we showed how to formulate the calculation of a two-cusp configuration with operator ordering as prescribed in the matrix element in~\eq{chi1_matrix_element}, and calculated it. Like in the one-cusp case, the crucial conceptual step was to identify a conserved momentum flow conjugate with the transverse separation between each cusp, providing a concrete way to match the string configurations on either copy of the path integral to each other, allowing us to determine the leading behavior of the extension of the string along the $x_\perp$ direction by solving linear equations of motion near each cusp without the need to solve the nonlinear equations of motion for the string away from them. The result, given by~\eq{chi1_bperp_cusp}, naturally follows from these considerations, features the real part of the cusp anomalous dimension we calculated earlier as its nontrivial ingredient.

Going forward, our results provide the setup and a proof of principle for the calculation of soft functions in theories with a holographic dual. For example, the introduction of a scale in the holographic background would mean that the result analogous to~\eq{chi1_bperp_cusp} will not only depend on the product between $b_\perp$ and the renormalization scale $\Lambda$, but also on the product of $b_\perp$ with the new scale. We anticipate that the same strategy, i.e., organizing the calculation in terms of Lagrange multipliers/integrals over conserved momenta to enforce the boundary conditions of the problem, will prove equally fruitful. In this scenario, the behavior of the matrix element at distances larger than the new distance scale set by the holographic background can be qualitatively different than its behavior in the conformal theory we have studied. We speculate that such a calculation could provide an interesting model to study flux tube breaking phenomena in gauge theories, which should map to a change in the characteristic behavior of~\eq{chi1_bperp_cusp}, because the cusp anomalous dimension can be related directly to flux tube string tension~\cite{Alday:2007he}. 
This connection is particularly natural given that the Wilson loop configuration we study here can be directly related to the energy-energy correlator (EEC) in the back-to-back limit, providing an experimentally accessible observable whose deviation from the perturbative Sudakov behavior would signal the onset of flux tube breaking~\cite{Jaarsma:2025tck}.
We note that flux tube breaking is sensitive to the matter content and dynamics of the particular gauge theory; the EEC, however, has been computed across a wide range of theories, which makes it a natural common observable through which such theory-dependence can be studied systematically.

Another interesting direction would be to investigate this Wilson loop configuration in nonzero temperature states, which would correspond to holographic backgrounds with a horizon at some finite value of the radial holographic coordinate. While similar to the introduction of an additional scale in the theory, this setup would allow one to study the interplay between production of a heavy quark and its subsequent dynamics in a thermal environment, completing the picture of~\cite{Rajagopal:2025ukd,Rajagopal:2026urq} to a description in which the details of the production mechanism are also taken into account. 
Lastly, we envision that it will be interesting to study related quantities, such as the same Wilson loop as we discussed here but with local operator insertions along the Wilson lines and/or at the cusp(s), using our calculation as a starting point. In this way, a much more expansive characterization of high-energy dynamics in strongly coupled gauge theories will take shape.

\acknowledgments
We gratefully acknowledge helpful conversations with Mart\'in Kruczenski, Johannes Michel, Ian Moult, Ignazio Scimemi, and Iain Stewart. 
This work was performed in part at the Aspen Center for Physics, which is supported by National Science Foundation grant PHY-2210452 and by a grant from the Simons Foundation (1161654, Troyer).
B.S. was supported in part by National Science Foundation grant PHY-2309135 to the Kavli Institute for Theoretical Physics (KITP) and by grant 994312 from the Simons Foundation.
Z.S. was supported in part by Department of Energy award DESC0025293 and National Science Foundation grant PHY-2515115. 
Z.S. would also like to thank the Erwin-Schrödinger International Institute for Mathematics and Physics at the University of Vienna for partial support during the Thematic Programme ``New Paradigms for Harnessing Quantum Field Theory at Colliders'', while this work was in its finishing stage.

\appendix

\section{Elements of complex analysis} \label{app:complex_analysis}

The purpose of this Appendix is to discuss in further detail some of the intermediate steps we carried out in~\sec{onecusp}. The starting point for this discussion is~\eq{W_Loop_step_3}, which determines the expectation value of a Wilson line with one-cusp
configuration in terms of an integral over Lagrange multipliers $C_\eta$ and $C_\theta$.

\subsection{Deforming the integration contours for $\eta',\theta'$} \label{app:complex_analysis_eta_theta}

It is convenient to introduce the function
\begin{equation}
    \tilde{W} (C_\eta,C_\theta) \equiv \int d u_{\rm max}  \int_{\substack{ \left.\frac{du}{d\eta}\right|_{u = u_{\rm max} } =\, 0 \\ \left.\frac{du}{d\theta}\right|_{u = u_{\rm max} } =\, 0  }} D\theta' D\eta' \exp \left( -  \tilde{\mathcal H}[\eta',\theta';C_\eta,C_\theta,u_{\rm max}] \right) \, , \label{eq:fourier_tf_loop}
\end{equation}
where
\begin{align} \label{eq:fourier_tf_action_real}
    &\tilde{\mathcal H}[\eta',\theta';C_\eta,C_\theta,u_{\rm max}] \nn \\ &=  \frac{i \sqrt{\lambda}}{ \pi} \int \frac{da}{a}\left\{ \int_0^{u_{\rm max}} \!\!\!\! du \left[ \frac{\sqrt{1 + (1-u^2) (\eta')^2 + u^2(1-u^2)(\theta')^2}}{u^2} 
    - C_\eta \eta' 
    - C_\theta \theta'  
    \right]
    \right\} \, ,
\end{align}
such that the expectation value associated with the one-cusp configuration is exactly given by a Fourier transform
\begin{equation} \label{eq:WfromWtilde}
    \langle W[\mathcal{C}_1] \rangle(\Delta \eta, \Delta \theta) = \int_{-\infty}^\infty dC_\eta dC_\theta e^{i \frac{\sqrt{\lambda}}{\pi}  \ln (\Lambda L) \left[ C_\eta \Delta \eta + C_\theta \Delta \theta  \right]/2 } \, \tilde{W} (C_\eta,C_\theta) \, .
\end{equation}
In~\eq{fourier_tf_action_real} we omitted the subtraction terms that render the result finite, as they will not play any role in what follows. Consistent with our discussion on regularization in~\sec{divergences_regularization}, we have used $\int \frac{da}{a} = \ln (\Lambda L)$.

\Eq{WfromWtilde} makes it clear that it is sufficient to characterize $\tilde{W}$ at real $C_\theta$, $C_\eta$ in order to determine $\langle W[\mathcal{C}_1] \rangle$, because this equation determines the latter in terms of an integral that only sums over real values of the Lagrange multipliers. And, as obvious as it may sound, for purposes of evaluating the Fourier transform of $\tilde{W}$ via the saddle point approximation (as we will do later in this Appendix and as we discussed in the main text in~\sec{saddle_point_solution}), it is the analytic continuation of $\tilde{W}(C_\eta, C_\theta)$ starting from its value on the real axis that which determines the value of the integral.

We therefore start by discussing the integration over $\eta'$ and $\theta'$ at real $C_\eta$ and $C_\theta$. This is an infinite product of integrals that are decoupled from each other, and can be evaluated independently. In the large $\sqrt{\lambda}$ limit, the stationary points of the exponent give the dominant contribution. Taking $u_{\rm max}$ as an arbitrary positive number, there are three qualitatively distinct cases for the integrations over $\eta'(u),\theta'(u)$ depending on the value of $u$ relative to $u_+(C_\eta,C_\theta)$ and $1$ (we abbreviate $u_+ = u_+(C_\eta,C_\theta)$ in what follows): 
\begin{enumerate}
    \item If $u < u_+$, the extremum of the integrand 
    is real for both components $(\eta',\theta')$. 
    \item If $u_+ < u < 1$, the extremum of the integrand is imaginary for both components $(\eta',\theta')$.
    \item If $1 < u$, the extremum is again real for both components.
\end{enumerate}
The last case is never relevant in the end, as the stationary points for $u_{\rm max}$ are always less than one (if $C_\eta$ and $C_\theta$ are real; see next subsection for the complex case). As mentioned in the main text, surfaces satisfying the extremum conditions and the derivative boundary conditions satisfy either $u_{\rm max} = u_+$ or $u_{\rm max } = 1$. In all cases the extremum for $\eta'(u),\theta'(u)$ is given by the same formulas
\begin{align} \label{eq:eta_sol_Ceta_Ctheta_app}
    \eta'(u) &= \frac{C_\eta u^2}{\sqrt{(1-u^2)(1-u^2(1+C_\theta^2) - C_\eta^2 u^4)}} \, ,
     \\
    \theta'(u) &= \frac{C_\theta}{\sqrt{(1-u^2)(1-u^2(1+C_\theta^2) - C_\eta^2 u^4)}} \, , \label{eq:theta_sol_Ceta_Ctheta_app}
\end{align}
which uniquely define the solution for $u < u_+$. However, one needs to specify the sign of the imaginary part of each equation if $u_+ < u < 1$. This is fixed by examining which sign choice is actually reachable via a deformation of the integration contour into the complex plane. To make the steps as explicit as possible, we first carry out the integral over $\theta'$ at a fixed value of $\eta'$, and then the integral over $\eta'$ once the former has been evaluated.

\begin{figure}
    \centering
    \includegraphics[width=0.49\linewidth]{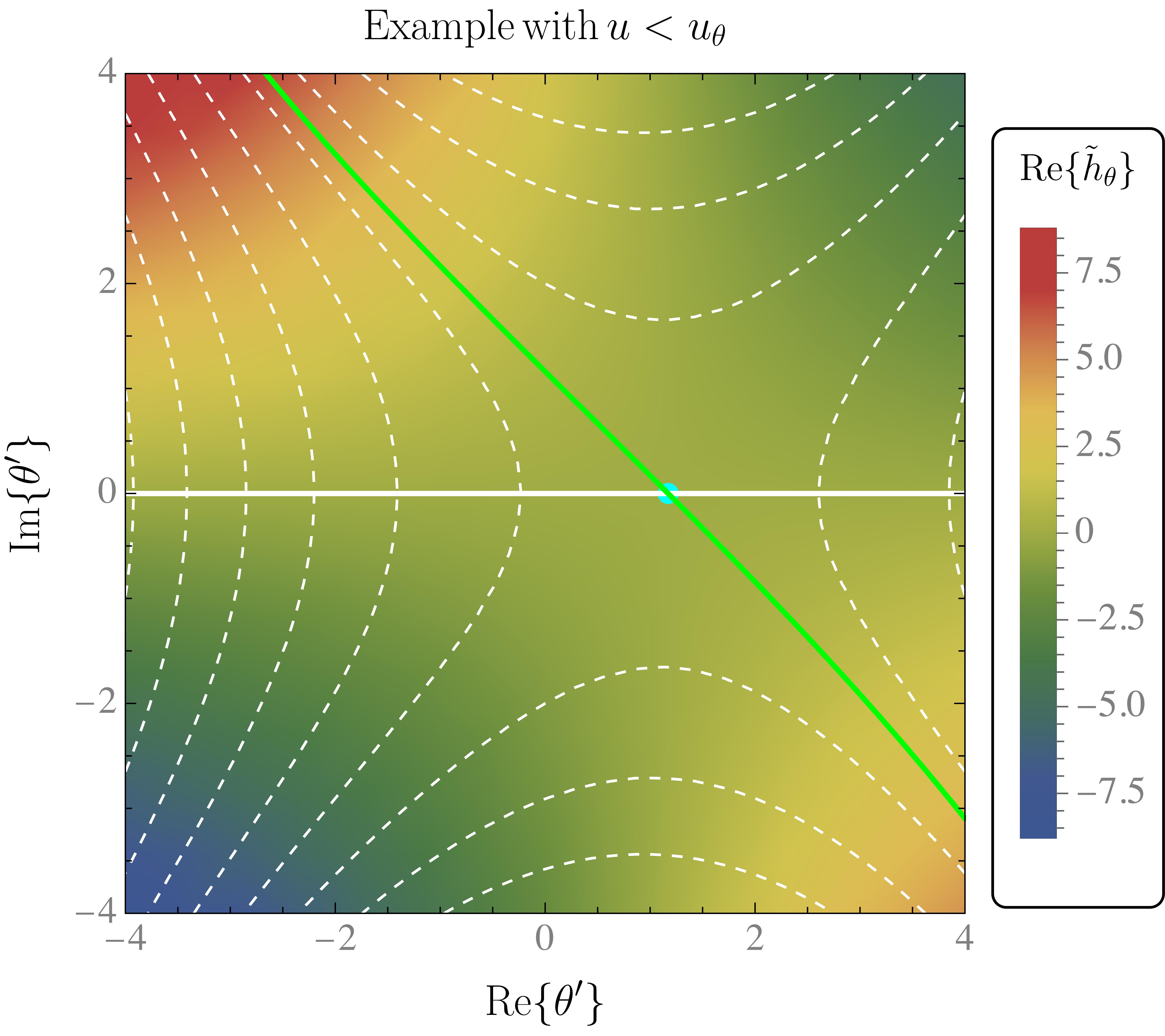}
    \includegraphics[width=0.49\linewidth]{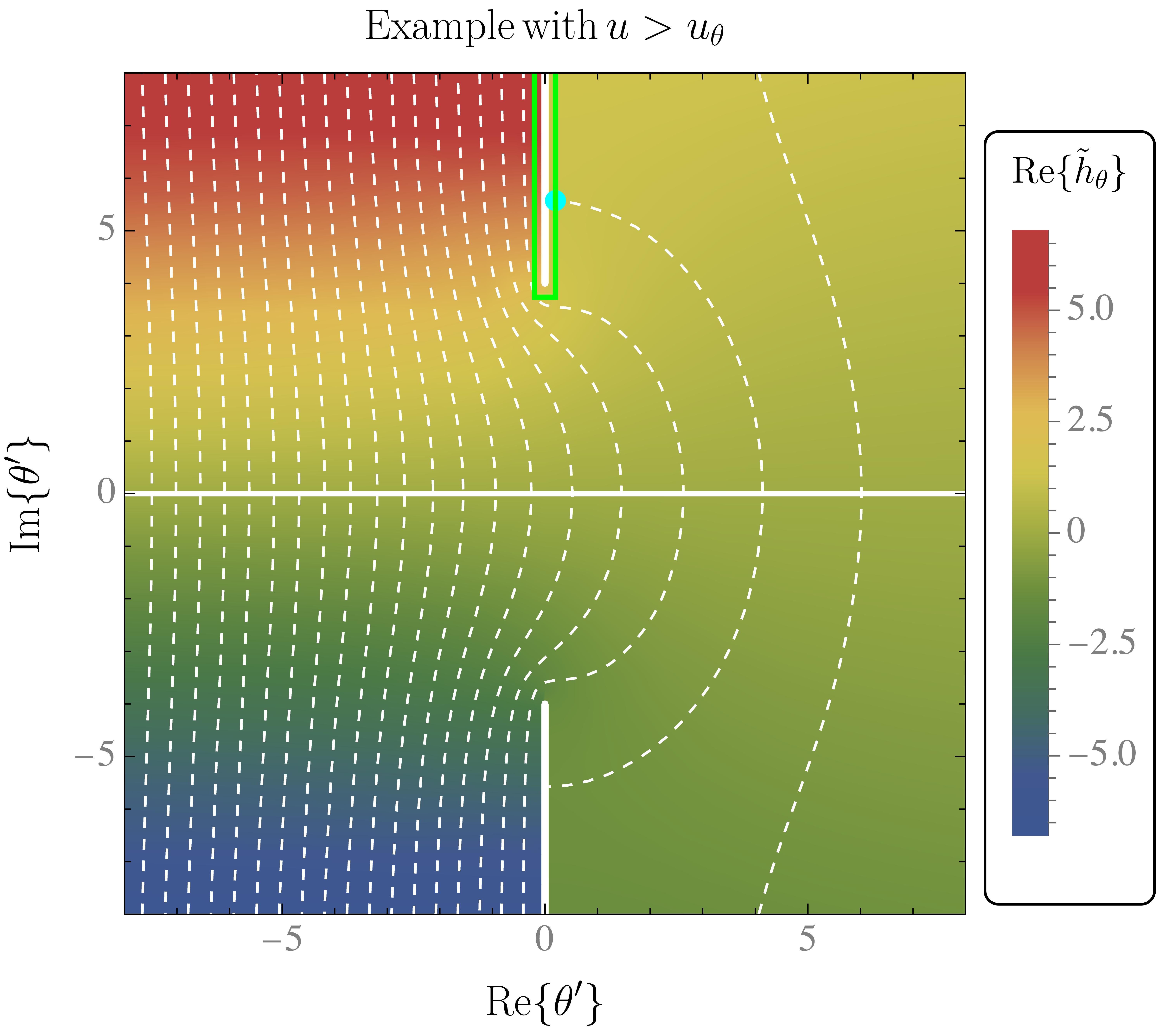}
    \caption{Original integration contour (white) and the deformation into the complex plane that allows one to take the integral over $\theta'$ with the steepest descent method (green). The saddle point is shown with a cyan circle. Here, $\eta'$ is held at a fixed, real value. The contour lines (dashed white) are lines with constant imaginary part of the action density relevant for the integration over $\theta'$, namely $\tilde{h}_\theta \equiv i [\sqrt{1 + (1-u^2) (\eta')^2 + u^2(1-u^2)(\theta')^2}/u^2 - C_\theta \theta']$; the density plot represents the real part of the action density $\tilde{h}_\theta$. Left: representative case of $u < u_\theta$. We chose $u = 1/4$, $C_\theta = 1/2$, $\eta'^2 = 4$; Right: representative case of $u > u_\theta$. We chose $u = 15/16 $, $ C_\theta = 1/2$, $\eta'^2 = 4$.}
    \label{fig:contours_theta_fixed_eta}
\end{figure}

\begin{figure}
    \centering
    \includegraphics[width=0.49\linewidth]{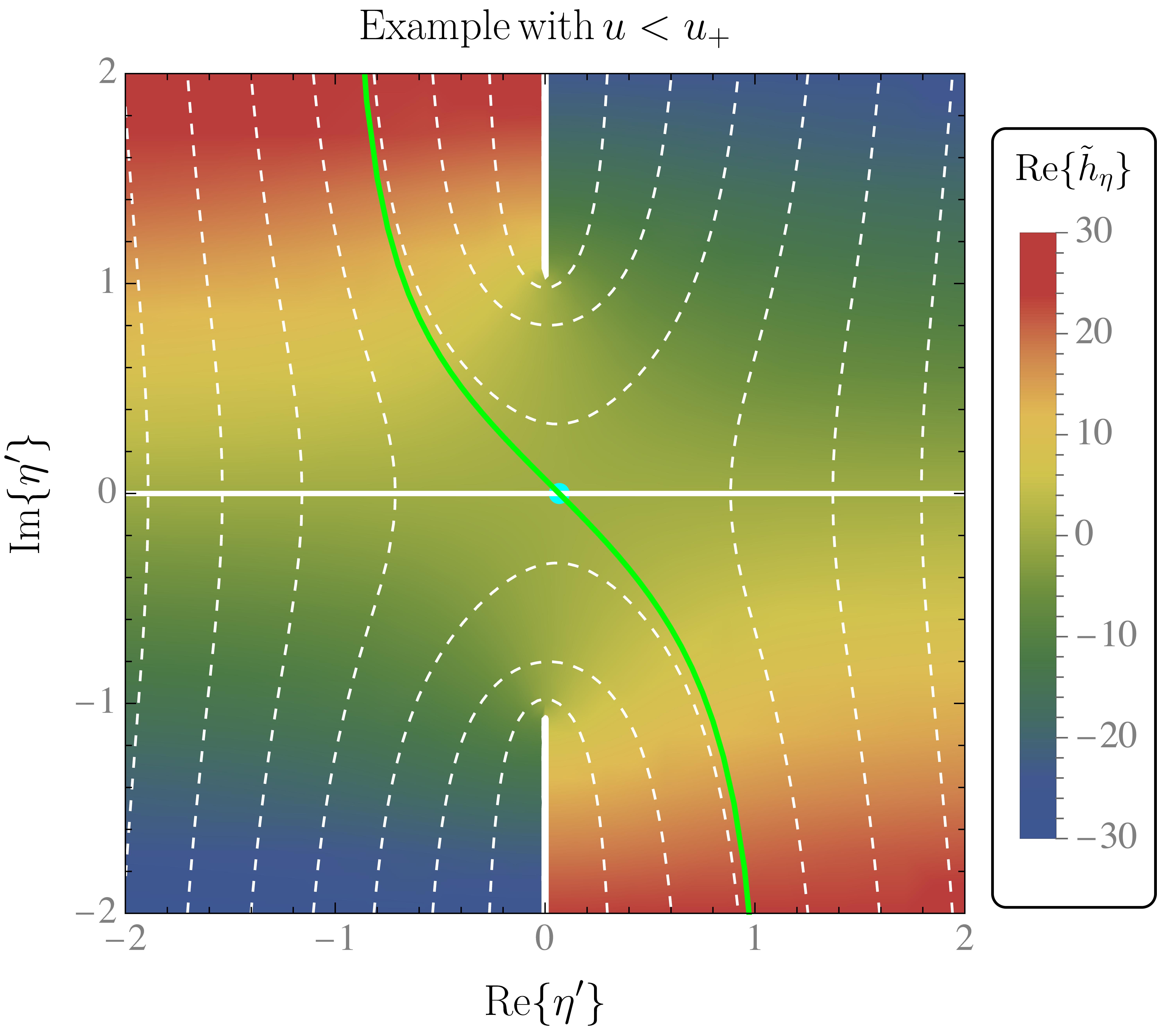}
    \includegraphics[width=0.49\linewidth]{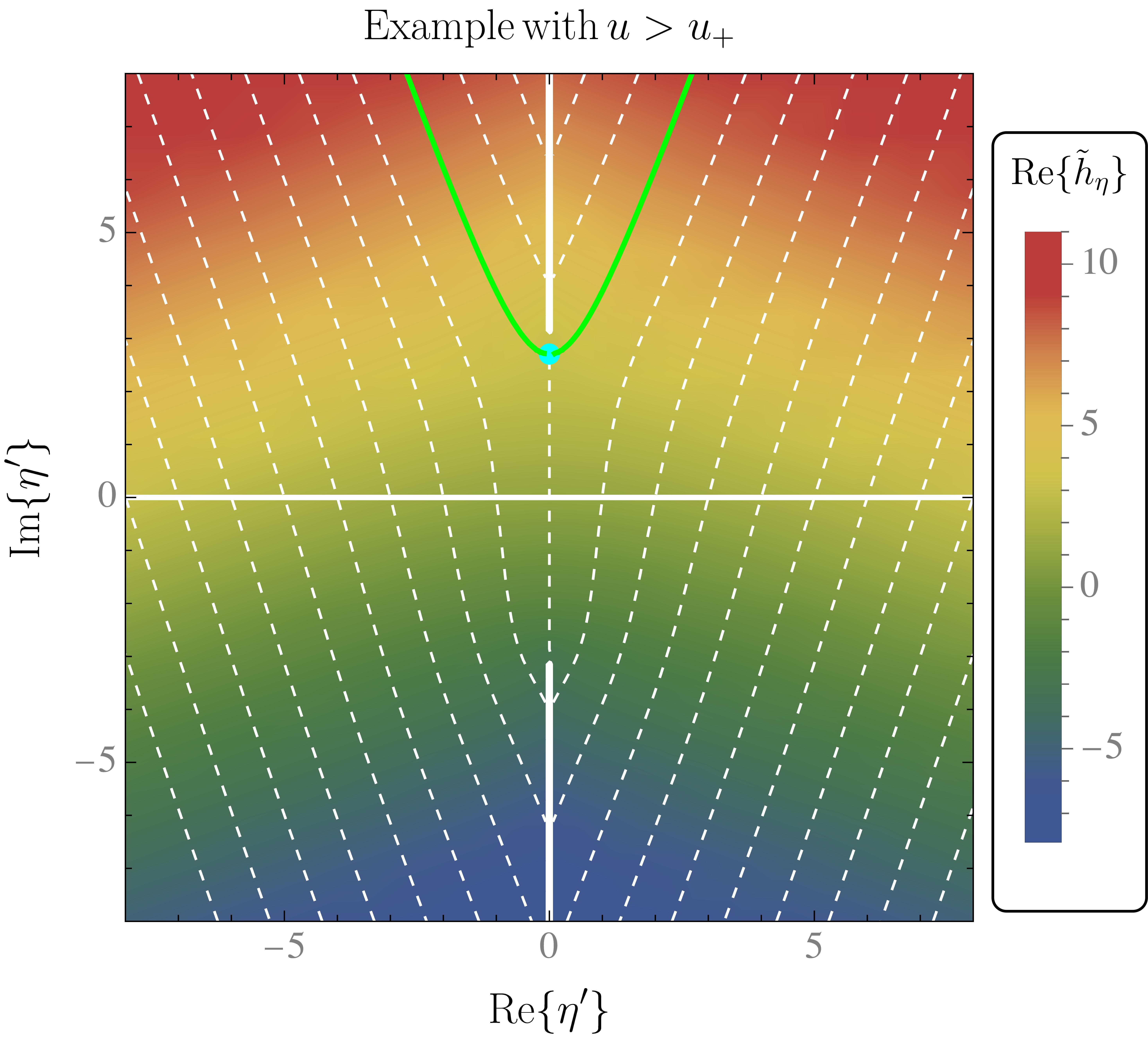}
    \caption{Original integration contour (white) and the deformation into the complex plane that allows one to take the integral over $\eta'$ with the steepest descent method (green), after having done the integral over $\theta'$. The saddle point is shown with a cyan circle. The contour lines (dashed white) are lines with constant imaginary part of the action density relevant for the integration over $\eta'$, namely $\tilde{h}_\eta \equiv i [\sqrt{1 - (1 + C_\theta^2)u^2 - i \epsilon } \sqrt{1 + (1 - u^2)\eta'^2 } /(u^2 \sqrt{1-u^2}) - C_\eta \eta']$; the density plot represents the real part of the action density $\tilde{h}_\eta$. Left: representative case of $u < u_+$. Concretely, $u = 1/4 $, $C_\theta = 1/2$, $C_\eta = 1$; Right: representative case of $u > u_+$. Concretely, $u = 15/16 $, $ C_\theta = 1/2$, $C_\eta = 1$.}
    \label{fig:contours_eta_after_theta}
\end{figure}

At a fixed value of $\eta'$, it is easy to verify that the extremum of the integrand is at
\begin{equation}
    \theta' = \frac{C_\theta \sqrt{1 + (1-u^2)\eta'^2} }{\sqrt{(1-u^2)(1-(1+C_\theta^2)u^2)}  } \, . \label{eq:theta_ext_fixed_eta}
\end{equation}
Note that the square root in the numerator always has a positive argument, and the one in the denominator changes sign at $u_\theta = (1+C_\theta^2)^{-1/2}> u_+$. When doing the integral over $\theta'$, the two relevant cases, exemplified in~\fig{contours_theta_fixed_eta}, are:
\begin{enumerate}[label=(\roman*)]
    \item For $u < u_\theta$, one has 
    \begin{equation}
        \frac{\sqrt{1-u^2 } }{u} > |C_\theta| \, ,
    \end{equation}
    meaning that the asymptotics of the action in~\eq{fourier_tf_action_real} at large $\theta'$ are determined by the asymptotics of the term coming from the Nambu-Goto action. As illustrated on the left panel of~\fig{contours_theta_fixed_eta}, this means that the contour deformation that allows the integral to be evaluated with the steepest descent method reaches complex infinity through the second and fourth quadrants.
    \item For $u > u_\theta$, the converse
    \begin{equation}
        \frac{\sqrt{1-u^2 } }{u} < |C_\theta| \,
    \end{equation}
    holds and as such, the asymptotics of the integrand at large $\theta'$ are solely determined by the sign of $C_\theta$. If $C_\theta > 0$, the action in~\eq{fourier_tf_action_real} grows going into the upper half $\theta'$-plane and decreases (becomes more and more negative) going into the lower half $\theta'$-plane. The opposite is true if $C_\theta < 0$. Since there are branch cuts in the integrand along the imaginary axis starting at $\theta'_{\rm br} = \pm i \sqrt{[1 + (1-u^2) \eta'^2]/[u^2(1-u^2)]}$ and extending to $\pm i \infty$ (and note that $|\theta'| > |\theta'_{\rm br}|$ for $\theta'$ as in~\eq{theta_ext_fixed_eta}), to evaluate this integral via the steepest descent method one needs to consider a path that goes around the cuts. Concretely, for $C_\theta > 0$, the integration path has to go from $+i\infty$ on the left side of the cut down to $\theta'_{\rm br}$ (with positive imaginary part) and then back up to $+i \infty$ on the right side of the cut. This is illustrated on the right panel of~\fig{contours_theta_fixed_eta}. The integrand is purely real along these segments, and the maximum along this path lies on the right side of the cut (first quadrant).
\end{enumerate}
These considerations can be summarized with an $i\epsilon$ prescription to select the point in the complex plane 
\begin{equation}
    \theta' = \frac{C_\theta \sqrt{1 + (1-u^2)\eta'^2} }{\sqrt{(1-u^2)(1-(1+C_\theta^2)u^2)- i\epsilon}  } \, , \label{eq:theta_ext_fixed_eta_with_iepsilon}
\end{equation}
which then can be replaced back into the action to obtain the integrand for the integration over $\eta'$. Explicitly, one obtains that the action density becomes
\begin{align}
    \frac{\sqrt{1 + (1-u^2) (\eta')^2 + u^2(1-u^2)(\theta')^2}}{u^2} - C_\eta \eta' - C_\theta \theta' & \nonumber \\ = \frac{\sqrt{1 - (1 + C_\theta^2)u^2 - i \epsilon } \sqrt{1 + (1 - u^2)\eta'^2 } }{u^2 \sqrt{1-u^2} } - C_\eta \eta' & \, , \label{eq:action_density_thetaprimereplaced}
\end{align}
and then one needs to repeat the same logic as when we compared $u$ and $u_\theta$, but now comparing $u$ and $u_+$. As illustrated in~\fig{contours_eta_after_theta}:
\begin{enumerate}[label=(\roman*)]
    \item If $u < u_+$ (which implies $u < u_\theta$), then
    \begin{equation}
        \frac{\sqrt{1 - (1 + C_\theta^2) u^2} }{u^2}  > |C_\eta| \, ,
    \end{equation}
    and as such, the asymptotics of the integral are controlled by the term with square roots, leading to a growth of the action into the second and fourth $\eta'$ quadrants, as before for $\theta'$. The integral may then be evaluated via the steepest descent method along the contour shown on the left panel of~\fig{contours_eta_after_theta}.
    \item If $u > u_+$, then 
    \begin{equation}
        \left|\frac{\sqrt{1 - (1 + C_\theta^2) u^2 - i\epsilon} }{u^2} \right|  < |C_\eta| \, ,
    \end{equation}
    and the asymptotics of the integral are controlled by the sign of $C_\eta$ (regardless of whether $u > u_\theta$ or not). Like before for $\theta'$, if $C_\eta > 0$ the integral over $\eta'$ may then be deformed to go over the upper half plane alone. The saddle will lie on the imaginary axis; depending on the values of the parameters it may be above or below the start of the cut. If it is below, then the steepest descent contour will go into the second and third quadrants. If it is above, then following the same logic as for $\theta'$, one arrives at the fact that the maximum of the integrand along this contour is at $\eta'$ as given by~\eq{eta_sol_Ceta_Ctheta_app}, on the right side of the branch cut (i.e., on the first quadrant).
\end{enumerate}

One concludes that, for real values of the Lagrange multipliers, the saddle points that determine the value of the integrals over $\eta'$ and $\theta'$ are at
\begin{align} \label{eq:eta_sol_Ceta_Ctheta_app_w_eps}
    \eta'(u) &= \frac{C_\eta u^2}{\sqrt{(1-u^2)(1-u^2(1+C_\theta^2) - C_\eta^2 u^4 - i\epsilon)}} \, ,
     \\
    \theta'(u) &= \frac{C_\theta}{\sqrt{(1-u^2)(1-u^2(1+C_\theta^2) - C_\eta^2 u^4 - i \epsilon)}} \, , \label{eq:theta_sol_Ceta_Ctheta_app_w_eps}
\end{align}
and therefore that
\begin{align}
    \tilde{W}(C_\eta, C_\theta) = \exp \left( - \tilde{\mathcal{S}}^{(+)}(C_\eta,C_\theta)  \right) + \exp \left( -\tilde{\mathcal{S}}^{(1)}(C_\eta,C_\theta) \right) \, , \label{eq:Wtilde_as_sum_saddles}
\end{align}
where $\tilde{\mathcal{S}}^{(+)}$ and $\tilde{\mathcal{S}}^{(1)}$ are given by
\begin{equation}
    \tilde{\mathcal{S}} = \frac{i \sqrt{\lambda} }{\pi} \int \frac{da}{a} \int_0^{u_{\rm max}}\frac{du}{u^2} \sqrt{\frac{1-(1+C_\theta^2)u^2 - C_\eta^2 u^4 - i\epsilon}{1-u^2}} \, ,
\end{equation}
with $u_{\rm max} = u_+$ and $u_{\rm max} = 1$, respectively. We note that the content of~\eq{Wtilde_as_sum_saddles} is mostly schematic, as the relative prefactors of the two terms are unknown, and, as we explain in the paragraph after the next, one should only trust the dominant saddle for quantitative purposes.

In summary, all of this led us to the same result as if one simply replaced the expressions for the extrema in~\eq{eta_sol_Ceta_Ctheta_app} and~\eq{theta_sol_Ceta_Ctheta_app} into the action and, whenever the argument of the square root becomes negative, simply chose to evaluate it in the only way consistent with unitarity (i.e., enforcing that the action has a positive real part).

One should remark, however, that down to this point there seems to be no reason to consider the saddle with $u_{\rm max} = 1$, since its contribution to $\tilde{W}(C_\eta, C_\theta)$ is manifestly exponentially smaller than the one for $u_{\rm max} = u_+$. One may further object that this saddle is not at all meaningful at real values of the Lagrange multipliers, because there will be an infinite number of corrections coming from subleading orders in the $1/\sqrt{\lambda}$ expansion around the $u_{\rm max} = u_+$ saddle that are quantitatively more important than the exponentially small contribution from the $u_{\rm max} = 1$ saddle. That being said, the situation is different if one considers the analytic continuation of $\tilde{W}(C_\eta,C_\theta)$ at the level of the integral that defines it. This is what we examine next.

\subsection{Extending $\tilde{\mathcal{S}}$ into the complex $C_\eta,C_\theta$ planes }\label{app:complex_analysis_complex_C}

In the previous section, we showed how to calculate $\tilde{W}(C_\eta,C_\theta)$ at real values of its arguments. Our goal, however, is to evaluate~\eq{WfromWtilde}, which we restate here:
\begin{equation}
    \langle W[\mathcal{C}_1] \rangle = \int_{-\infty}^\infty dC_\eta dC_\theta e^{i \frac{\sqrt{\lambda}}{\pi}  \ln (\Lambda L) \left[ C_\eta \Delta \eta + C_\theta \Delta \theta  \right]/2 } \tilde{W} (C_\eta,C_\theta) \, . \label{eq:WfromWtilde_rep}
\end{equation}
Provided one knows the analytic continuation of 
$\tilde{W} (C_\eta,C_\theta)$ into the complex planes of its arguments, these integrals can be carried out via the saddle point method as before.

The simplest point of view would be to take the action from the two saddles we found and analytically continue those directly. However, one could legitimately worry that the analytic continuation of the exponentially small saddle ($u_{\rm max} = 1$) might not be meaningful, precisely because it is never relevant for real values of the Lagrange multipliers.

\begin{figure}
    \centering
    \includegraphics[width=0.49\linewidth]{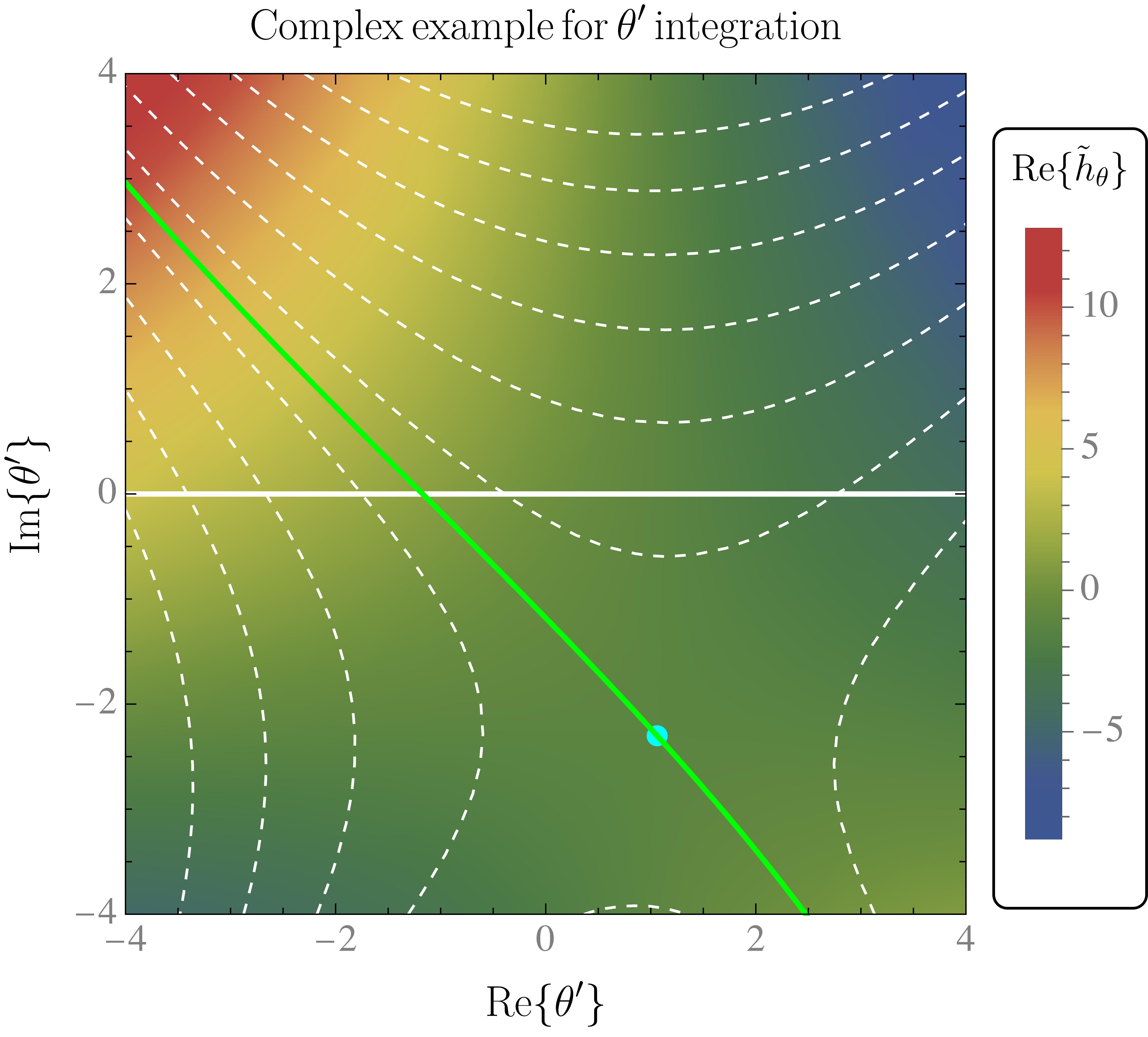}
    \includegraphics[width=0.49\linewidth]{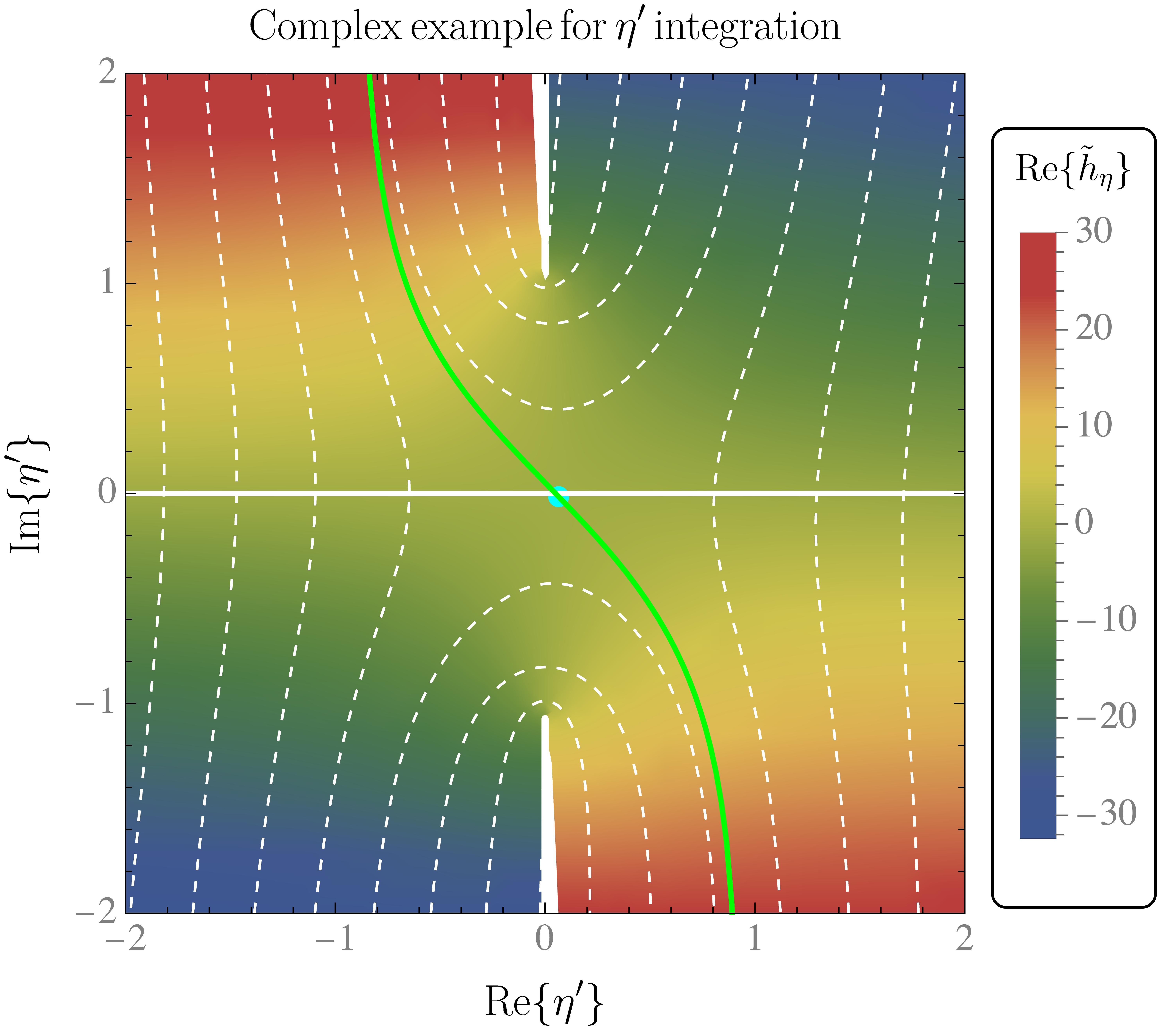}
    \caption{Left: $\theta'$ integration contour that defines the analytic continuation of the Fourier transform of the Wilson loop for complex Lagrange multipliers at fixed $\eta'^2 = 2$, with $u = 1/4$, $C_\theta = 1/2 - i$. Right: $\eta'$ integration contour that defines the analytic continuation of the Fourier transform of the Wilson loop for complex Lagrange multipliers, where $u = 1/4$, $C_\theta = 1/2 - i$, $C_\eta = 1 - i/5$. The notation and labeling is the same as in~\fig{contours_theta_fixed_eta} and~\fig{contours_eta_after_theta}. }
    \label{fig:contours_eta_real_complex_example}
\end{figure}

However, one can convince oneself that this last objection is not a problem. The reason is that from~\eq{fourier_tf_loop} and~\eq{fourier_tf_action_real} one may directly evaluate $\tilde{W}(C_\eta,C_\theta)$ at \textit{complex} values of the Lagrange multipliers, provided the integration contours over $\eta'$, $\theta'$, and $u_{\rm max}$ have been suitably deformed into the complex plane. The only subtle point about these deformations is to be able to write the constraints $du/d\eta|_{u=u_{\rm max}} = 0$ and $du/d\theta|_{u=u_{\rm max}} = 0$ as analytic contributions to the integrand. This can be done, as elsewhere in this work, via additional Lagrange multipliers. Once this addition is made, there is no obstruction to deforming the integration contours for $\eta'(u)$ and $\theta'(u)$ ($u = u_{\rm max}$ is somewhat subtle; for concreteness we consider real $0 < u < u_{\rm max}$ in this discussion) such that the integrand decreases in magnitude towards infinity. Then there is no obstruction to explicitly input complex values of $C_\eta$ and $C_\theta$ into~\eq{fourier_tf_loop}. See~\fig{contours_eta_real_complex_example} for an example of how a deformation of the integration contour allows for this. Crucially, this determines the analytic continuation of~\eq{fourier_tf_loop} into the complex plane.

Because all of the functions involved are analytic, this means that the values of the saddles in~\eq{Wtilde_as_sum_saddles} are meaningful even outside the real axis. At any given pair of complex values $(C_\eta, C_\theta)$, one should calculate the value of the action at each saddle, and keep the dominant one.

\subsection{Deforming the integration contours for $C_\eta,C_\theta$}\label{app:complex_analysis_C}

With~\eq{Wtilde_as_sum_saddles} established, all one needs to do is evaluate~\eq{WfromWtilde_rep} by using the analytic continuation of the action on each saddle ($\tilde{\mathcal{S}}^{(+)}$ and $\tilde{\mathcal{S}}^{(1)}$) and finding the contributing saddle points of each integral.

There are two qualitatively different cases, depending on where the candidate saddle points for the integral over the Lagrange multipliers lie for each saddle over the string configurations. The first, which is arguably the simplest case to handle, is the $u_{\rm max} = u_+$ saddle at small values of $\Delta \eta$. In this case, for a given pair $(\Delta \eta,\Delta \theta)$ at sufficiently small $\Delta \eta$ there are two candidate saddles, both of which lie on the real axes for the Lagrange multipliers. In effect, they are all the ones described by~\fig{dtheta_deta_plot_real_constants}, from where it is clear (for all contours drawn) that the constant $\Delta \theta$ contours intersect the constant $\Delta \eta$ contours exactly twice. For such values of $\Delta \eta$, both candidate pairs of values of $(C_\eta,C_\theta)$ are picked up by the original  integration contour, and so both contribute without any need to make a contour deformation (albeit only the one with smaller imaginary part is typically kept, as it corresponds to a lower energy state).

Formally, if one does not deform the contour integral and the saddles lie on the real axis, the integral is carried out via the stationary phase method instead of the steepest descent method. This is sufficient for the saddles with no exponential suppression. For the others, because of their exponential suppression relative to their value on the original integration contour, it is important to show that the saddles can be reached via a steepest descent path, so that there is a global bound on the magnitude of the integrand --- and hence the saddle is indeed a good approximation of the integral at large $\sqrt{\lambda}$.

\begin{figure}
    \centering
    \includegraphics[width=0.47\linewidth]{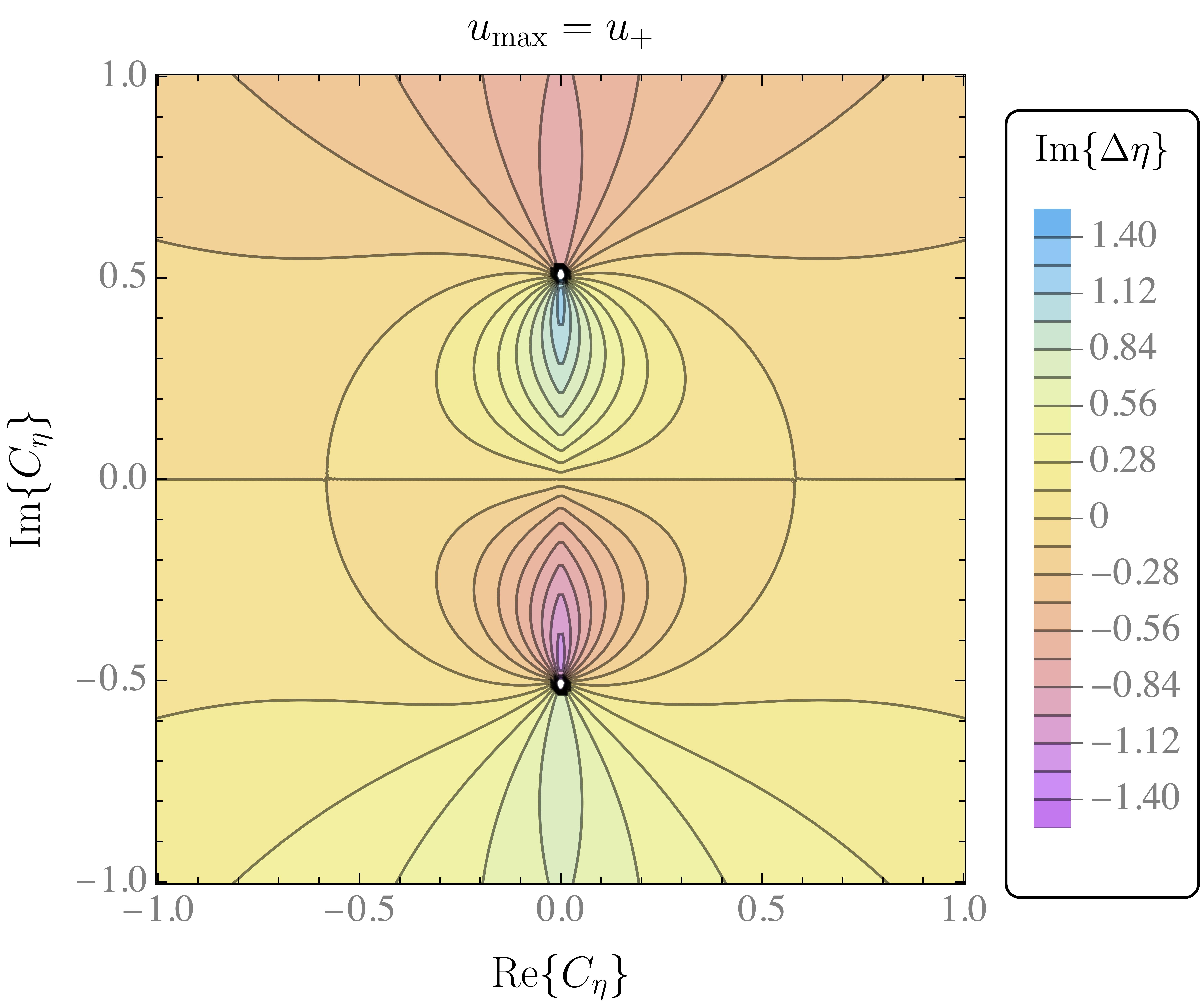}
    \includegraphics[width=0.47\linewidth]{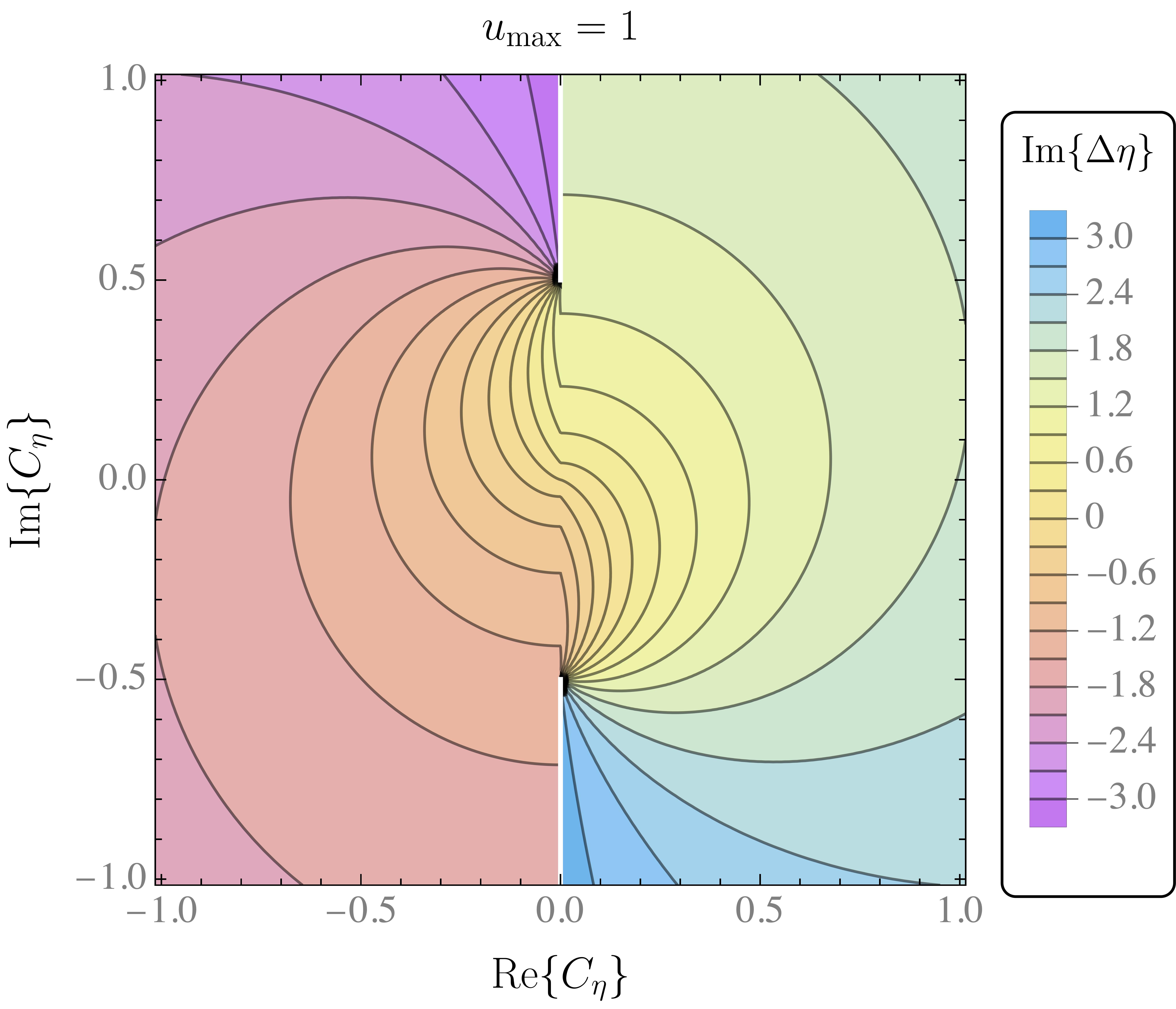}
    \includegraphics[width=0.47\linewidth]{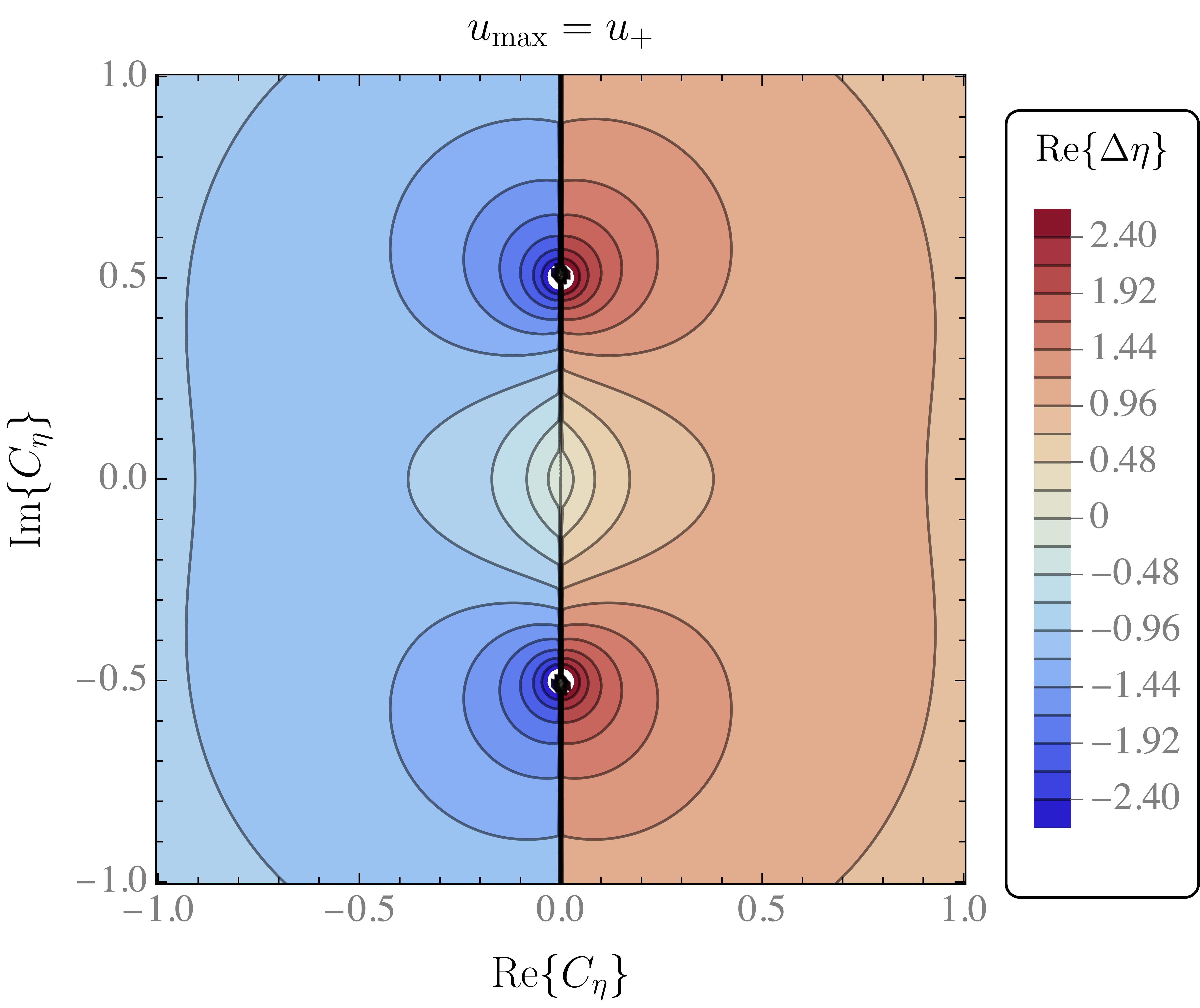}
    \includegraphics[width=0.47\linewidth]{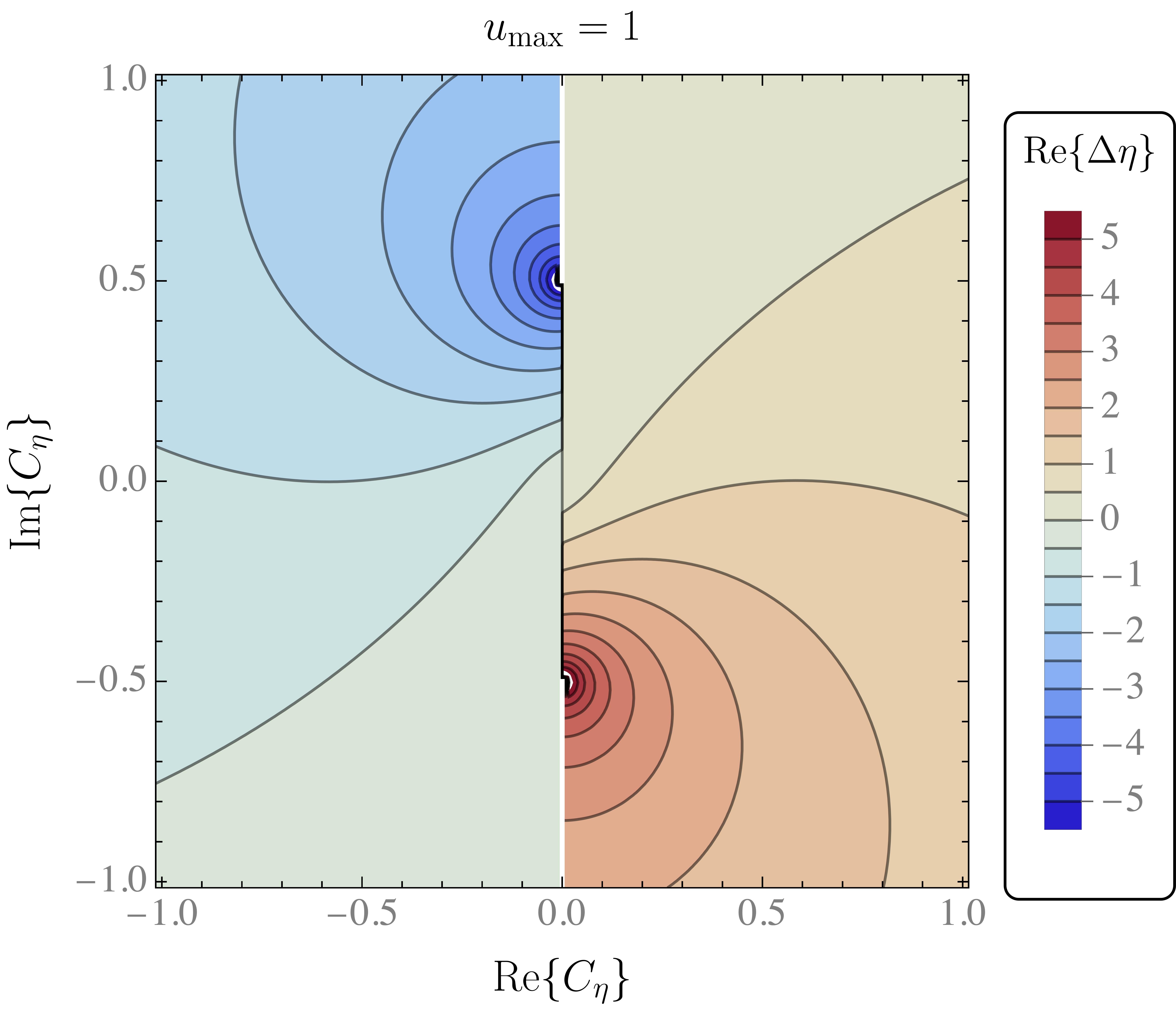}
    \includegraphics[width=0.47\linewidth]{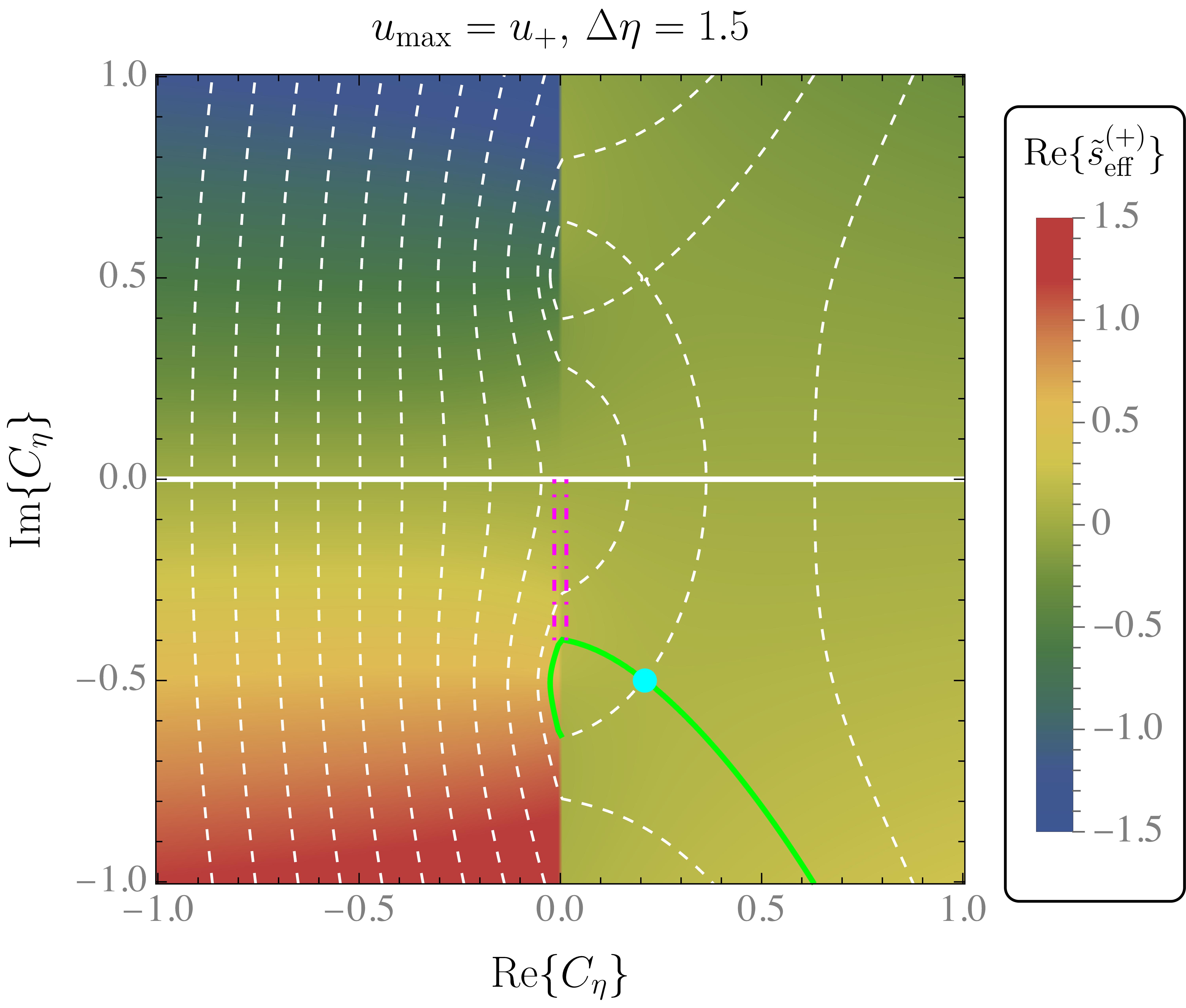}
    \includegraphics[width=0.47\linewidth]{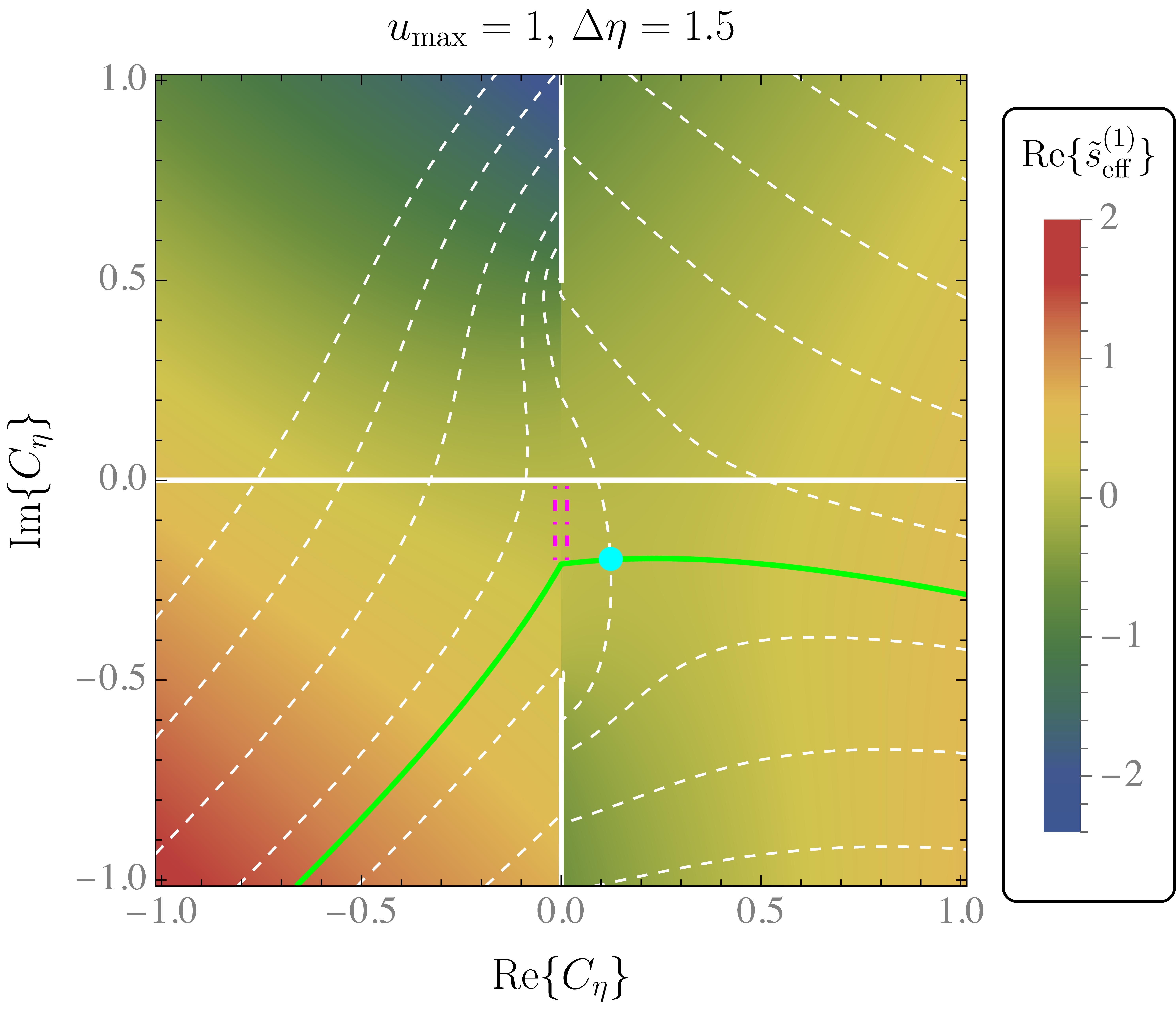}
    \caption{Top panels: contours of constant imaginary part of the expression that determines $\Delta \eta$. Physical solutions lie on the ${\rm Im}\{\Delta \eta\} = 0$ contour level. Middle panels: contours of constant real part of the expression that determines $\Delta \eta$. Lower panels: density plot of the real part of the action $\tilde{s}_{\rm eff} = \tfrac{\pi}{\sqrt{\lambda} \ln(\Lambda L)  } \tilde{\mathcal{S}}_{\rm eff} + i C_\eta \Delta \eta/2$, with $\Delta \eta = 3/2$. The white dashed lines are  constant imaginary part contours of the action. The green line is the steepest descent contour. The cyan circle is the contributing saddle point. The white line over the real axis is the original integration contour. Because the action is discontinuous along the imaginary axis (except at the origin), the magenta dot-dashed lines are necessary to close the contour. Left panels: $u_{\rm max} = u_+$. Right panels: $u_{\rm max} = 1$. The white lines along the imaginary axis lie along the discontinuity of ${\rm Im}\{\Delta \eta\}$.}
    \label{fig:contours_umax_uplus_one}
\end{figure}

The second case does exhibit the subtlety we just discussed. Furthermore, in this case a direct search for $(C_\eta,C_\theta)$ candidate saddles for a given pair $(\Delta \eta,\Delta \theta)$ can also have two solutions. However, contrary to the first case, only one candidate saddle actually contributes. When there are two potential saddles, which one contributes is decided by whether the integration contour can be deformed to go over the saddle in a steepest descent path. In practice, it is easy to identify which is which as one leads to an action with a negative real part and the other to an action with a positive real part. The one with a negative real part is inadmissible because one can bound the actual value of the integral by the integral of unity, which is formally smaller than the exponential of a positive number proportional to $\sqrt{\lambda}$. It turns out that the saddle with positive real part coincides with the one that is reachable via an admissible steepest descent contour. We illustrate these considerations in~\fig{contours_umax_uplus_one} with an example for $u_{\rm max} = u_+$ (left panels) and another for $u_{\rm max} = 1$ (right panels). One can see that for $u_{\rm max} = u_+$ there are two candidate saddles for any given $\Delta \eta$, but only one of them contributes because the real part of the action becomes more and more negative going into the upper half of the complex plane. In~\fig{latte} and~\fig{sunset} we only plotted the contributing ones.

There is one additional subtlety in the contour deformation procedure, which turns out to be inconsequential as far as we have checked, illustrated by the magenta dot-dashed lines in the lower panels of~\fig{contours_umax_uplus_one}. Namely, the action actually has a branch cut along the imaginary axis for both cases, and is only continuous from left to right crossing the imaginary axis at the origin. As such, a valid deformation of the original integration contour (the white line over the real axis) must pass through the origin, meaning that if we deform the contour to pass over the saddle point following the steepest descent contour that includes it, the magenta lines must also be included. While the contribution from an infinitesimal vicinity around the origin manifestly cancels (because the function is continuous), there could in principle be a contribution from the rest of these segments. To show that they do not modify the asymptotic behavior given by the saddle point, one has to find an integration contour on the left side of the complex plane (or its analytic continuation past the branch cut) from the origin to a point where the action has a bigger real part than at the saddle point which equals the contribution from the magenta segment on the right side of the complex plane. As of the time of writing of this work, we have not found a simple, systematic way of finding such a contour.
That being said, there is an easier way to verify that the contribution from the saddle point governs the behavior of the integral: to carry out the integration in~\eq{W_Loop_step_5} directly. We do so in~\app{numbers}.

\section{A direct approach to the $C_\eta$ integral in the $\Delta \theta = 0$ case} \label{app:numbers}

\begin{figure}
    \centering
    \includegraphics[width=0.8\linewidth]{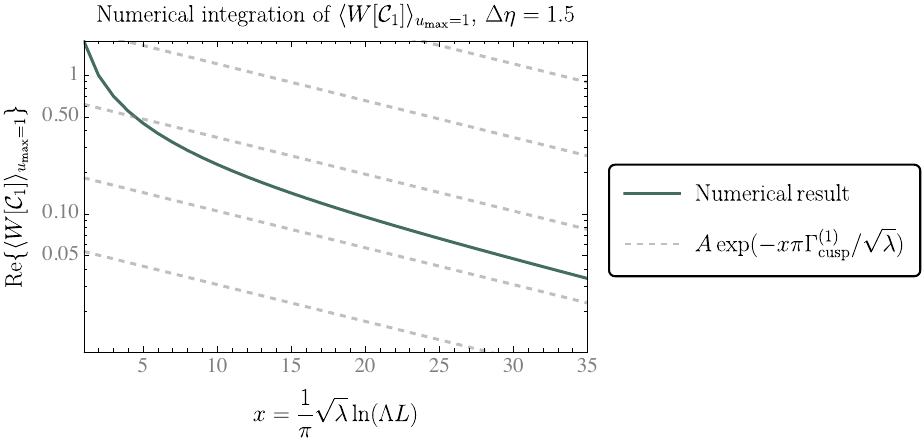}
    \includegraphics[width=0.8\linewidth]{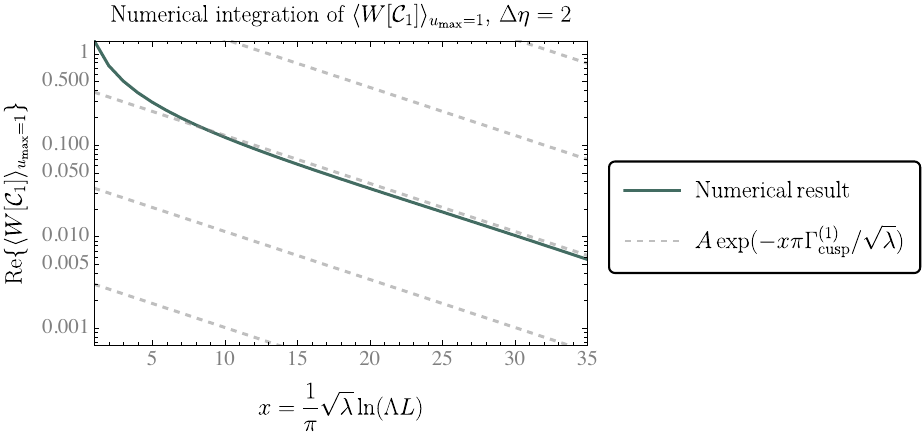}
    \caption{Comparison of the real part of the numerical evaluation of the integral~\eq{W_Loop_step_6_app} (solid line) with the ``prediction'' from the saddle point approximation calculation, represented by the slope of the gray dashed lines. Top: $\Delta \eta = 1.5$. Bottom: $\Delta \eta = 2$.}
    \label{fig:numerical_real_part}
\end{figure}

\begin{figure}
    \centering
    \includegraphics[width=0.9\linewidth]{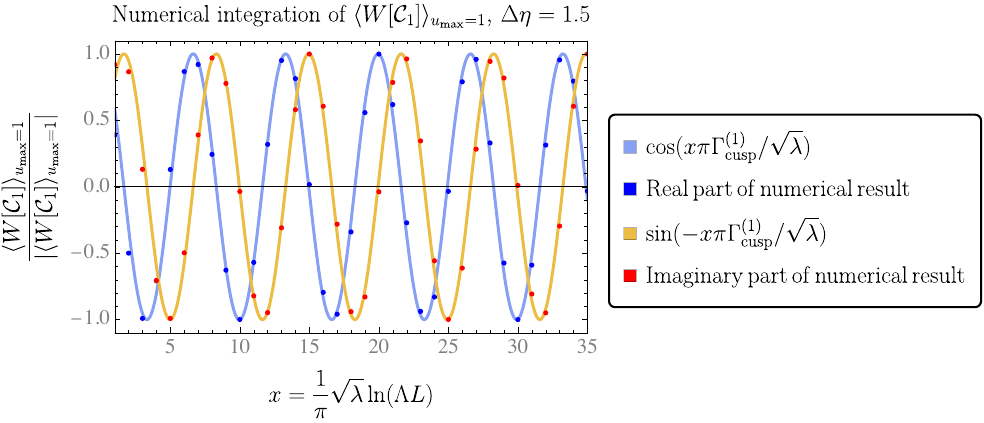}
    \includegraphics[width=0.9\linewidth]{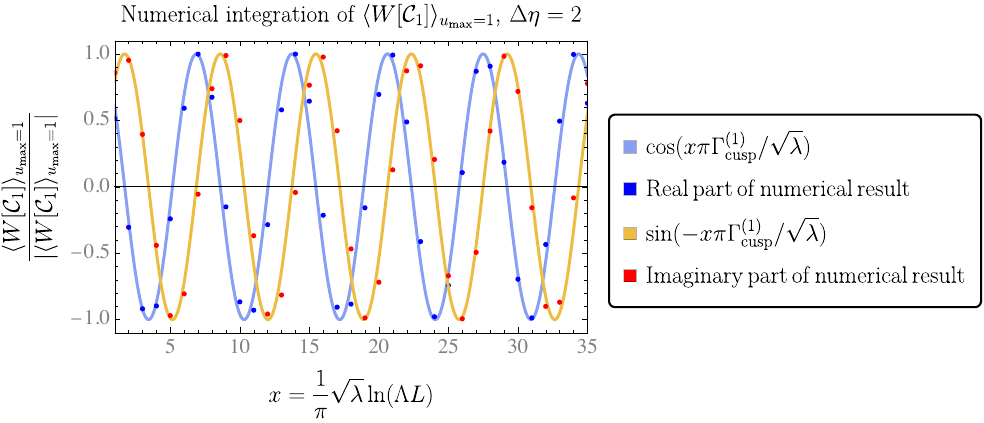}
    \caption{Comparison of the phase of the integral~\eq{W_Loop_step_6_app} evaluated numerically (points) with the ``prediction'' from the saddle point approximation calculation (solid lines), represented by the slope of the gray dashed lines. The quantity $A$ in the legend is a constant that takes different values for each dashed line. Top: $\Delta \eta = 1.5$. Bottom: $\Delta \eta = 2$.}
    \label{fig:numerical_phase_part}
\end{figure}

In this Appendix we present a complementary approach to the calculation of the Wilson loop, starting from~\eq{W_Loop_step_5}. We focus on the case with $\Delta \theta = 0$, which corresponds to setting $C_\theta = 0$. As such, one is left with the one-dimensional integral
\begin{align}
    \left\langle W[\mathcal{C}_1] \right\rangle 
    &= \int_{-\infty}^\infty dC_\eta  \left[ \exp \left( -  \tilde{\mathcal S}^{(+)}_{{\rm eff}}(C_\eta,C_\theta = 0 ) \right) + \exp \left( -  \tilde{\mathcal S}^{(1)}_{{\rm eff}}(C_\eta,C_\theta = 0 ) \right) \right] \, . \label{eq:W_Loop_step_5_app}
\end{align}
While the $u_{\rm max} = u_+$ saddle has an integrand with unit absolute value along the real axis, the action from the $u_{\rm max} = 1$ saddle yields an integrand which is exponentially decaying along the real axis. Therefore, we may study it directly via numerical integration, without any contour deformation.

In~\fig{numerical_real_part} and~\fig{numerical_phase_part} we show the result of carrying out the integral
\begin{align}
    \left\langle W[\mathcal{C}_1] \right\rangle_{u_{\rm max}=1} 
    &= \int_{-\infty}^\infty dC_\eta \, \exp \left( -  \tilde{\mathcal S}^{(1)}_{{\rm eff}}(C_\eta,C_\theta = 0 ) \right) \, , \label{eq:W_Loop_step_6_app}
\end{align}
for $\Delta \eta = 1.5$ and $\Delta \eta = 2$. We can see the ``prediction'' from the saddle point contribution describes the value of the integral very well. In practice, going to larger and larger values of $ \sqrt{\lambda} \ln( \Lambda L) \Delta \eta$ becomes numerically more and more difficult due to the need to evaluate an integrand given by the exponential of bigger and bigger numbers.

While the integral over $C_\eta$ of the saddle at $u = u_+$ cannot be carried out without deforming the integration contour, and while we do not explore this further here, we note that it is also possible to study it numerically after a contour deformation that makes the integrand decay exponentially towards infinity.

\addcontentsline{toc}{section}{References}
\bibliographystyle{jhep}
\bibliography{refs}

\end{document}